\documentclass[traditabstract]{aa} 
\pdfoutput=1
\usepackage{graphicx}
\usepackage{subfig}
\usepackage{natbib}
\usepackage{subfig}
\usepackage{tabularx}
\usepackage{lscape}
\usepackage{color}
\usepackage{rotating}
\usepackage{threeparttable}
\usepackage{ulem}
\bibpunct{(}{)}{;}{a}{}{,}

\usepackage[colorlinks=true, citecolor=blue, linkcolor=blue, breaklinks=true]{hyperref}

\usepackage[fleqn]{amsmath} 
\usepackage{amsfonts} 
\usepackage{amstext}
\usepackage{amssymb} 

\newcommand{\U}[1]{\ensuremath{\mathrm{#1}}}

\newcommand{\kpc}{\U{kpc}}

\newcommand{\msun}{\U{M}_{\odot}}
\newcommand{\Lsun}{\U{L}_{\odot}}
\newcommand{\kms}{\U{km\,s^{-1}}}

\newcommand{\cmd}{\U{cm^{-2}}}

\newcommand{\OIIa}{\mbox{[OII]$_{\lambda 3727}$}}
\newcommand{\OIIIa}{\mbox{[OIII]$_{\lambda 4959}$}}
\newcommand{\OIIIb}{\mbox{[OIII]$_{\lambda 5007}$}}
\newcommand{\Ha}{\mbox{H$_{\alpha}$}}
\newcommand{\Hb}{\mbox{H$_{\beta}$}}
\newcommand{\Hg}{\mbox{H$_{\gamma}$}}
\newcommand{\Hd}{\mbox{H$_{\delta}$}}
\newcommand{\NIIa}{\mbox{[NII]$_{\lambda 6548}$}}
\newcommand{\NIIb}{\mbox{[NII]$_{\lambda 6584}$}}
\newcommand{\SIIa}{\mbox{[SII]$_{\lambda 6717}$}}

\newcommand{\HII}{H\,{\sc ii} \,}
\newcommand{\HIIA}{H\,{\sc ii}}
\newcommand{\NII}{[N\,{\sc ii}]}
\newcommand{\OII}{[O\,{\sc ii}]}
\newcommand{\OIII}{[O\,{\sc iii]}}

\newcommand{\XCO}{{X}_{\rm CO}}

\newcommand{\hi } {{\rm H}\,{\tiny \rm I} \,}

\begin{document}

\title{Gas dynamics in tidal dwarf galaxies: disc formation at $z=0$\thanks{Based on observations made with ESO telescopes at Paranal Observatory under programme 65.O-0563, 67.B-0049, and  083.B-0647.}}
\titlerunning{Gas dynamics in tidal dwarf galaxies}
\author{Federico~Lelli\inst{1}
\and Pierre-Alain~Duc\inst{2}
\and Elias~Brinks\inst{3}
\and Fr\'ed\'eric~Bournaud\inst{2}
\and Stacy~S.~McGaugh\inst{1}
\and Ute~Lisenfeld\inst{4}
\and Peter~M.~Weilbacher\inst{5}
\and M\'ed\'eric~Boquien\inst{6,7}
\and Yves~Revaz\inst{8}
\and Jonathan~Braine\inst{9}
\and B\"{a}rbel~S.~Koribalski\inst{10}
\and Pierre-Emmanuel~Belles\inst{2,3}}
\authorrunning{F. Lelli et al.}
\institute{Astronomy Department, Case Western Reserve University, 10900 Euclid Avenue, Cleveland, OH 44106, USA \\
\email{federico.lelli@case.edu}
\and Laboratoire AIM, CNRS, CEA/DSM, Universit\'e Paris Diderot, F--91191 Gif-sur-Yvette, France
\and Centre for Astrophysics Research, University of Hertfordshire, College Lane, Hatfield, AL10 9AB, UK
\and Departamento de F\'isica Te\'orica y del Cosmos, Universidad de Granada, Granada, Spain
\and Leibniz--Institut f\"ur Astrophysik Potsdam (AIP), An der Sternwarte 16, D--14482 Potsdam, Germany
\and Laboratoire d'Astrophysique de Marseille, Observatoire Astronomique Marseille--Provence, Universit\'e de Provence \& CNRS, 2 Place Le Verrier, 13248 Marseille Cedex 4, France
\and Unidad de Astronomia, Facultad de Ciencias Basicas, Universidad de Antofagasta, Avenida Angamos 601, Antofagasta, Chile
\and Laboratoire d'Astrophysique, \'Ecole Polytechnique F\'ed\'erale de Lausanne (EPFL), 1290 Sauverny, Switzerland
\and Laboratoire d'Astrophysique de Bordeaux, Observatoire de Bordeaux, 2 rue de l'Observatoire, 33270 Floirac, France
\and Australia Telescope National Facility, CSIRO, PO Box 76, Epping, NSW 1710, Australia}

\date{Received 201? ?? ??; Accepted 201? ?? ??}
 
\abstract{
Tidal dwarf galaxies (TDGs) are recycled objects that form within the collisional debris of interacting/merging galaxies. They are expected to be devoid of non-baryonic dark matter, since they can form only from dissipative material ejected from the discs of the progenitor galaxies. We investigate the gas dynamics in a sample of six bona-fide TDGs around three interacting and post-interacting systems: NGC~4694, NGC~5291, and NGC~7252 (``Atoms for Peace''). For NGC~4694 and NGC~5291 we analyse existing \hi data from the Very Large Array (VLA), while for NGC~7252 we present new \hi observations from the Jansky VLA together with long-slit and integral-field optical spectroscopy. For all six TDGs, the \hi emission can be described by rotating disc models. These \hi discs, however, have undergone less than a full rotation since the time of the interaction/merger event, raising the question of whether they are in dynamical equilibrium. Assuming that these discs are in equilibrium, the inferred dynamical masses are consistent with the observed baryonic masses, implying that TDGs are devoid of dark matter. This puts constraints on putative ``dark discs'' (either baryonic or non-baryonic) in the progenitor galaxies. Moreover, TDGs seem to systematically deviate from the baryonic Tully-Fisher relation. These results provide a challenging test for alternative theories like MOND.}
 

\maketitle

\section{Introduction}
\label{sec:intro}

During the collision of two disc galaxies, tidal forces strip material out of the discs and form long tails of gas and stars \citep{Toomre72}. \citet{Zwicky56} was the first to suggest that gas can concentrate in the tidal debris and collapse under self-gravity, leading to the formation of a new object: a tidal dwarf galaxy (TDG). Tidal debris around galaxy mergers often contains gas and stellar condensations that have masses, sizes, and star-formation rates (SFRs) comparable to those of dwarf galaxies \citep[e.g.][]{Weilbacher00, Boquien10, Dabringhausen13}. To be considered as \textit{bona-fide} TDGs, these condensations must be \textit{self-gravitating} entities that can survive for several Gyr against internal or external disruption \citep[e.g.][]{Duc00, Bournaud06, Recchi07, Ploeckinger14, Ploeckinger15}. For example, they should show a velocity gradient that is kinematically decoupled from the surrounding tidal debris, pointing to a local gravitational potential well. Such kinematic decoupling has actually been observed in some TDG candidates, e.g. \citet[][hereafter B07]{Bournaud07} and \citet[][hereafter D07]{Duc07}. Since TDGs are formed out of material pre-enriched in the progenitor galaxies, they are found to have higher metallicities than typical dwarfs \citep{Duc98b, Weilbacher03} and are often detected by CO observations \citep{Braine01, Duc07, Boquien11}. The systematic deviation of young TDGs from the mass-metallicity relation has also been used to identify older TDG candidates, which are no longer embedded in collisional debris \citep{Hunter00, Sweet14, Duc14}.

\begin{table*}
\centering
\caption{Galaxy Sample}
\resizebox{18cm}{!}{
\begin{tabular}{lcccccll}
\hline
System   & RA         & Dec         & V$_{\rm sys}$ & Dist  & $L_{\rm K}$& Short Descrption                      & Bona-fide TDGs \\
         & (J2000)    & (J2000)     & ($\kms$)      & (Mpc) & (10$^{10}$~$L_{\odot}$)&                                       &      \\ 
\hline
NGC 4694 & 12 48 15.1 & $+$10 59 01 & 1160          & 17    & 1.5 & disturbed lenticular: post-merger     & VCC~2062 \\
NGC 5291 & 13 47 24.5 & $-$30 24 25 & 4378          & 62    & 14  & peculiar lenticular: head-on collision& NGC 5291N, 5291S, 5291SW\\
NGC 7252 & 22 20 44.7 & $-$24 40 42 & 4792          & 66.5  & 17  & advanced merger: spiral+spiral        & NGC~7252E, 7252NW \\
\hline
\end{tabular}
}
\label{tab:sample}
\end{table*}
A key aspect of TDGs is that they are expected to be free of non-baryonic dark matter \citep[DM;][]{Barnes92, Elmegreen93}. This occurs because of two basic dynamical principles: (i) tidal forces have different effects on the dynamically-cold disc and on the dynamically-hot halo, thus they effectively segregate baryons in the disc (which end up forming tails, debris, and TDGs) from DM in the halo (which is too hot to form narrow tails); and (ii) once a TDG has formed, it has a shallow potential well with a typical escape velocity of a few tens of km~s$^{-1}$, thus it cannot accrete dynamically-hot DM particles with velocity dispersions of $\sim$200 km~s$^{-1}$. High-resolution numerical simulations, including gas and star formation, confirm that TDGs should be nearly devoid of non-baryonic DM \citep{Duc04, Bournaud06, Wetzstein07}.

Few observational attempts have been made to measure the DM content of TDGs. \citet{Braine01} used CO linewidths to estimate the dynamical masses of eight TDGs and found no evidence of DM. These CO observations do not resolve the gas distribution and kinematics, thus it remains unclear whether the CO linewidths can be interpreted as rotation in a local potential well. B07 presented high-resolution \hi observations of the collisional ring around NGC~5291, and identified three TDGs\footnote{Stricktly speaking, the recycled objects around NGC~5291 are not tidal dwarf galaxies because they are not formed in tidal debris but in a collisional ring. For simplicity, we refer to them as TDGs since they are expected to share similar properties.} that are rotating and gravitationally bound. Assuming that these TDGs and the collisional ring have the same inclination angle (as inferred from numerical models), B07 measured dynamical masses a factor of $\sim$3 higher than the baryonic masses, suggesting that TDGs may contain some unconventional type of DM (either baryonic or non-baryonic). They hypothesized that the missing mass consists of very cold molecular gas, which is not traced by CO observations and is not actively forming stars, but that may constitute a large fraction of \textit{baryonic} DM in galaxy discs \citep[][]{Pfenniger94a}. On the other hand, \citet{Milgrom07} and \citet{Gentile07b} showed that MOdified Newtonian Dynamics (MOND) could reproduce the observed rotation velocities of these TDGs, providing an alternative explanation to the missing mass without any need of DM. As pointed out by \citet{Kroupa12}, TDGs may help to discriminate between MOND and the standard model of cosmology, given that these two paradigms predict different internal dynamics for these recycled galaxies.

\begin{figure*}
\centering
\includegraphics[width=\textwidth]{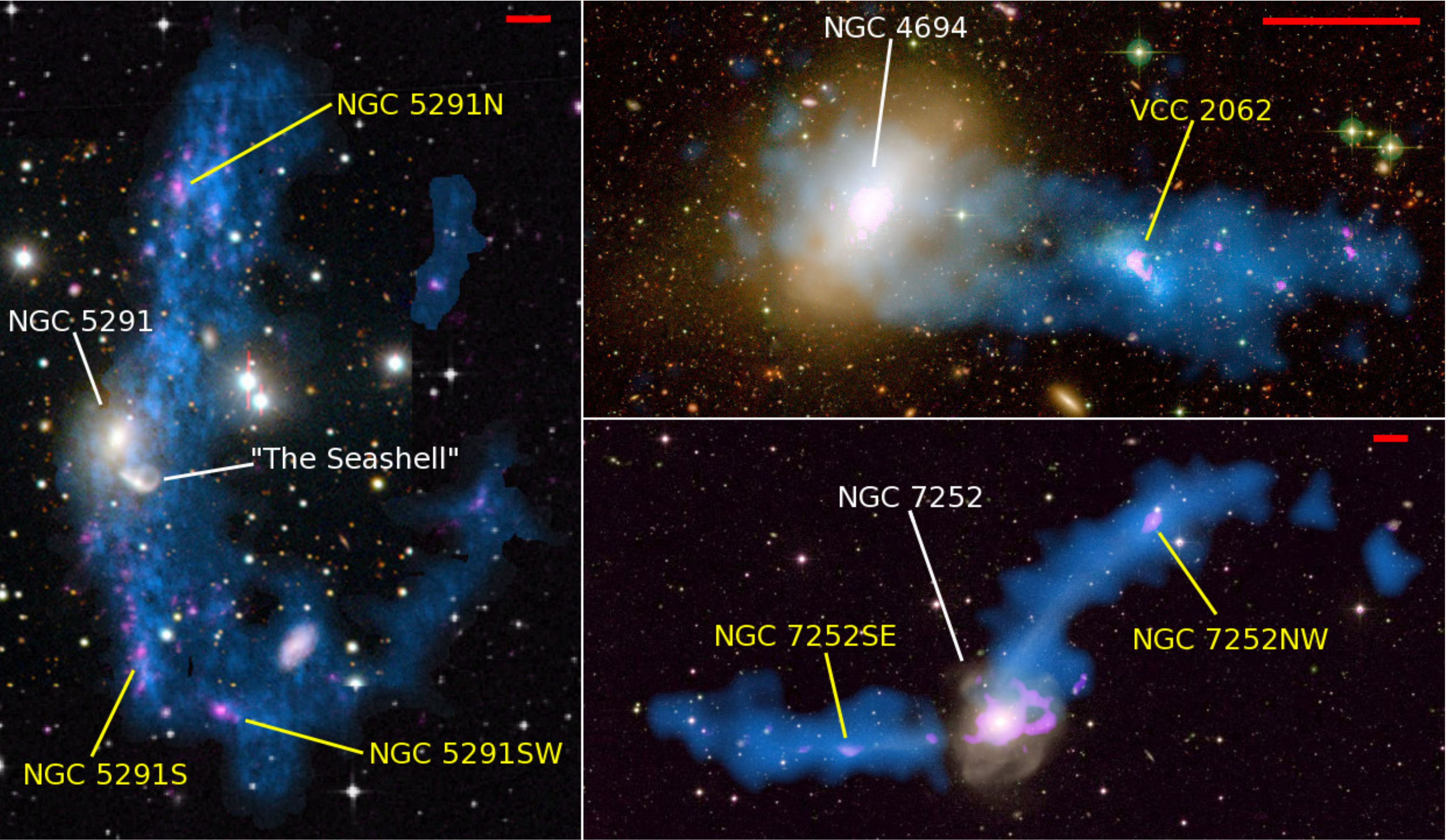}
\caption{Optical images of the selected systems overlaid with FUV emission from GALEX (pink) and \hi emission from the VLA (blue). TDGs are indicated in yellow. North is up; East is left. The bar to the top-right corner corresponds to $\sim$10 kpc for the assumed distances (as given in Table~\ref{tab:sample}). NGC~5291 (\textit{left}): composite ESO-NTT ($BVR$) and SDSS image from B07. NGC~4694 (\textit{top-right}): MegaCam/CFHT image obtained as part of NVGS \citep{Ferrarese12}. NGC~7252 (\textit{bottom-right}): composite ESO-MPG ($BR$) image from \citet{Baade99}.}
\label{fig:color}
\end{figure*}

In this paper we present a systematic study of a sample of six \textit{bona-fide} TDGs around three nearby systems: the late-stage merger NGC~7252 (``Atoms for Peace''), the collisional galaxy NGC~5291, and the disturbed lenticular NGC~4694 (see Fig.~\ref{fig:color}). Specifically, we build 3D disc models to study the \hi kinematics of the TDGs and constrain their inclination angles. These techniques have been widely used and validated for typical dwarf galaxies \citep{Swaters99, Swaters09, Gentile07a, Gentile10, Lelli12a, Lelli12b, Lelli14}. For NGC~7252, we present new \hi data from the Karl G. Jansky Very Large Array (VLA) together with long-slit and integral-field optical spectroscopy from the Very Large Telescope (VLT). For NGC~5291 and NGC~4694, we re-analyse existing \hi data from the VLA. In particular, we revisit the DM content estimated by B07 for the TDGs around NGC~5291.

\section{Galaxy sample}

Table~\ref{tab:sample} lists the basic properties of the systems in our sample and their bona-fide TDGs. All six TDGs satisfy the following requirements: (i) they are ``genuine'' condensations of gas and young stars, which are \textit{not} due to projection effects along the debris \citep[see e.g.][]{Bournaud04}; (ii) they have higher nebular metallicities than galaxies of similar mass, indicating that they are not pre-existing dwarf galaxies; (iii) they display a velocity gradient that appears kinematically decoupled from the tidal debris, pointing to a local potential well. For the TDGs around NGC~7252, the validity of these requirements is demonstrated in Sect.~\ref{sec:overview}. For the TDGs around NGC~4694 and NGC~5291, we refer to D07 and B07, respectively. Here we provide a concise description of these systems.

\subsection{The disturbed lenticular NGC~4694}
\label{sec:N4694}

NGC~4694 is a disturbed lenticular galaxy that inhabits the outskirts of the Virgo cluster. It shows asymmetric optical isophotes, dust lanes, and signs of recent star formation. A long \hi feature stretches to the West of the galaxy for $\sim$40 kpc and contains a TDG candidate: VCC~2062 (see Fig.~\ref{fig:color}). D07 concluded that VCC~2062 is a recycled galaxy based on the following observations: (i) a high metallicity (12+log(O/H)$\simeq$8.6), (ii) a strong CO detection, and (iii) an \hi velocity gradient that is kinematically decoupled from the debris. Recently, deep optical images have been obtained as part of the Next Generation Virgo Cluster Survey \citep[NGVS,][]{Ferrarese12}, revealing that NGC~4694 has a strongly perturbed halo and a prominent tidal feature to the West (Lisenfeld et al. in prep.). This confirms that NGC~4694 is the result of an old merger ($\gtrsim$0.5~Gyr ago). The \hi kinematics of VCC~2062 is modelled in Sect.~\ref{sec:3Dmodels}.

\subsection{The collisional galaxy NGC~5291}
\label{sec:N5291}

NGC~5291 is a peculiar early-type galaxy that lies in the outskirts of the cluster Abell~3574. It is surrounded by a giant \hi ring with a diameter of $\sim$160~kpc and \hi mass of $\sim$5$\times$10$^{10}$~$M_{\odot}$ \citep[][]{Malphrus97}. A disturbed galaxy (the ``Seashell'') lies at a projected distance of $\sim$13~kpc (see Fig.~\ref{fig:color}), but is a fly-by object with a higher systemic velocity (by $\sim$400~$\kms$, \citealt{Duc98b}). According to the numerical model of B07, the \hi ring was formed during a head-on collision $\sim$360~Myr ago. Notably, B07 found that the ``Seashell'' galaxy is not massive enough to reproduce the \hi ring and identified the elliptical IC~4329 as a more likely interloper. The \hi ring hosts several TDG candidates with nearly solar metallicities \citep{Duc98b}. The three largest TDG candidates (NGC~5291N, 5291S, 5291SW) are associated with \hi velocity gradients that are kinematically decoupled from the underlying debris (B07). In Sect.~\ref{sec:3Dmodels} we model the \hi kinematics of these TDGs.

\subsection{The advanced merger NGC~7252}
\label{sec:N7252}

NGC~7252 (``Atoms for Peace'') is a prototype late-stage merger from the Toomre sequence of interacting galaxies \citep{Toomre77}. Numerical models have been able to reproduce its morphology and kinematics by simulating the merger of two disc galaxies falling together $\sim$600-700 Myr ago \citep{Hibbard95, Chien10}. The central remnant has a single nucleus and a de Vaucouleurs luminosity profile, suggesting the morphological transformation into an elliptical galaxy \citep{Schweizer82, Hibbard94}. Using VLA observations, \citet{Hibbard94} identified two TDG candidates along the tidal tails, but the spatial and velocity resolutions of the \hi data were too low to study their internal kinematics. In this paper we show that (i) both TDG candidates are associated with a steep \hi velocity gradient that represents a kinematic discontinuity along the tail (Sect.~\ref{sec:N7252kin}), (ii) both galaxies have relatively high metallicities, confirming that they are not pre-existing dwarfs but are forming out of pre-enriched tidal material (Sect.~\ref{sec:N7252metal}), and (iii) the TDG to the North-West has a complex H$\alpha$ kinematics, which is likely dominated by turbulent motions (Sect.~\ref{sec:N7252ha}). For both TDGs we present 3D disc models in Sect.~\ref{sec:3Dmodels}.

\section{Observations and data analysis}
\label{sec:observations}

In this section we describe the collection, reduction, and analysis of new multiwavelength data of NGC~7252: \hi observations from the NRAO-VLA\footnote{The National Radio Astronomy Observatory is a facility of the National Science Foundation operated under cooperative agreement by Associated Universities, Inc.} (Sect.~\ref{sec:HIobs}) and optical spectroscopy using both FORS1 multi-object spectrograph and GIRAFFE integral-field unit (IFU) at ESO-VLT (Sect.~\ref{sec:OptObs}). In Appendix~\ref{app:ArchObs} we compare the new \hi data with archival VLA data from \citet{Hibbard94} and single-dish observations. For the existing VLA data of NGC~5291 and NGC~4694, we refer to B07 and D07, respectively.

\subsection{\hi observations}
\label{sec:HIobs}

Spectral-line \hi observations were carried out with multiple configurations of the VLA. We used the hybrid DnC, BnC, and BnA configurations to compensate for the foreshortening due to the low declination of NGC~7252 ($\delta = -24\fdg7$). For simplicity, we will refer to them as D, C, and B configurations. In order to achieve a velocity resolution of 2.6 $\kms$ across a velocity range of $\sim$500 $\kms$, we used the correlator in 2AD mode covering a 1.56~MHz-wide band with 128 spectral channels and tuning the two intermediate frequencies (IFs) at different central velocities. One IF was set at $\sim$4825 $\kms$, while the other one was set at $\sim$4570 $\kms$. The two IFs were merged during the data reduction to form a single band centered at 4700 $\kms$. The VLA observing parameters are given in Table~\ref{tab:VLAobs}.

\begin{table}
\centering
\caption{VLA observing parameters for NGC~7252}
\resizebox{9cm}{!}{
\begin{tabular}{lccc}
\hline
Array configuration & DnC               & CnB               & BnA\\
Observations Date   & Jun 2008          & Feb 2008          & Sep--Oct 2007\\
Bandwidth (MHz)     & 1.56 ($\times 2$) & 1.56 ($\times 2$) & 1.56 ($\times 2$)\\
Number of channels  & 128 ($\times 2$)  & 128 ($\times 2$)  & 128 ($\times 2$)\\
Time on source (hr) & 2.9               & 5.0               & 16.0\\
\hline
\end{tabular}
}
\label{tab:VLAobs}
\end{table}

The \textit{uv}--data were calibrated and combined using AIPS \citep{Greisen03} and following standard procedures. Self-calibration was required for each configuration to remove low-level artefacts due to a 1.16 Jy-bright source (PKS~2217-251) located to the South of the field at ($\alpha_{2000}$, $\delta_{2000}$) = (22$^{h}$ 20$^{m}$ 37.65$^{s}$, $-$24$^{\circ}$53$^{'}$58.0$^{''}$). This continuum source was subtracted using {\sc uvsub}, after combining the B+C+D data and merging the two IFs. The residual radio continuum was fitted with a first order polynomial considering line-free channels, which were determined by inspecting channel maps based on the D+C data. The continuum was then removed from the B+C+D data using {\sc uvlsf}. Some faint residual emission is still observable in several channels at the location of the bright source; this emission was systematically discarded during the masking of the channel maps (described below). The \textit{uv}--data were mapped using multi-scale cleaning \citep{Cornwell08, Greisen09} and different weighting schemes: (i) robust (RO) weighting \citep{Briggs95} with the RO parameter set to 0.5, giving a beam size of 15.8$''$~$\times$~10.7$''$; and (ii) natural (NA) weighting, giving a beam size of 21.1$''$~$\times$~17.8$''$.

After the Fourier transform, the \hi cubes were analysed using the Groningen Imaging Processing Systems \citep[Gipsy,][]{Vogelaar01}. Total \hi maps were built by summing masked channel maps. After various trials, we used a mask obtained by smoothing the cubes to $25''$ in space and 12.5~$\kms$ in velocity, and clipping at 2.5$\sigma_{\rm s}$ (where $\sigma_{\rm s}$ is the rms noise in the smoothed cube). Velocity fields were built from masked channel maps estimating an intensity-weighted mean velocity. Since the \hi line profiles are complex and asymmetric, these velocity fields provide only an overall view of the \hi kinematics. Our kinematical analysis is based on full 3D models of the observations, as we describe in Sect.~\ref{sec:3Dmodels}.

\subsection{Optical spectroscopy}
\label{sec:OptObs}

\begin{figure}
\centering
\includegraphics[angle=-90, width=0.45\textwidth]{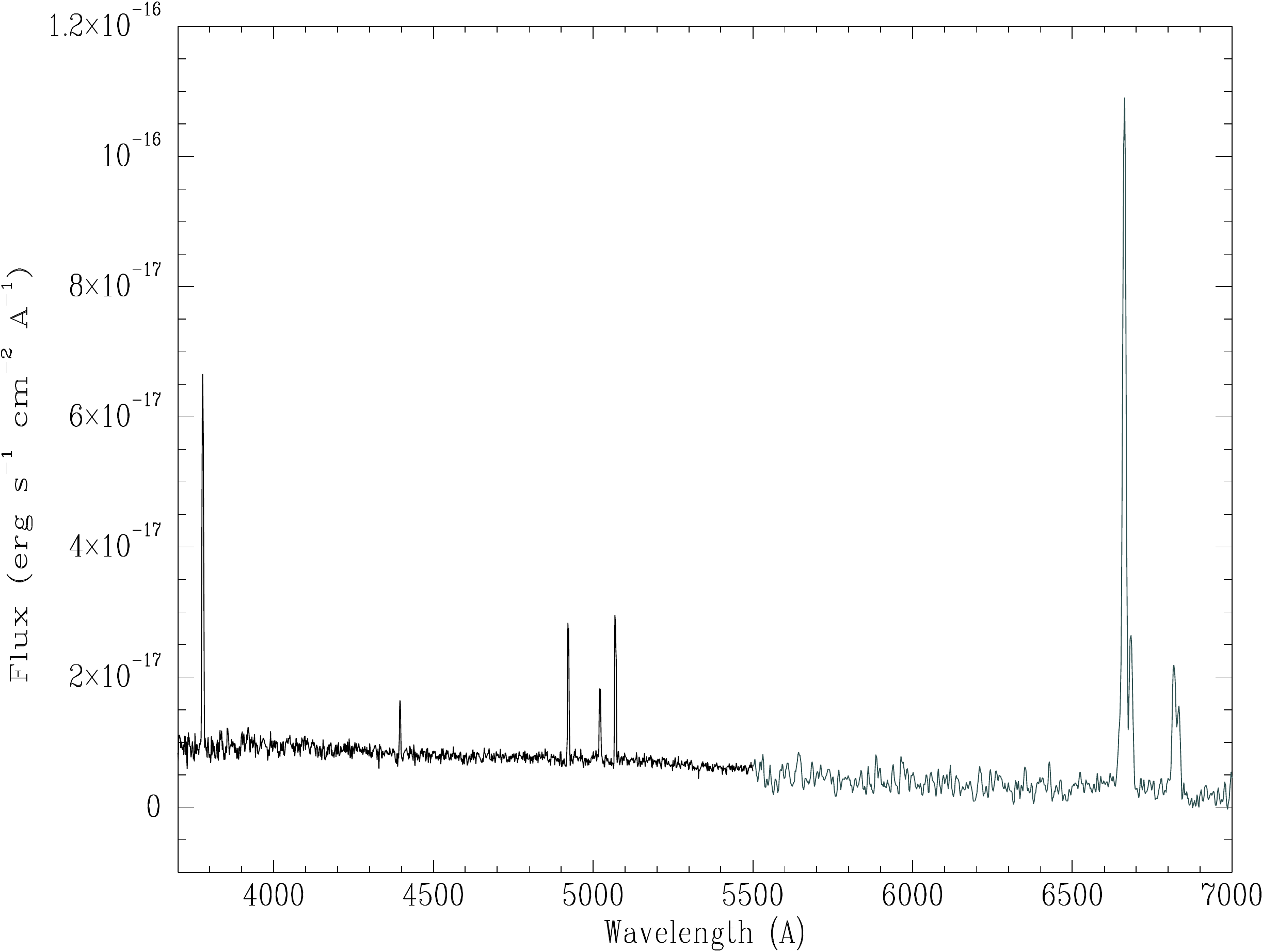}
\caption{
Optical spectrum of NGC 7252NW obtained with VLT/FORS1 (black, $\lambda \le 5500$~\AA) and NTT/EMMI (grey, $\lambda > 5500$~\AA). The FORS1 spectrum has been rescaled to match the H$\beta$ line flux measured from the NTT spectrum.}
\label{fig:FORS_EMMI}
\end{figure}
Optical spectroscopy was obtained as part of several observing runs at ESO facilities. For both NGC~7252NW and NGC~7252E, low-resolution long-slit spectra (covering the 4500-7500~\AA\ wavelength domain) were obtained in July 1994 with the red channel of the EMMI camera installed on the New Technology Telescope (NTT) at ESO/La Silla. Details on observing conditions and data reduction are given in \citet{Duc98b}. The spectrum of NGC~7252NW is shown in Fig.~\ref{fig:FORS_EMMI} (grey line).

New observations were carried out in September 2000 with the FORS1 multi-object spectrograph installed on the Unit Telescope 1 (UT1) of the VLT. We simultaneously obtained the spectra of 19 objects, including NGC~7252E and NGC~7252NW. The position angle of the slit-lets was fixed to $72.9^{\circ}$. We used grism 600B (1.2\,\AA\ per pixel) with a total exposure time of 4$\times$1800 sec. The resulting spectrum has higher spectral resolution and sensitivity than the EMMI one, but only covers the wavelength domain 3700-5500~\AA. Data reduction was carried out using IRAF. The spectrum of NGC\,7252NW is shown in Fig.~\ref{fig:FORS_EMMI} (black line). Spectrophotometric data and resulting metallicities are discussed in Sect.~\ref{sec:N7252metal} and listed in Table~\ref{tab:Oabun}. As is common practice, the line intensities are normalized relative to \Hb=100 and the errors are estimated using the IRAF splot procedure which relies on Poisson statistics.

\begin{table}
\centering
\caption{VLT observing parameters for NGC~7252}
\label{tab:IFUobs}
\begin{tabular}{lr}
\hline
Date & 22 June - 24 August 2009 \\
Instrument & GIRAFFE/ARGUS \\
Filter name & HR15 \\
Grating $\lambda_{\rm central}$ (nm) & 665 \\
Spectral sampling (nm) & 0.005 \\
Velocity resolution ($\kms$) & 2.3 \\
Field of view & $4.2'' \times 6.6''$ \\
Pixel size & $0.3'' \times 0.3''$ \\
Seeing & $\sim$0.65$''$\\
Integration time (s) &  $55 \times 1365 = 75075$ \\
\hline
\end{tabular}
\end{table}

We also obtained IFU observations of NGC~7252NW to investigate its H$\alpha$ kinematics. We used GIRAFFE on the UT2 of the VLT \citep{Pasquini02}, obtaining 55 exposures of $\sim$23 minutes each. The observations are summarized in Table~\ref{tab:IFUobs}. The data were reduced using the ESO pipeline. Additionally, we corrected each reduced cube to heliocentric velocity, before combining all 55 exposures into a variance-weighted mean cube. This final cube was then inspected and analysed using Gipsy. To isolate the H$\alpha$ emission, we subtracted the continuum in the wavelength range 6650$-$6680~\AA, avoiding \NII\ emission lines. The continuum was determined by fitting a first-order polynomial to line-free channels bordering the H$\alpha$ line. A total H$\alpha$ map was built by integrating the cube in velocity and clipping at 3$\sigma$.

\begin{figure*}
\centering
\includegraphics[width=0.9\textwidth]{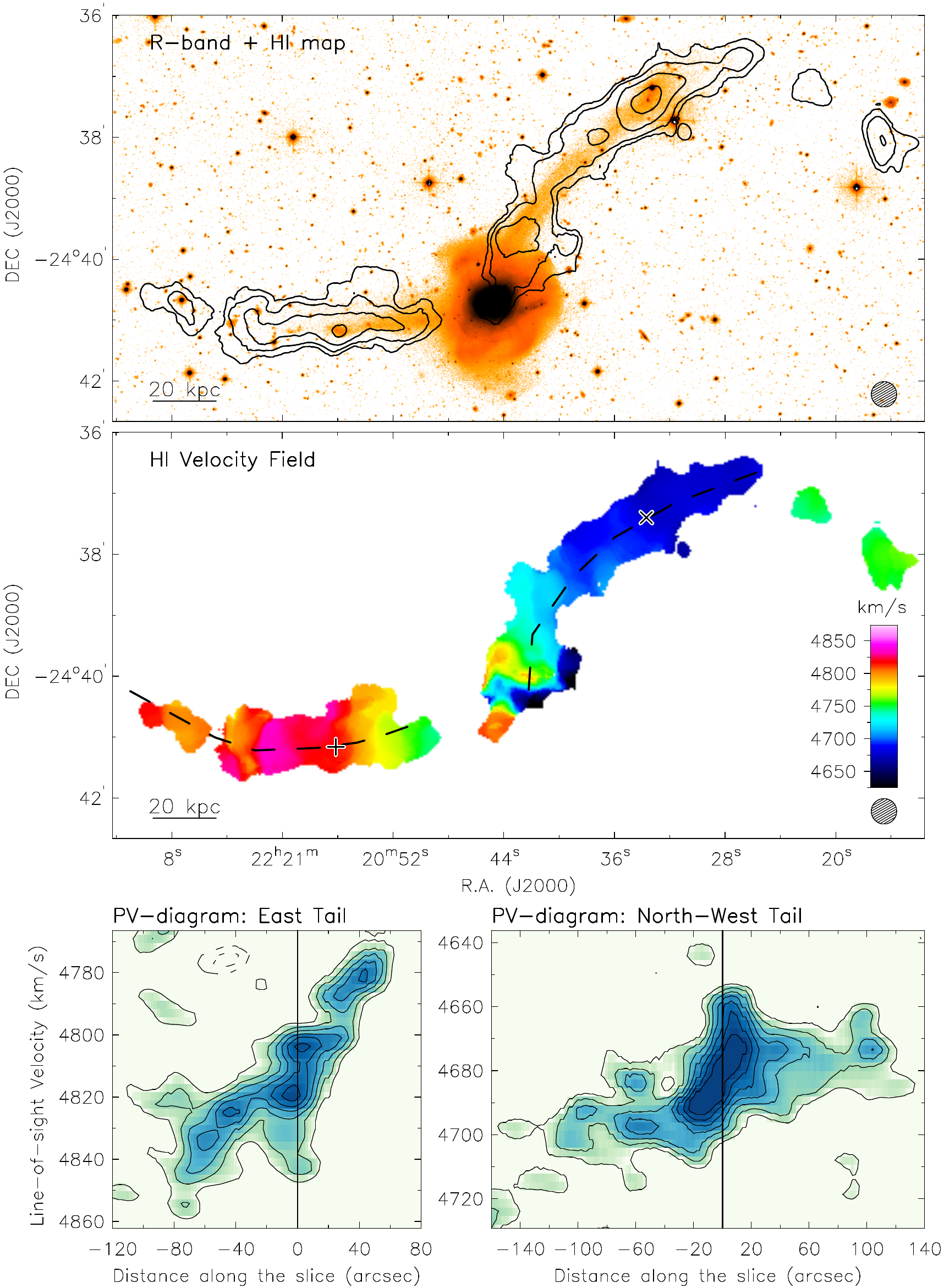}
\caption{Overview of the new \hi observations of NGC~7252. \textit{Top}: total \hi map at 25$''$ resolution overlaid on the ESO~MPG $R$-band image. Contours are at 0.3 ($\sim$3$\sigma$), 0.6, 1.2, and 2.4 M$_{\odot}$~pc$^{-2}$. The \hi beam is shown to the bottom-right. The bar to the bottom-left corresponds to 20 kpc. \textit{Middle}: \hi velocity field at 25$''$ resolution. The crosses mark the location of the putative TDGs. The dashed lines show the path followed to obtain the PV-diagrams. \textit{Bottom}: PV-diagrams along the East tail (\textit{left}) and Nort-West tail (\textit{right}). The vertical lines correspond to the crosses in the velocity field. Contours range from 2$\sigma$ to 8$\sigma$ in steps of $1\sigma = 0.7$ mJy~beam$^{-1}$. See Sect.~\ref{sec:N7252kin} for details.}
\label{fig:overview}
\end{figure*}
\begin{table}
\centering
\caption{Properties of NGC\,7252 and its tidal tails}
\resizebox{9cm}{!}{
\begin{tabular}{lccccc}
\hline
\noalign{\smallskip}
Component& \hi velocities & M$_{\hi}$   & M$_{\rm H_2}$     & $L_{\U{B}}$       \\
         & ($\kms$) & ($10^{9}\,\msun$) & ($10^{9}\,\msun$) & ($10^{9}\,\Lsun$) \\
\hline
Remnant  &4596$-$4618 & 0.25 & 3.5     & 51.6 \\
ET       &4745$-$4835 & 2.0  & -       & 0.9 \\
NWT      &4660$-$4830 & 3.2  & $>0.02$ & 2.9 \\
Global   &4596$-$4837 & 5.45 & $>3.52$ & 55.4 \\
\hline
\end{tabular}
}
\tablefoot{The \hi mass and velocities of the remnant refer to the so-called Western Loop (see Fig.~\ref{completeness} in Appendix~\ref{app:ArchObs}). Molecular masses are calculated assuming $\XCO = 2.0 \times 10^{20}\, (\cmd \, / \, \U{K}\,\kms$) and using the CO fluxes from \citet{Dupraz90} and \citet{Braine01}. For the NWT the molecular mass is a lower limit because single-dish observations cover only a fraction of the tail. $B$-band luminosities are taken from \citet{Hibbard94} and rescaled to a distance of 66.5~Mpc.}
\label{tab:GlobProp}
\end{table}

\section{New results for NGC~7252}
\label{sec:overview}

In the following we describe the results from the new multiwavelength observations of NGC~7252. Specifically, we use (i) \hi data at 25$''$ resolution to study the overall gas distribution and kinematics, (ii) long-slit optical spectroscopy to determine the metallicity of several \HII regions along the tails, and (iii) IFU spectroscopy to investigate the H$\alpha$ kinematics of the TDG to the North-West.

\subsection{\hi distribution and kinematics}
\label{sec:N7252kin}

Fig.~\ref{fig:overview} (top panel) shows an $R$-band image of NGC~7252 overlaid with the total \hi map at 25$''$ resolution. In agreement with \citet{Hibbard94}, we find that the \hi gas is mostly located in the outer regions of the system. The \hi tails closely follow the stellar tails and are more extended than the stellar tails in the available $R$-band image. The East tail (ET) has a linear geometry with a twist in the outer parts, while the North-West tail (NWT) has a curved morphology with detached \hi clouds at its tip. In both tails, large \hi clumps match the location of the putative TDGs, which are identified as blue stellar condensations with \HII regions and UV emission (see Fig.~\ref{fig:color}). None of the TDG candidates are located at the tip of the \hi tails, thus they are not affected by projection effects but represent genuine condensations of gas and stars \citep[cf.][]{Bournaud04}. Other \hi clumps are present in the tails but they are not associated with stellar overdensities.

The middle panel of Fig.~\ref{fig:overview} shows the \hi velocity field at 25$''$ resolution. The ET and NWT display a regular velocity gradient and are, respectively, red-shifted and blue-shifted with respect to the central remnant, having $V_{\rm sys} = 4740 \pm 14$~$\kms$ \citep[from CO observations;][]{Dupraz90}. The velocity field in the NW part of the merger remnant is uncertain because two different kinematic components overlap at the same spatial position along the line of sight (see Fig.~\ref{completeness} in Appendix~\ref{app:ArchObs}): (i) the base of the NWT ranging from $\sim$4700 to $\sim$4850 $\kms$, and (ii) the so-called Western loop \citep[][]{Hibbard94} ranging from $\sim$4500 to $\sim$4650 $\kms$. The masses of these different components are given in Table~\ref{tab:GlobProp}. We associate the Western loop to the remnant as it likely corresponds to tidal material that is falling back on the galaxy \citep{Hibbard95}.

The bottom panels of Fig.~\ref{fig:overview} show position-velocity (PV) diagrams obtained along the two tails (dashed lines on the velocity field). Remarkably, steep \hi velocity gradients are observed at the location of the putative TDGs (corresponding to the vertical lines in the PV diagrams). In the NWT, the \hi velocity gradient appears to be decoupled from the underlying tail, pointing to a distinct kinematical structure. In the ET, the \hi velocity gradient follows the general trend of the tail, although it becomes slightly steeper at the location of the TDG. These anomalous kinematical structures are further investigated in Sect.~\ref{sec:TDG} using data at higher spatial resolution and 3D disc models.

\subsection{Metallicities along the tidal tails}
\label{sec:N7252metal}

Here we use the spectroscopic observations described in Sect.~\ref{sec:OptObs} to determine the nebular metallicity of the two putative TDGs and other \HII regions along the tidal tails. Direct estimates of the Oxygen abundance depend on the electron temperature of the ionised gas, which is probed by the [OIII] line at 4363~\AA. Unfortunately, this weak line could not be detected in any of our spectra, thus we estimate Oxygen abundances using semi-empirical methods.

For the EMMI red spectra, the oxygen abundance can be estimated from the \NII/H$\alpha$ flux ratio. We adopt the calibration of \citet[][]{Denicolo02} and find \mbox{12+log(O/H) = 8.6 $\pm$ 0.1} for both TDGs. For the FORS1 blue spectra, we use the $R_{23}$ method involving the \OII, \OIII, and H$\beta$ lines (after correcting all fluxes for internal extinction using the H$\beta$/H$\gamma$ decrement). We adopt the independent $R_{23}$ calibrations of \citet[][]{McGaugh91}, as updated by \citet{KdN04}, and \citet[][]{Pilyugin00}. The $R_{23}$ method is degenerate and gives two values of the abundance for one observable. This degeneracy is broken by the \NII/\OII\ ratio \citep{McGaugh94}. This line-ratio is only available for NGC 7252NW and clearly indicates that the higher abundance is the correct one. The resulting abundance is remarkably consistent with that from the \NII/H$\alpha$ method (see Table~\ref{tab:Oabun}). The high metallicities confirm that both TDG candidates are \textit{not} pre-existing dwarfs, but they are currently forming out of pre-enriched material ejected from the progenitor galaxies.

Using additional spectra from the FORS1 run, we estimate metallicities at two other locations: a giant \HII region near the tip of the ET and one \HII knot to the West of NGC\,7252. We find Oxygen abundances ranging between 8.65 and 8.8, as summarized in Table~\ref{tab:Oabun}. The metallicities appear to be uniform along the tidal tails and close to the solar value, with no evidence of strong gradients over distances of 80 kpc from the remnant. For comparison, the metallicity in the outskirts of isolated spiral galaxies may show gradients of several dex and reach 1/10~$Z_{\odot}$ \citep{Ferguson98}. This suggests that radial gas mixing occurs during galaxy mergers, in agreement with numerical simulations \citep{Rupke10b, Montuori10} and recent observations (\citealt{Kewley10, Rupke10a}; but see also \citealt{Werk11}). The latter works emphasise the gas dilution due to the transportation of metal-poor gas from the outskirts to the inner regions. Here, instead, we have evidence of an outwards gas transfer, where metal-rich gas has been brought to the outer regions.

\begin{figure*}[t!]
\centering
\includegraphics[width=\textwidth]{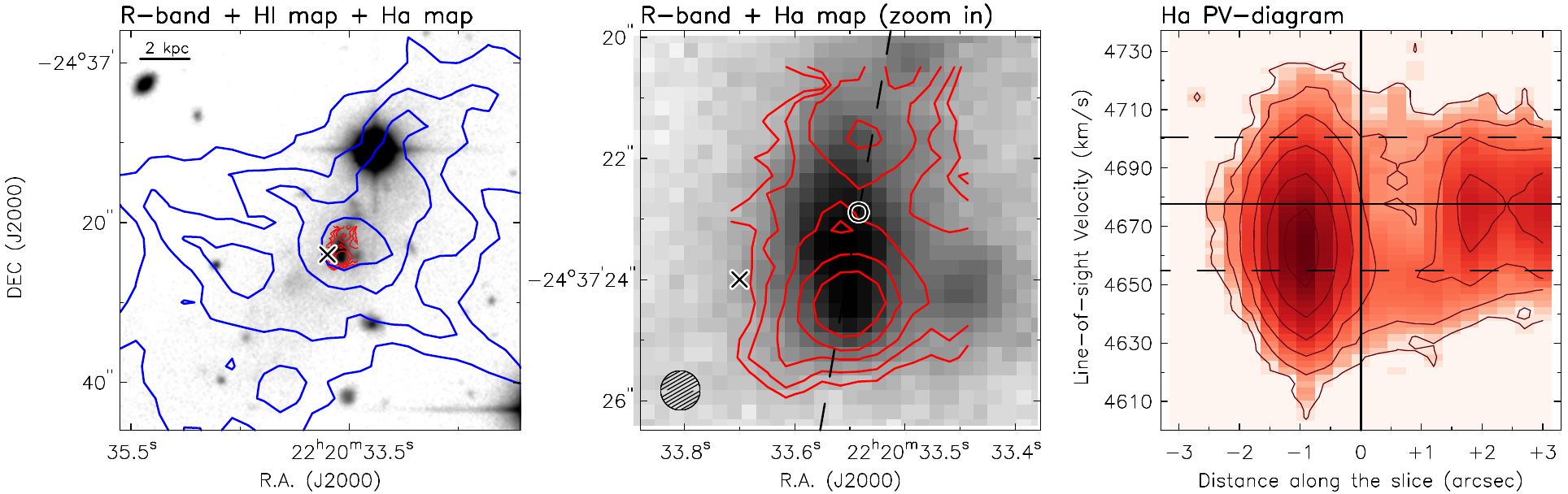}
\caption{H$\alpha$ distribution and kinematics in NGC~7252NW. \textit{Left}: $R$-band image (gray scale) overlaid with \hi (blue) and H$\alpha$ (red) emission. \hi contours are the same as in Fig.~\ref{fig:N7252NW} (top middle). The outer H$\alpha$ contour corresponds to a 10$\sigma$ detection. The cross shows the \hi dynamical centre determined in Sect.~\ref{sec:3Dmodels}. \textit{Middle}: enlarged view of the H$\alpha$ emission. The circle to the bottom-left shows the H$\alpha$ beam. \textit{Right}: PV-diagram obtained from the H$\alpha$ datacube along the two main peaks (dashed line in the middle panel). The vertical line corresponds to the circle in the middle panel. Horizontal lines indicate the \hi systemic velocity of NGC~7252NW (solid) and the velocity range covered by the \hi emission (dashed). The H$\alpha$ velocity resolution is $\sim$2.3 km~s$^{-1}$ (see Table \ref{tab:IFUobs} for details).}
\label{fig:N7252_Ha}
\end{figure*}

\begin{table*}
\centering
\caption{Spectrophotometric data and resulting metallicities in the tidal tails of NGC~7252.}
\begin{tabular}{lcccc}
\hline
Region    & NGC 7252NW   & NGC 7252E & \HIIA$-$ET     & \HIIA$-$W   \\
\hline
RA(J2000) & 22$^{\circ}$ 20$^m$ 33.6$^s$ & 22$^{\circ}$ 20$^m$ 56.1$^s$ & 22$^{\circ}$ 20$^m$ 59.3$^s$ & 22$^{\circ}$ 20$^m$ 41.2$^s$\\
Dec(J2000)& $-$24$^{\circ}$ 37$'$ 24$''$ & $-$24$^{\circ}$ 41$'$ 9$''$  & $-$24$^{\circ}$ 41$'$ 11$''$ & $-$24$^{\circ}$ 40$'$ 45$''$\\
\OIIa  & 331$\pm$66   & ...       & 237$\pm$40 & 256$\pm$49 \\
\Hd    & 28$\pm$5     & ...       & 28$\pm$5   & 24$\pm$4 \\
\Hg    & 47$\pm$5     & ...       & 52$\pm$6   & 47$\pm$5 \\
\Hb    & 100$\pm$2    & 100$\pm$6 & 100$\pm$5  & 100$\pm$2 \\
\OIIIa & 37$\pm$2     & 53$\pm$5  & 41$\pm$5   & 29$\pm$2 \\
\OIIIb & 109$\pm$4    & 150$\pm$9 & 125$\pm$5  & 89$\pm$4  \\
\NIIa  & 25$\pm$5     & 23$\pm$6  & ...        & ...  \\
\Ha    & 285$\pm$30   & 233$\pm$28& ...        & ...  \\
\NIIb  & 62$\pm$10    & 58$\pm$11 & ...        & ...  \\
\SIIa  & 51$\pm$10    & 42$\pm$8  & ...        & ...  \\ 
\SIIa  & 33$\pm$9     & 33$\pm$9  & ...        & ...  \\ 
12+log(O/H):\\
- \NII/\Ha\ \citep{Denicolo02} & 8.6  & 8.6  & ...        & ...  \\
- R23 \citep{Pilyugin00}       & 8.55 & ...  & 8.65       & 8.7  \\
- R23 \citep{KdN04}            & 8.7  & ...  & 8.8        & 8.8  \\
\hline
\end{tabular}
\label{tab:Oabun}
\end{table*}

\subsection{H$\alpha$ kinematics of NGC~7252NW}
\label{sec:N7252ha}
For NGC~7252NW we also obtained IFU spectroscopy of the H$\alpha$ line (Sect.~\ref{sec:OptObs}). The GIRAFFE field of view covers only the innermost region of NGC~7252NW, corresponding to the peak of the \hi distribution (see Fig.~\ref{fig:N7252_Ha}, left panel). Since the VLT has a pointing accuracy of $\sim$3$''$, the astrometry was determined assuming that the H$\alpha$ emission spatially coincides with the $R$-band emission, as suggested by their similar morphologies (see Fig.~\ref{fig:N7252_Ha}, middle panel). The H$\alpha$ distribution is elongated towards the North-South direction and shows two peaks separated by $\sim$3$''$.

The H$\alpha$ emission is resolved in both space and velocity, but it does not show a coherent kinematic structure. Fig.~\ref{fig:N7252_Ha} (right panel) shows a PV-diagram obtained along the two H$\alpha$ peaks (dashed line in Fig.~\ref{fig:N7252_Ha}, middle panel). The H$\alpha$ line profiles are broad: the main H$\alpha$ peak has a FWHM of $\sim$80 $\kms$ (more than twice the \hi linewidth), while the average FWHM over the entire field is $\sim$45 $\kms$. The central velocity of the main H$\alpha$ peak ($\sim$4665 $\kms$) is signficantly lower than the \hi systemic velocity of NGC~7252NW ($4678\pm 3$ $\kms$), but it is consistent with the \hi velocities on its approaching side. The H$\alpha$ emission, indeed, does not coincide with the dynamical centre of NGC~7252NW but is shifted towards the Western, approaching side (see cross in Fig.~\ref{fig:N7252_Ha}, middle panel). In general, the ionized gas appears more asymmetric and disturbed than the \hi gas. Most likely, we are observing \HII regions dominated by turbulence and/or non-circular motions due to stellar feedback.

\section{Dynamics of Tidal Dwarf Galaxies}
\label{sec:TDG}
\begin{figure*}
\centering
\includegraphics[width=0.9\textwidth]{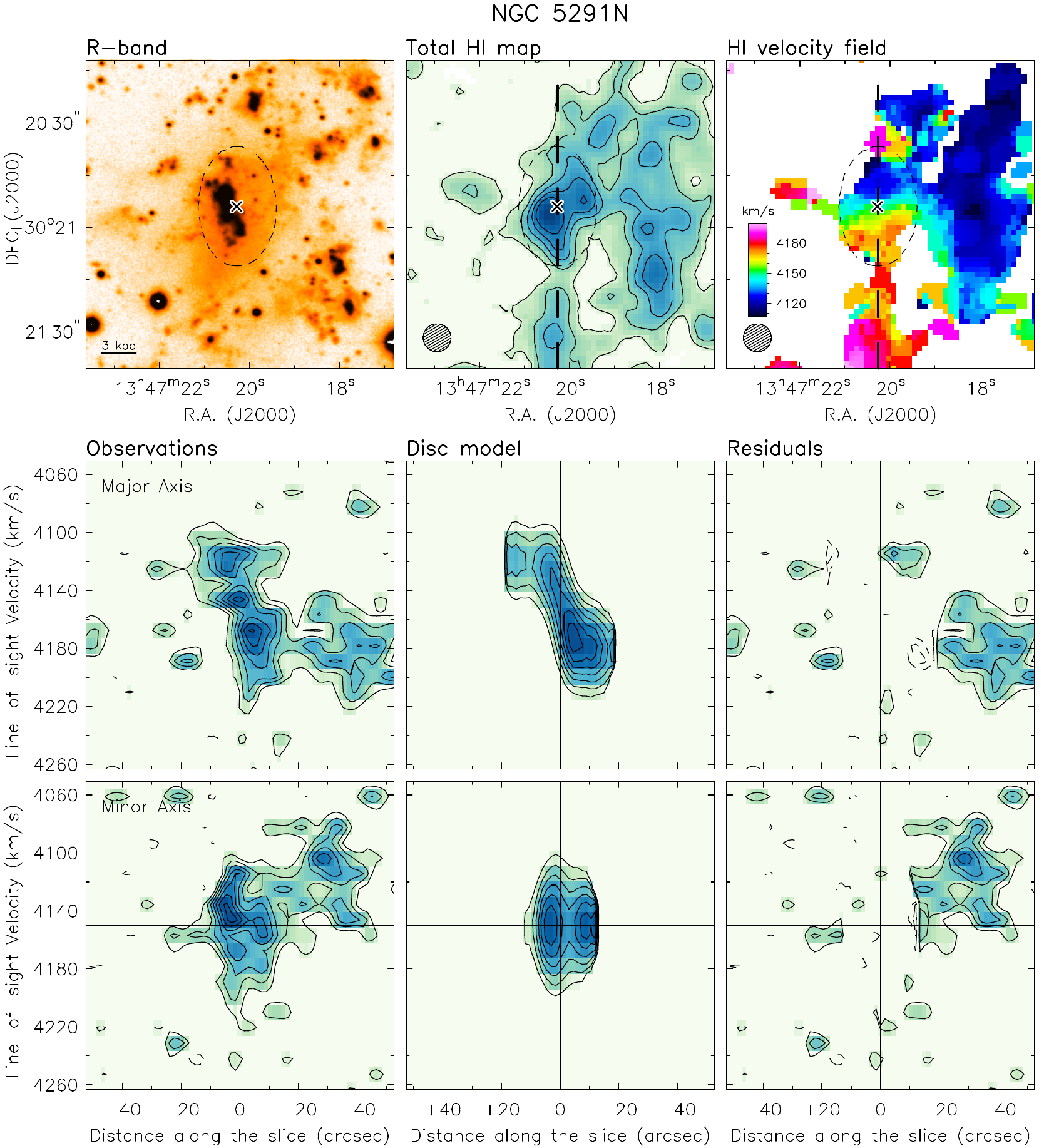}
\caption{\textit{Top panels}: optical image (\textit{left}), total \hi map (\textit{middle}), and \hi velocity field (\textit{right}). The dashed ellipse corresponds to the disc model described in Sect.~\ref{sec:3Dmodels}. The cross and dashed line illustrate the kinematical centre and major axis, respectively. In the bottom-left corner, we show the linear scale (optical image) and the \hi beam (total \hi map and velocity field) as given in Table~\ref{tab:HIcubes}. In the total \hi map, contours are at $\sim$4.5, 9, 13.5, 18, and 22.5 $\msun$~pc$^{-2}$. \textit{Bottom panels}: PV-diagrams obtained from the observed cube (\textit{left}), model-cube (\textit{middle}), and residual-cube (\textit{right}) along the major and minor axes. Solid contours range from 2$\sigma$ to 8$\sigma$ in steps of $1\sigma$. Dashed contours range from $-$2$\sigma$ to $-$4$\sigma$ in steps of $-$1$\sigma$. The horizontal and vertical lines correspond to the systemic velocity and dynamical centre, respectively.}
\label{fig:N5291N}
\end{figure*}

\begin{figure*}
\centering
\includegraphics[width=0.9\textwidth]{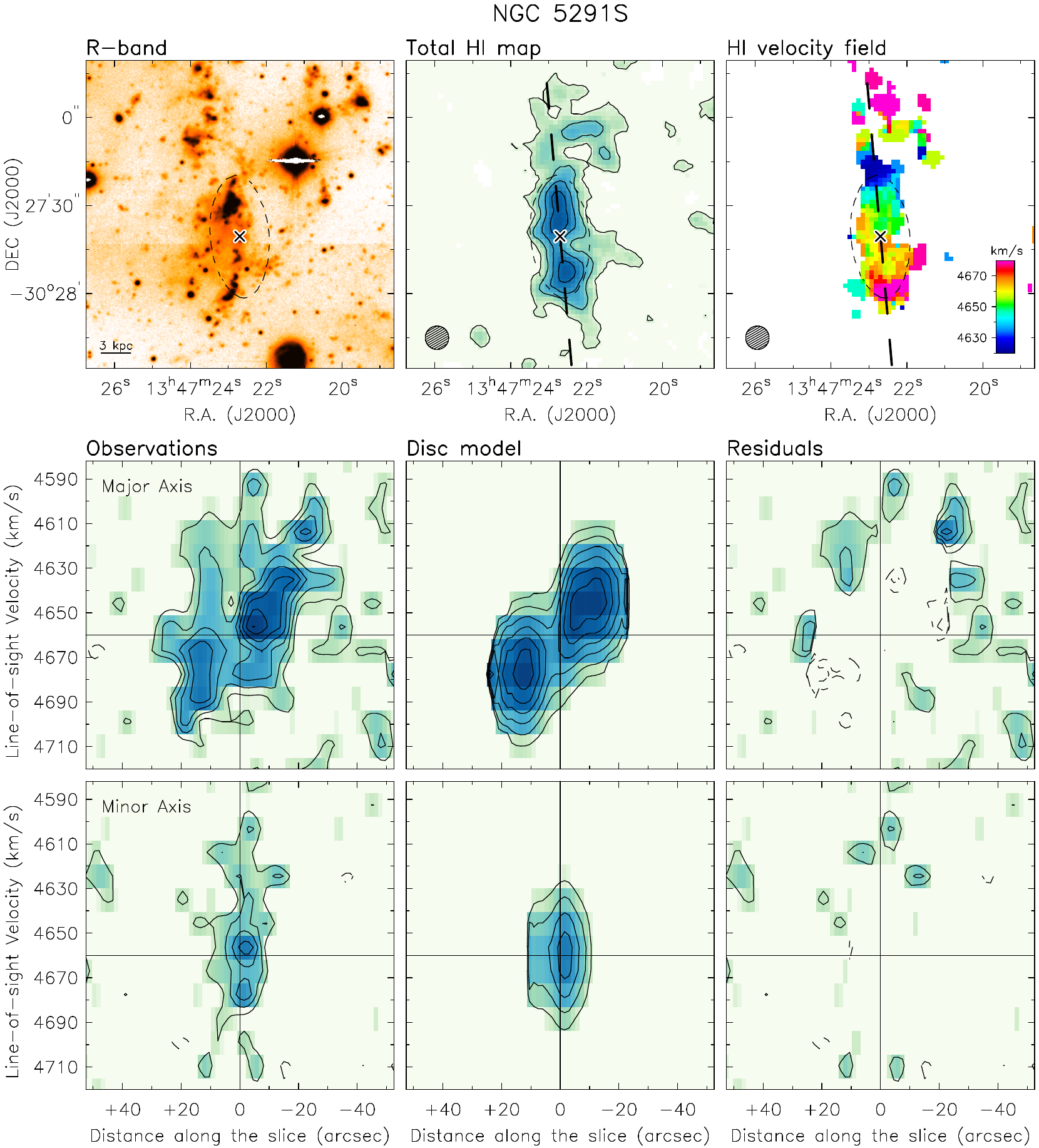}
\caption{Same as Fig.~\ref{fig:N5291N} for NGC~5291S. In the total \hi map, contours are at $\sim$4.5, 9, 13.5, and 18 $\msun$~pc$^{-2}$.}
\label{fig:N5291S}
\end{figure*}

\begin{figure*}
\centering
\includegraphics[width=0.9\textwidth]{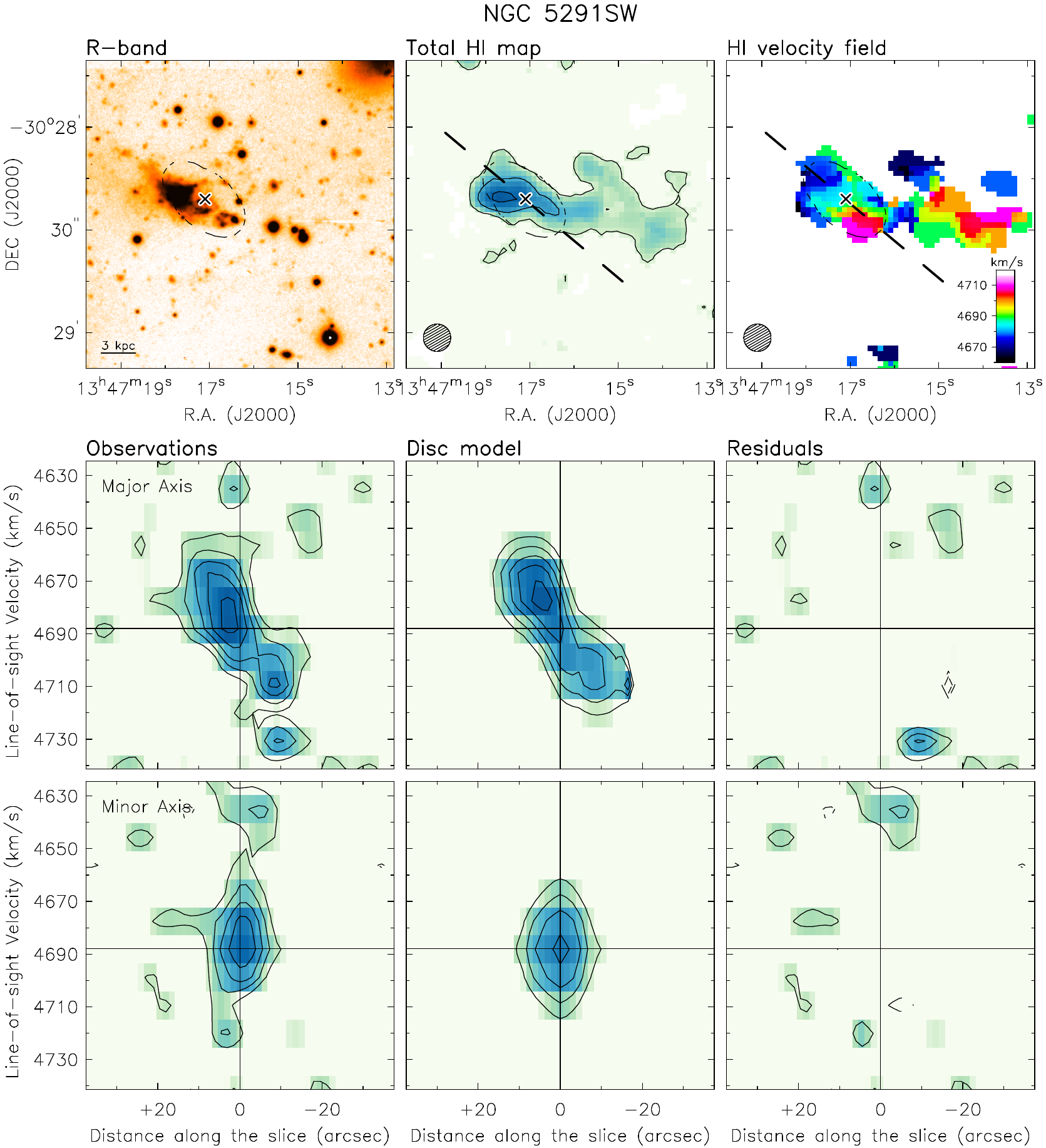}
\caption{Same as Fig.~\ref{fig:N5291N} for NGC~5291SW. In the total \hi map, contours are at $\sim$4.5, 9, and 13.5 $\msun$~pc$^{-2}$.}
\label{fig:N5291SW}
\end{figure*}

\begin{figure*}
\centering
\includegraphics[width=0.9\textwidth]{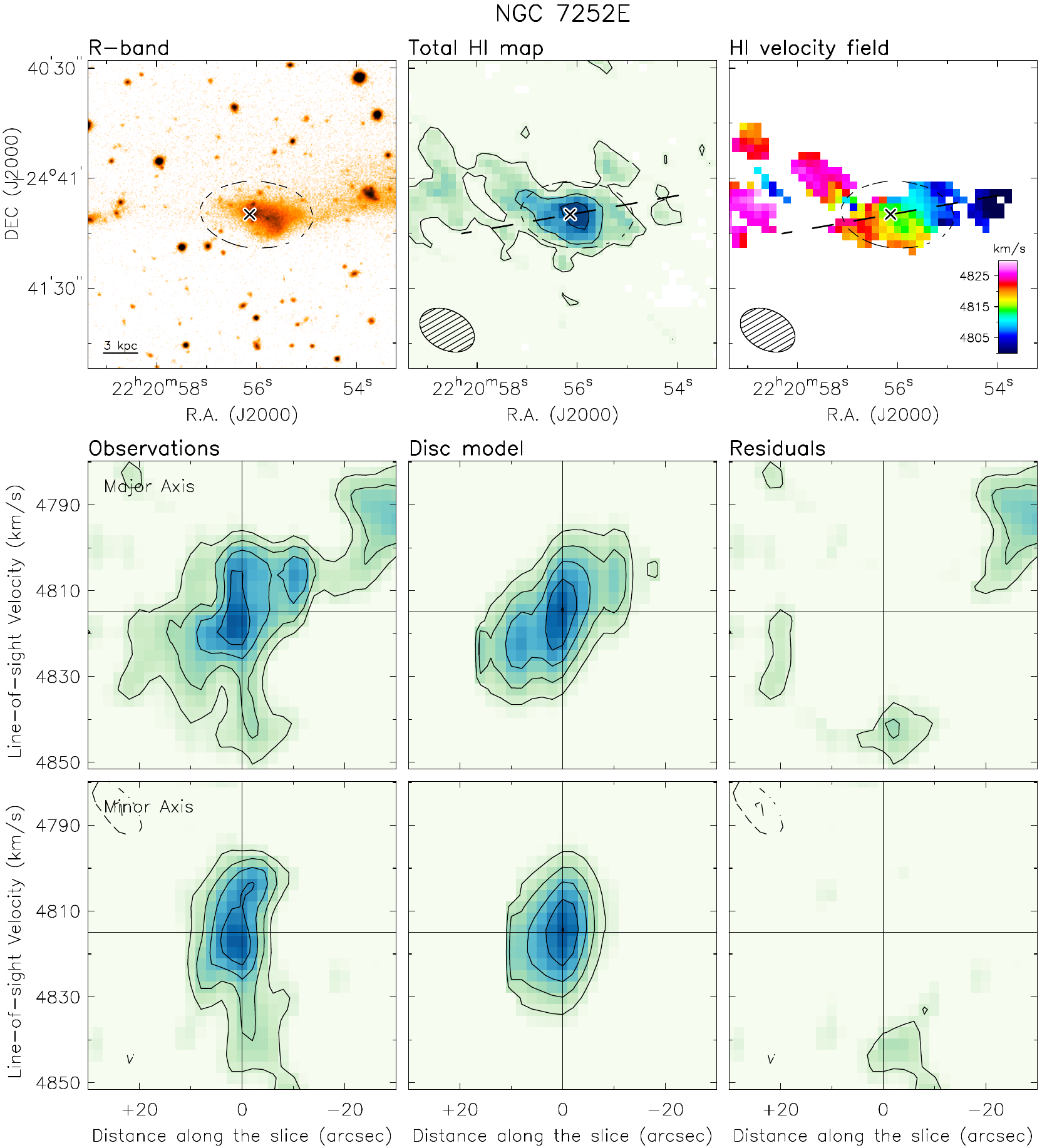}
\caption{Same as Fig.~\ref{fig:N5291N} for NGC~7252E. In the total \hi map, contours are at $\sim$1.5, 3.0, and 4.5 $\msun$~pc$^{-2}$.}
\label{fig:N7252E}
\end{figure*}

\begin{figure*}
\centering
\includegraphics[width=0.9\textwidth]{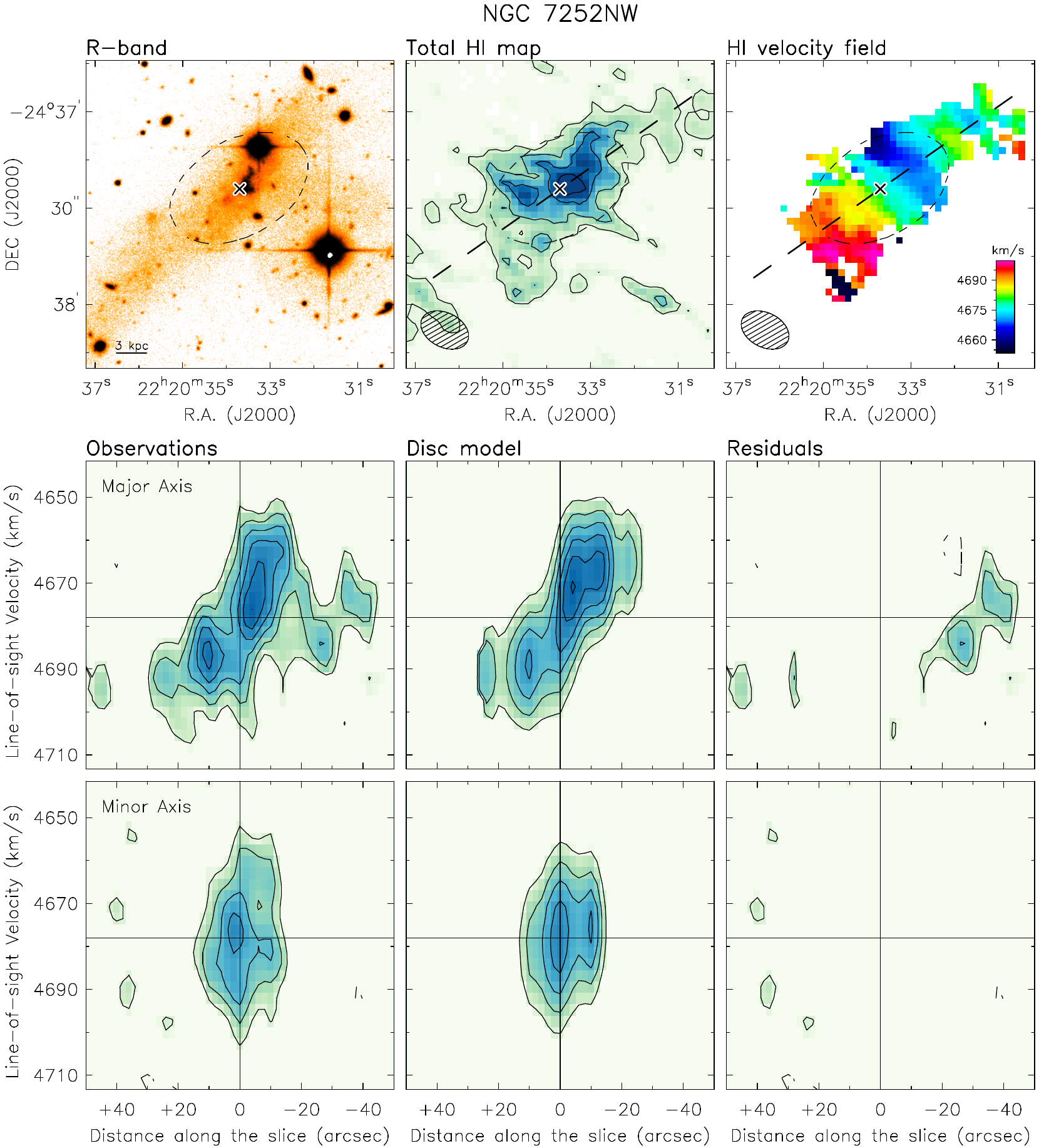}
\caption{Same as Fig.~\ref{fig:N5291N} for NGC~7252NW. In the total \hi map, contours are at $\sim$1.5, 3.0, 4.5, and 6 $\msun$~pc$^{-2}$.}
\label{fig:N7252NW}
\end{figure*}

\begin{figure*}
\centering
\includegraphics[width=0.9\textwidth]{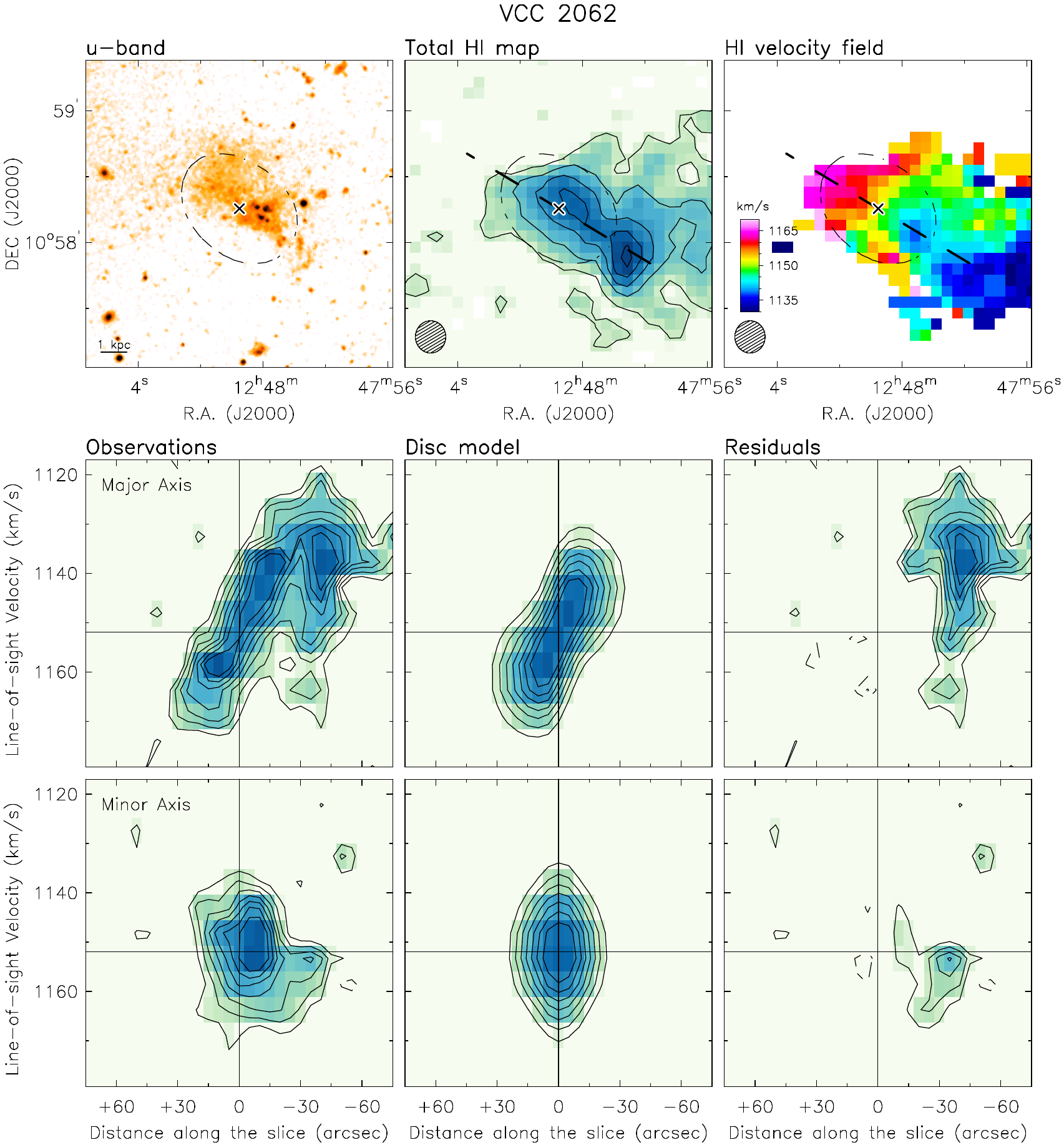}
\caption{Same as Fig.~\ref{fig:N5291N} for VCC~2062 In the total \hi map, contours are at $\sim$1.5, 3.0, 4.5, 6, and 7.5 $\msun$~pc$^{-2}$.}
\label{fig:VCC2062}
\end{figure*}

In the following, we investigate the internal dynamics of the six TDGs in our sample using high-resolution \hi cubes (see Table~\ref{tab:HIcubes}). These cubes are obtained using robust weighting during the Fourier transform and have higher resolutions than those obtained with natural weighting (Sect.~\ref{sec:HIobs}), but at a slight cost in terms of signal-to-noise (S/N) ratio. Figures~\ref{fig:N5291N} to \ref{fig:VCC2062} provide an overview of each TDG: top panels show an optical image (left), a total \hi map (middle), and a \hi velocity field (right); bottom panels show PV-diagrams along the major and minor axes from the observed cube (left), a model-cube (middle; see Sect.~\ref{sec:3Dmodels}), and a residual-cube (right). The \hi maps were built by integrating the cube over a narrow velocity range (corresponding to the kinematically decoupled velocity gradients) and clipping at 1$\sigma$. The \hi velocity fields were built by estimating an intensity-weighted-mean (IWM) velocity considering the same narrow spectral range, and clipping at 3$\sigma$.

Using the high-resolution data, all six TDGs are spatially resolved with at least 2-3 beams along the major axis. The diffuse \hi emission associated with the underlying debris is resolved out by the VLA at these high spatial resolutions, but some \hi clumps are still visible near the TDGs. We warn that the \hi maps and velocity fields are uncertain due to the low S/N ratio, low number of beams across the galaxy, and possible contamination from the underlying \hi debris. Despite these limitations, the \hi morphology and kinematics resemble a rotating disc. These putative discs are studied in detail in the next section using 3D kinematical models.

\begin{table}
\centering
\caption{Properties of the high-resolution \hi datacubes}
\resizebox{9cm}{!}{
\setlength{\tabcolsep}{4pt}
\begin{tabular}{lccccc}
\hline
System   & Beam        & PA           & $\Delta V$ & Rms Noise  & Ref.\\
& (asec $\times$ asec) & ($^{\circ}$) & ($\kms$)   & (mJy beam$^{-1}$) &     \\ 
\hline
NGC 4694 & 14.7$\times$14.3 &-70.65 & 7.3        & 0.6        & a \\
NGC 5291 & 8.0$\times$8.0   & 45.76 & 10.4       & 0.5        & b \\
NGC 7252 & 15.8$\times$10.7 & 62.96 & 10.6       & 0.6        & c \\
\hline
\end{tabular}
}
\tablefoot{To increase the signal-to-noise ratio, the robust cube of NGC~5291 (with a beam $6.6''\times4.36''$) has been smoothed to $8''\times8''$, whereas the robust cube of NGC~7252 has been smoothed in velocity to $\sim$10.6 $\kms$. References: (a)~\citet{Duc07}; (b)~\citet{Bournaud07}; (c)~this work.}
\label{tab:HIcubes}
\end{table}

\begin{table*}
\centering
\caption{Properties of the TDGs from 3D kinematical models}
\begin{tabular}{lcccccccccc}
\hline
Object     & R.A.    & Dec.    & $V_{\rm sys}$ & PA          & $i$        & $R_{\rm out}$ & $V_{\rm rot}$ & $\sigma_{\hi}$ & $t_{\rm orb}$ & $t_{\rm merg}/t_{\rm orb}$ \\
           & (J2000) & (J2000) & ($\kms$)      & ($^{\circ}$)&($^{\circ}$)& (kpc)         & ($\kms$)      & ($\kms$)       & (Gyr) & \\
\hline
NGC~5291N  & 13 47 20.3 & $-$30 20 54 & 4150 & 0   & 55 & 4.8$\pm$1.2 & 40$\pm$9 & 20$\pm$2$^{*}$ & 0.7$\pm$0.2 & 0.5 \\
NGC~5291S  & 13 47 22.7 & $-$30 27 40 & 4660 & 5   & 65 & 7.2$\pm$1.2 & 20$\pm$6 & 20$\pm$2$^{*}$ & 2.2$\pm$0.7 & 0.2 \\
NGC~5291SW & 13 47 17.1 & $-$30 28 21 & 4688 & 50  & 60 & 4.8$\pm$1.2 & 22$\pm$7 & 12$\pm$2 & 1.3$\pm$0.4 & 0.3 \\
NGC~7252E  & 22 20 56.0 & $-$24 41 10 & 4815 & 100 & 80 & 4.5$\pm$1.9 & 11$\pm$5 & 10$\pm$2 & 2.5$\pm$1.6 & 0.3 \\
NGC~7252NW & 22 20 33.7 & $-$24 37 24 & 4678 & 125 & 60 & 7.7$\pm$1.9 & 16$\pm$6 & 10$\pm$2 & 3.0$\pm$1.2 & 0.2 \\
VCC~2062   & 12 48 00.8 & $+$10 58 15 & 1152 & 55  & 45 & 2.6$\pm$0.6 & 13$\pm$7 & 7$\pm$1  & 1.2$\pm$0.5 & 0.4$-$0.8\\
\hline
\end{tabular}
\tablefoot{$^{*}$These high values of $\sigma_{\hi}$ may be either physical or due to unresolved non-circular motions within the \hi beam.}
\label{tab:TDGkin}
\end{table*}

\subsection{Kinematical models}
\label{sec:3Dmodels}
To test the hypothesis that TDGs are associated with an \hi disc, we built 3D kinematical models following similar procedures as \citet{Lelli12a, Lelli12b, Lelli14}. As a first step, we inspected channel maps and PV-diagrams to obtain initial estimates of the dynamical centre ($x_{0}, \, y_{0}$), position angle (PA), inclination ($i$), systemic velocity ($V_{\rm sys}$), and rotation velocity ($V_{\rm rot}$). These initial estimates were then used as inputs to build model-cubes and subsequently refined iteratively until we obtained a model-cube that closely matches the observed cube. Given the low S/N ratio and possible contamination from the underlying \hi debris, the best models were not determined by a blind $\chi^{2}$ minimization, but by visual inspection of channel maps and PV-diagrams.

The intial models are built assuming a thin disc with an exponential surface density profile, a flat rotation curve, and a constant \hi velocity dispersion ($\sigma_{\hi}$). The disc is projected on the sky (according to the values of $x_{0}$, $y_{0}$, PA, $i$, and $V_{\rm sys}$) to generate a model-cube, which is subsequently smoothed to the same spatial and spectral resolutions of the observations. Finally, the model-cube is renormalized on a pixel-by-pixel basis to the observed \hi map, i.e. we reproduce the asymmetric gas distribution by imposing that, at every spatial pixel, the flux along each \hi line profile in the model-cube is equal to that of the corresponding line profile in the observations \citep[see][for details]{Lelli12a, Lelli12b}. As a result of this normalization, the assumption of an exponential \hi surface density profile has little effect on the final model, but is still useful to (i) reproduce beam-smearing effects due to a varying \hi flux on spatial scales smaller than the \hi beam, and (ii) obtain an estimate of $i$ by comparing model \hi maps (before normalization) to the observed \hi maps (see dashed ellipses on the top-middle panels of Figs.~\ref{fig:N5291N} to \ref{fig:VCC2062}). For VCC~2062, the model-cube was \textit{not} normalized to the observed \hi map because two distinct kinematic components overlap in both space ($R\simeq15''$ to 30$''$ towards the South-West) and velocity ($V_{\rm l.o.s.}\simeq 1110$ to 1160~$\kms$); a simple exponential disc model allows disentangling these two kinematic components (see Fig.~\ref{fig:VCC2062}).

Once the geometrical parameters of the \hi disc are fixed, the free parameters of the model are $\sigma_{\hi}$, $V_{\rm rot}$, and the outermost disc radius $R_{\rm out}$. These are largely independent as they have different effects on the model: $V_{\rm rot}$ and $R_{\rm out}$ increases/decreases the extent of PV-diagrams in the vertical and horizontal directions, respectively, whereas $\sigma_{\hi}$ changes the overall shape of PV-diagrams and increases/decreases the distance between density contours. 

\subsubsection{Comparing models and observations}

The bottom panels of Figs.~\ref{fig:N5291N} to \ref{fig:VCC2062} compare PV-diagrams from the observed cubes and the model-cubes. For all six TDGs, a rotating disc with \textit{axisymmetric} kinematics can reproduce the \hi velocity gradient along the major axis. The \hi emission along the minor axis is also well reproduced, suggesting that any non-circular motion should be relatively small. Some radial motions may be present in NGC~5291N and VCC~2062, as hinted to by the tilt of their minor-axis PV-diagrams. For these two objects, we built additional models including a global radial component ($V_{\rm rad}$). It is unclear whether the inclusion of $V_{\rm rad}$ substantially improves the models, but we can constrain the ratio $V_{\rm rad}/V_{\rm rot}$ to be smaller than 0.3. Hence, the disc kinematics is dominated by rotation and radial motions can be neglected to a first approximation. We stress that the inclusion of $V_{\rm rad}$ has no effect on the measured values of $V_{\rm rot}$. The quality of our modelling is further demonstrated in Appendix~\ref{sec:ChanMaps}, where we compare \hi channel maps from the observed cubes and the model-cubes.

In general, our initial estimates of the kinematical parameters gave a good description of the observations, but we have also built models with different parameter values to estimate the corresponding uncertainties. The parameters of the final models are listed in Table~\ref{tab:TDGkin}. For $\sigma_{\hi}$ we find typical values between 7 and 12 $\kms$ apart from NGC~5291N and NGC~5291S. For these two TDGs, a velocity dispersion of $\sim$20~$\kms$ is required to reproduce the observed cube: this high value may be either physical (high turbulence in the ISM) or due to unresolved non-circular motions within the beam. In NGC~5291 a higher turbulence may be expected because the \hi ring is the result of a fast head-on collision with a massive companion.

Although a rotating disc provides a good description of the \hi kinematics, there is an important issue with this model. The orbital times at the outer edge of the disc ($t_{\rm orb}$) are much longer than the estimated ages of the TDGs, implying that the discs would have undergone less than a full orbit. In Table~\ref{tab:TDGkin}, we provide both $t_{\rm orb}$ and the ratio $t_{\rm merg}/t_{\rm orb}$, where $t_{\rm merg}$ is the characteristic dynamical time of the interaction/merger event. In particular, we assume that (i) the \hi ring around NGC~5291 has formed $\sim$360~Myr ago (B07), (ii) the progenitor galaxies forming NGC~7252 fell together $\sim$700 Myr ago \citep{Hibbard95, Chien10}, and (iii) the interaction/merger forming VCC~2062 and NGC~4694 has happened between 0.5 and 1 Gyr ago. Note that $t_{\rm merg}$ is a hard upper limit for the TDG ages, since it takes time for the tidal debris to be ejected and for the gas within them to collapse under self-gravity. It is unclear, therefore, whether the \hi discs associated with the TDGs have had enough time to reach dynamical equilibrium. This issue is further discussed in Sect.~\ref{sec:simu} with the aid of numerical simulations.

Motivated by the long orbital times, we have explored alternative models and found that a collision between two large \hi clouds could also reproduce the \hi data. These models, however, need to be highly fine-tuned, given that the two \hi clouds must have sizes, spatial separation, and velocity separation mimicking the symmetric sides of a rotating disc. This scenario, therefore, appears contrived. We conclude that we are likely observing the formation of rotating \hi discs in the nearby Universe.

\subsubsection{Model assumptions and uncertainties}

Dwarf galaxies can show both flat and rising rotation curves \citep[e.g.][]{Swaters09, Lelli12a, Lelli12b, Lelli14, Oh15}. We assume that $V_{\rm rot}$ is constant with $R$ to minimize the number of free parameters in our models. This assumption has little effects on our results because we are interested in the rotation velocity at the outermost radius ($V_{\rm out}$). The actual shape of the rotation curve is difficult to determine due to the low number of resolution elements ($\lesssim4$ beams along the disc major axis): in this situation a flat rotation curve produces similar model-cubes as a solid-body one (cf. \citealt{Lelli12a}, Figure~5, second row). As a further test, we built additional models assuming a solid-body rotation curve (see Appendix~\ref{sec:solidbody}). For NGC~5291N and NGC~7252NW, a flat rotation curve is preferable to a solid-body one, as indicated by the observed \hi emission at high rotation velocities near the disc center. For all TDGs, solid-body and flat rotation curves return similar values of $V_{\rm out}$ but there is a weak systematic effect leading to higher $V_{\rm out}$ for solid-body rotation curves by $\sim$2-3 km~s$^{-1}$. This is well within our assumed errors. Hence, the dynamical masses estimated in Sect.~\ref{sec:mass} are largely independent of the assumed rotation curve shape.

The uncertainties on ($x_{0}$, $y_{0}$), PA, and $V_{\rm sys}$ have negligible effect on the measured values of $V_{\rm rot}$. PV diagrams from model-cubes and observed cubes show that these parameters have been properly determined (see Figs.~\ref{fig:N5291N} to \ref{fig:VCC2062}, bottom panels). The value of $i$ requires special attention because $V_{\rm rot} \propto V_{\rm l.o.s.}/\sin(i)$, where $V_{\rm l.o.s.}$ is the projected velocity along the line-of-sight. We constrain the values of $i$ by comparing model \hi maps (before pixel-to-pixel renormalization) to the observed \hi maps (see dashed ellipses in Figs.~\ref{fig:N5291N} to \ref{fig:VCC2062}). In general, we find that inclinations higher/lower than the assumed ones by $\sim$30$^{\circ}$ are excluded by the data. This is conservatively considered as a 2$\sigma$ error. The error $\sigma_{V_{\rm rot}}$ on $V_{\rm rot}$ is then given by
\begin{equation}\label{eq:error}
 \sigma_{V_{\rm rot}} = \sqrt{\bigg[\dfrac{\sigma_{V_{\rm l.o.s}}}{\sin(i)}\bigg]^2 + \bigg[V_{\rm rot} \dfrac{\sigma_{i}}{\tan(i)}\bigg]^{2}}
\end{equation}
where $\sigma_{i}=15^{\circ}$ and $\sigma_{V_{\rm l.o.s}}$ is estimated as half the velocity resolution of the observations.

\subsection{Mass Budget}
\label{sec:mass}

In the previous section, we showed that rotating disc models provide a good description of the \hi gas associated with TDGs. It is unclear, however, whether these \hi discs are in dynamical equilibrium, given that they did not have enough time to complete a full orbit since the interaction/merger event. In the following, we assume that the TDGs in our sample are in dynamical equilibrium and estimate their dynamical masses from the \hi rotation velocities. The inferred dynamical masses are then compared with the observed baryonic masses to determine the DM content of TDGs. As we discuss in Sect.~\ref{sec:simu}, numerical simulations suggest that these dynamical masses may be quite reliable despite the long orbital times.

\subsubsection{Dynamical mass}

For the TDGs in our sample, the rotation velocities ($V_{\rm rot}$) are comparable with the \hi velocity dispersions ($\sigma_{\hi}$), thus the former must be corrected for pressure support in order to trace the gravitational potential. We apply the asymmetric-drift correction using the following equation:
\begin{equation}\label{eq:drift}
V_{\rm circ} = \sqrt{V_{\rm rot}^{2} + \sigma_{\hi}^2 (R/h_{\hi})},
\end{equation}
where $V_{\rm circ}$ is the circular velocity tracing the potential well. This equation assumes that (i) the velocity dispersion is isotropic, (ii) the velocity dispersion and scale height of the disc are constant with radius, and (iii) the \hi surface density profile is an exponential function with scale length $h_{\hi}$ \citep[see Appendix A of][for details]{Lelli14a}. Using 3D disc models, we estimate that our TDGs have $R_{\rm out}/h_{\hi}\simeq 2$, where $R_{\rm out}$ is the outermost radius of the \hi disc. The dashed ellipses in Figs.~\ref{fig:N5291N} to \ref{fig:VCC2062} (top-middle panels) show the model \hi maps at $R\simeq2h_{\hi}$ (before being renormalized to the observed HI map, see Sect.~\ref{sec:3Dmodels}). The only exception is NGC~5291N, which have $R_{\rm out}\simeq h_{\hi}$.

\begin{table*}[t!]
\centering
\caption{Mass budget in TDGs assuming dynamical equilibrium}
\label{tab:budget}
\begin{tabular}{lccccccc}
\hline
Object    & $V_{\rm circ}$ & $M_{\rm dyn}$     & $M_{\rm atom}$    & $M_{*}$ & $M_{\rm mol}$ & $M_{\rm bar}$ & $M_{\rm dyn}/M_{\rm bar}$ \\
          & ($\kms$)       & $(10^{8}\,\msun)$ & $(10^{8}\,\msun)$ & $(10^{8}\,\msun)$ & $(10^{8}\,\msun)$ & $(10^{8}\,\msun)$\\
\hline
NGC~5291N & 45$\pm$9 & 23$\pm$11   & 12.3$\pm$1.2  & 1.1$\pm$0.5   & 2.2$\pm$1.1   & 15.6$\pm$1.7      & 1.5$\pm$0.7 \\
NGC~5291S & 35$\pm$6 & 20$\pm$8    & 11.8$\pm$1.2  & 0.8$\pm$0.4   & 3.3$\pm$1.7   & 15.9$\pm$2.1      & 1.3$\pm$0.5 \\
NGC~5291SW& 28$\pm$7 & 8.7$\pm$4.9 & 4.9$\pm$0.5   & 0.3$\pm$0.2   & 1.9$\pm$1.0$^{*}$   & 7.1$\pm$1.2 & 1.2$\pm$0.7 \\
NGC~7252E & 18$\pm$5 & 3.4$\pm$2.4 & 2.9$\pm$0.3   & 0.85$\pm$0.43 & 0.20$\pm$0.10$^{*}$ & 3.9$\pm$0.5 & 0.9$\pm$0.6 \\
NGC~7252NW& 21$\pm$6 & 7.9$\pm$4.9 & 7.0$\pm$0.7   & 0.94$\pm$0.47 & 0.22$\pm$0.11 & 8.2$\pm$0.8       & 1.0$\pm$0.6 \\
VCC~2062  & 16$\pm$7 & 1.5$\pm$1.3 & 0.75$\pm$0.07 & 0.47$\pm$0.24 & 0.23$\pm$0.12 & 1.4$\pm$0.3       & 1.0$\pm$0.9 \\
\hline
\end{tabular}
\tablefoot{$^{*}$For these two galaxies CO observations are not available, thus $M_{\rm mol}$ is indirectly estimated from the SFR assuming that TDGs have the same SFE as spiral galaxies ($\sim5\times10^{-10}$ yr$^{-1}$). See Sect.~\ref{sec:baryonic} for details.}
\end{table*}
The dynamical mass within $R_{\rm out}$ is then given by
\begin{equation}
M_{\rm dyn} = \varepsilon \, \frac{ R_{\rm out} V_{\rm circ}^{2} }{G}
\end{equation}
where $G$ is Newton's constant and $\varepsilon$ is a parameter of the order of unity that depends on the 3D distribution of the \textit{total} mass. For simplicity, we assume $\varepsilon = 1$ (spherical geometry). For a thin exponential disc, one can calculate that $\varepsilon \simeq 0.95$ at $R = h$ and $\varepsilon \simeq 0.77$ at $R = 2 h$ \citep[cf.][]{Freeman70}. The resulting dynamical masses are given in Table~\ref{tab:budget}. Besides the uncertainty due to the dynamical equilibrium of TDGs, the uncertainty on $M_{\rm dyn}$ is driven by errors on $V_{\rm circ}$ and $R_{\rm out}$. The former error is estimated using Eq.~\ref{eq:error}, while the latter is conservatively estimated as half the spatial resolution of the \hi data.

\subsubsection{Baryonic mass}
\label{sec:baryonic}

The baryonic mass of a galaxy ($M_{\rm bar}$) is given by three main components \citep[e.g.][]{McGaugh12}: atomic gas ($M_{\rm atom}$), molecular gas ($M_{\rm mol}$), and stars ($M_{*}$). In TDGs $M_{\rm atom}$ largely dominates the baryonic mass budget, thus the uncertainties on $M_{*}$ and $M_{\rm mol}$ due to the choice of, respectively, a stellar mass-to-light ratio ($\Upsilon_{*}$) and CO-to-H$_{2}$ conversion factor ($\XCO$) have minimal impact. On the other hand, measuring $M_{\rm atom}$ is not straightforward because TDGs are still embedded within gas-rich tidal debris. It is reasonable to assume that all the \hi gas within $R_{\rm out}$ is associated with the TDGs, since they represent a local potential well that ``clears out'' the gas from the surroundings. The values of $M_{\rm atom}$ in Table~\ref{tab:budget} are based on this assumption and adopt a typical error of 10$\%$ (the \hi masses are multiplied by a factor of 1.37 to take the contributions of Helium and heavier elements into account). Alternatively, one may assume that, at the TDG location, some gas is associated with the underlying tidal debris. In the extreme hypothesis that tidal debris has a uniform \hi surface density of 1 $M_{\odot}$~pc$^{-2}$ over the TDG area, the values of $M_{\rm atom}$ in Table~\ref{tab:budget} would decrease by $\sim$30$\%$ to 40$\%$.

As we discussed in previous sections, TDGs have higher metallicities than typical dwarfs, thus CO lines are readily detected and H$_{2}$ masses can be estimated using a typical value of $\XCO = 2.0 \times 10^{20}$~cm$^{2}$~(K~\kms)$^{-1}$. For three TDGs (NGC~5291N, NGC~5291S, and NGC~7252NW), we use the molecular masses from \citet{Braine01}, rescaled to the distances in Table~\ref{tab:sample}. For VCC~2062, we adopt the molecular mass from D07. For the remaining two TDGs (NGC~5291SW and NGC~7252E), CO observations are not available, thus we estimated $M_{\rm mol}$ assuming that TDGs and spiral galaxies have the same star-formation efficiency: ${\rm SFE} = {\rm SFR}/M_{\rm mol} \simeq5 \times 10^{-10}$~yr$^{-1}$ \citep{Leroy08}. This is observationally motivated by the results of \citet{Braine01} for a sample of eight TDGs. For these two TDGs, SFRs are estimated using FUV fluxes from \citet{Boquien09} and the calibration from \citet{Kennicutt12}, which is based on Starburst99 models \citep{Leitherer99} and a Kroupa initial mass function \citep{Kroupa03}. For all TDGs, we adopt a conservative error of 50$\%$ on $M_{\rm mol}$ to consider possible variations in $\XCO$ and/or SFE among galaxies. The molecular masses of these TDGs are from $\sim$4 to $\sim$30 times smaller than $M_{\rm atom}$, thus they marginally contribute to $M_{\rm bar}$.

The stellar masses of TDGs are tricky to estimate because they are formed by a complex mix of young stars (comprising the dominant stellar population) and possibly older stars ejected from the parent galaxies. For the three TDGs around NGC~5291, we adopt the stellar masses from B07, which were derived using $K$-band luminosities and adopting $\Upsilon_{*} = 0.3$. For the other three TDGs, we adopt the values from Table~4 of \citet{Duc14}, which were obtained by re-analysing the photometric data of \citet{Boquien09, Boquien10} and fitting spectral-energy distributions with stellar population models. We adopt a conservative error of 50$\%$ on $M_{*}$. The stellar masses of these TDGs are from $\sim$3 to $\sim$20 times smaller than $M_{\rm atom}$, thus have little impact on $M_{\rm bar}$. The only exception is VCC~2062 that has $M_{\rm atom}\simeq 1.5 M_{*}$ and represents the oldest and most evolved TDG in our sample.

The mass budget of our six TDGs is summarized in Table~\ref{tab:budget}. Assuming that these TDGs are in dynamical equilibrium, we find that their dynamical masses are fully consistent with the observed baryonic masses, implying that they are nearly devoid of DM. Thus, we do \textit{not} confirm the results of B07, who reported values of $M_{\rm dyn}/M_{\rm bar}$ between 2 and 3 for three TDGs around NGC~5291. The reasons for this discrepancy are discussed in Appendix~\ref{app:compaB07}. We note that the ratio $M_{\rm dyn}/M_{\rm bar}$ depends on the assumed distance as $1/D$. NGC~4068 and VCC~2062 lie in the Virgo Cluster: assuming $D = 17 \pm 1$~Mpc \citep[cf.][]{Mei07}, the resulting error on $M_{\rm dyn}/M_{\rm bar}$ is smaller than 6$\%$. NGC~5291 and NGC~7252 are relatively far away ($V_{\rm sys}\gtrsim4300$~km~s$^{-1}$) and the Hubble flow is a reliable distance indicator: even assuming an extreme peculiar velocity of 500~km~s$^{-1}$ along the line of sight, the resulting error on $M_{\rm dyn}/M_{\rm bar}$ is smaller than 12$\%$.

\begin{figure}
\centering
\includegraphics[width=0.45\textwidth]{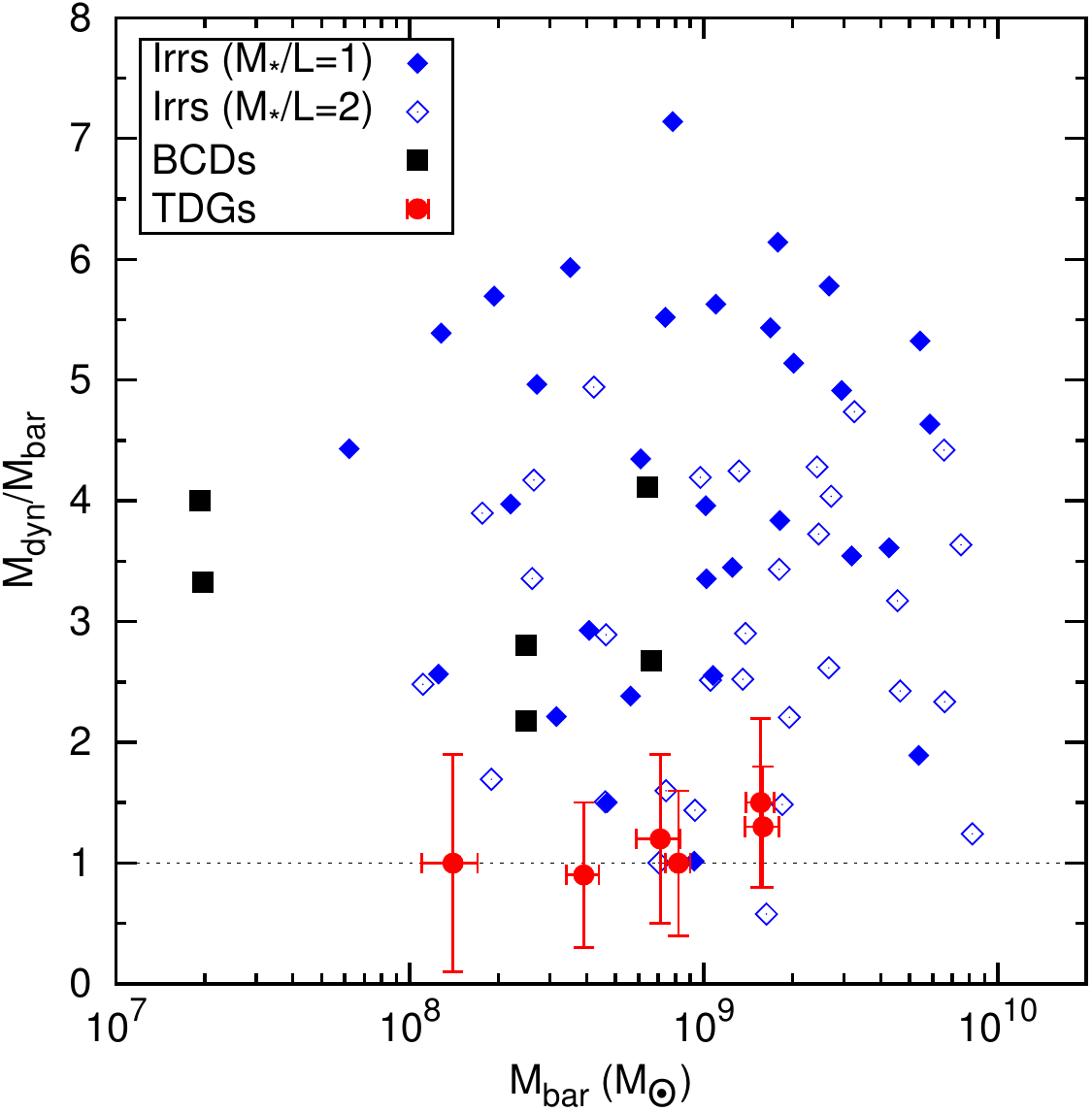}
\caption{Mass budget within $R\simeq3~h_{*}$ in different types of dwarf galaxies: Irrs \citep[blue diamonds, from][]{Swaters09}, BCDs \citep[black squares, from][]{Lelli14}, and TDGs (red points). For Irrs $M_{\rm bar}$ is estimated assuming both a ``heavy disc'' with $\Upsilon_{*}^{R}=2$ (open symbols) and a ``light disc'' with $\Upsilon_{*}^{R}=1$ (filled symbols). For BCDs the stellar mass is estimated from color-magnitude diagrams of resolved stellar populations \citep[see][]{Lelli14}.}
\label{fig:mass}\label{tab:MOND}
\end{figure}
\subsection{Comparison with other dwarf galaxies}
\label{sec:compa}
To assess whether the DM content of TDGs is really anomalous, we use a control-sample of typical star-forming dwarfs formed by 30 irregulars (Irrs) and 6 blue compact dwarfs (BCDs). The rotation velocities of these galaxies have been measured by \citet{Swaters09} and \citet{Lelli14} building model-cubes of the \hi observations, thus they are fully comparable with those derived here. TDGs are heavily gas-dominated ($M_{\rm gas}/M_{\rm *}\simeq 10$), thus $M_{\rm dyn}/M_{\rm bar}$ is estimated at roughly $R \simeq 2 h_{\hi}$, where $h_{\hi}$ is the scale-length of the \hi disc (Sect.~\ref{sec:mass}). To have a meaningful comparison with typical dwarfs, we need to compute $M_{\rm dyn}/M_{\rm bar}$ within a similar radius. In general, these galaxies have $M_{\rm gas}/M_{\rm *}\lesssim 1-2$, thus it is more natural to consider the scale-length of the stellar disc $h_{*}$. Given that typical dwarfs have $h_{\hi}/h_{*} \simeq 1.5$ \citep[cf.][]{Swaters02}, we estimate $M_{\rm dyn}/M_{\rm bar}$ at $R \simeq 3 h_{*}$ (the edge of the stellar disc). 

The Irrs are drawn from \citet{Swaters09} using the following criteria: (i) the galaxies have $30^{\circ} \lesssim i \lesssim 80^{\circ}$, (ii) the \hi rotation curves have quality flag $q > 2$ and are traced out to $R\simeq3 h_{*}$, and (iii) $V_{\rm rot} < 100$~$\kms$ at the last measured point. The BCDs are drawn from \citet{Lelli14} considering only galaxies with reliable estimates of $V_{\rm rot}$ (see their Sect.~7). 

Contrary to TDGs, typical dwarfs often have comparable stellar and gaseous masses, thus the values of $M_{\rm dyn}/M_{\rm bar}$ strongly depend on the assumed stellar mass-to-light ratio $\Upsilon_{*}$. For Irrs we compute two different values of $M_{\rm dyn}/M_{\rm bar}$ assuming a ``heavy disc'' ($\Upsilon^{R}_{*}=2$) and a ``light disc'' ($\Upsilon^{R}_{*}=1$). We do not provide an error bar on $M_{\rm dyn}/M_{\rm bar}$ because this is fully dominated by the assumed variation in $\Upsilon^{R}_{*}$. For BCDs, we use the stellar masses obtained from modelling color-magnitude diagrams of resolved stellar populations \citep[see][for details and references]{Lelli14}. Our results are illustrated in Fig.~\ref{fig:mass}. Despite the large error bars, it is clear that TDGs systematically have smaller values of $M_{\rm dyn}/M_{\rm bar}$ than typical dwarfs, pointing to an apparent lack of DM.

\section{Discussion}
\label{sec:discussion}

In this paper we studied the gas dynamics in a sample of six TDGs. These galaxies have higher metallicities than dwarfs of similar mass and appear kinematically decoupled from the surrounding tidal debris, indicating that they are \textit{bona-fide} TDGs (Sect.~\ref{sec:overview}). We built 3D kinematical models and found that the \hi emission associated with TDGs can be described by rotating discs (Sect.~\ref{sec:3Dmodels}). The orbital times of these discs, however, are much longer than the merger timescales (as inferred from numerical simulations), indicating that they have undergone less than a full orbit since the TDG formation. This raises the question as to whether TDGs are in dynamical equilibrium and the observed velocities can be used to estimate dynamical masses. If one assumes that these TDGs are in dynamical equilibrium, the inferred dynamical masses are fully consistent with the observed baryonic masses, implying that TDGs are nearly devoid of DM (Sect.~\ref{sec:mass}). In the following, we start by scrutinizing the hypothesis of dynamical equilibrium with the aid of numerical simulations. Then, we discuss the implications of our results for the distribution of non-baryonic DM in galaxies and the possible amount of dark baryons. Finally, we investigate the location of TDGs on the baryonic Tully-Fisher relation (BTFR, \citealt{McGaugh00}) and discuss alternative theories like MOND.

\subsection{Dynamical equilibrium: clues from numerical simulations}
\label{sec:simu}
\begin{figure}
\centering
\includegraphics[width=0.45\textwidth]{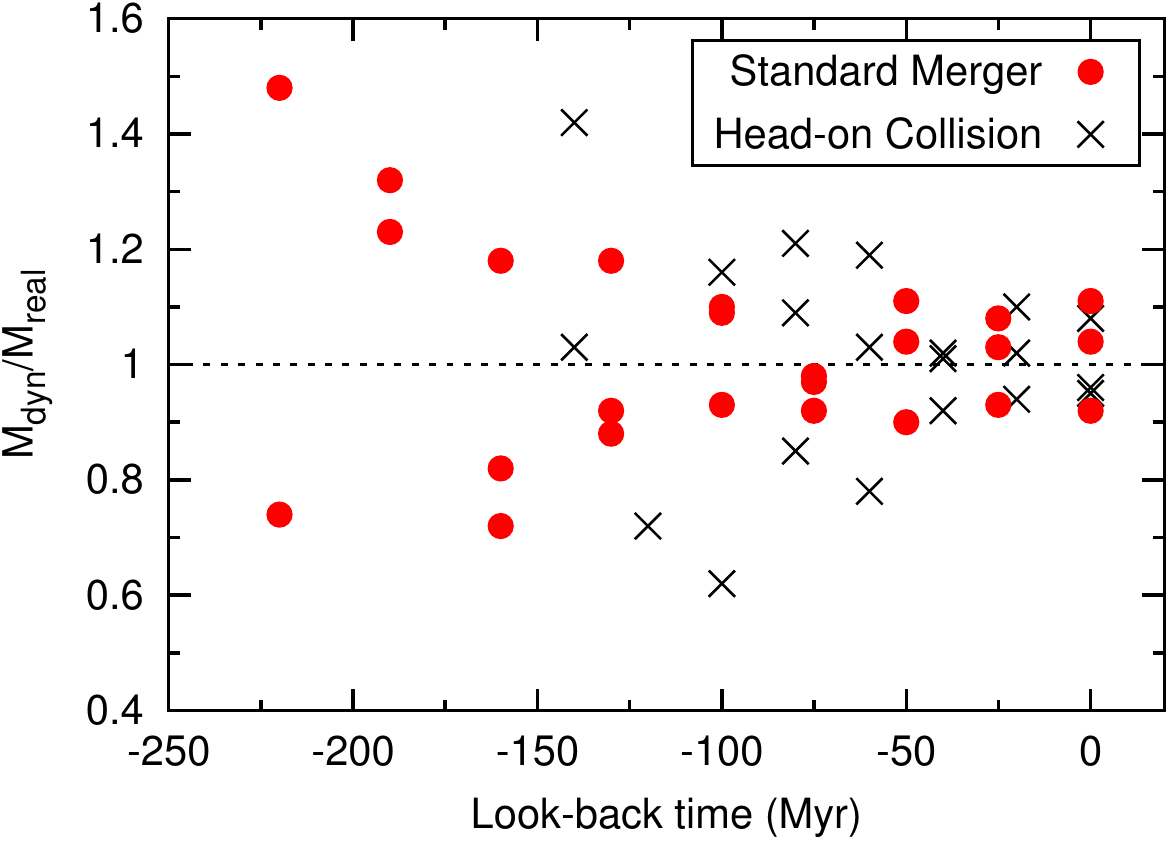}
\caption{The ratio $M_{\rm dyn}/M_{\rm real}$ versus look-back time for simulated TDGs. $M_{\rm dyn}$ is the dynamical mass inferred from the mean rotation velocity of gas particles, while $M_{\rm real}$ is the mass obtained by adding gas, stars, and DM particles within the TDG radius. The time $t=0$ corresponds to the snapshot of the simulation that reproduces the present-day large-scale properties of the interacting system. Red dots are TDGs formed in the tidal tails of a standard disc-disc merger that resembles NGC~7252. Black crosses are TDGs formed in a collisional \hi ring that matches the properties of NGC~5291 (B07).}
\label{fig:simu}
\end{figure}

\citet{Bournaud06} and B07 presented hydro-dynamical simulations of interacting/merging systems, investigating the detailed process of TDG formation. We have re-analysed the output of these simulations to check the hypothesis that the relaxation time in TDGs could be shorter than the orbital time. For different snapshot of the simulations, we measured (i) the dynamical mass $M_{\rm dyn}$ of the simulated TDGs inferred from gas rotation curves, i.e. considering the rotation velocities of individual gas particles with respect to the mean spin axis, and (ii) the real mass $M_{\rm real}$ within the TDG radius, obtained by summing gas, star, and DM particles (the latter ones constitute a negligible amount of mass). For a galaxy in dynamical equilibrium, we expect $M_{\rm dyn}/M_{\rm real} = 1$ within $\sim$10$\%$ (the typical errors associated with the estimate of $M_{\rm dyn}$ from simulated gas rotation curves). In Fig.~\ref{fig:simu}, the ratio $M_{\rm dyn}/M_{\rm real}$ is plotted versus the look-back time, where $t=0$ corresponds to the snapshot of the simulation that reproduces the present-day properties of the observed system. We show two different simulations: red dots are TDGs formed in the tidal tails of a standard disc-disc merger (similar to NGC~7252), black crosses are TDGs formed in a collisional ring (matching the properties of NGC~5291). These simulated TDGs have dynamical masses comparable to those of observed TDGs (cf. B07). At early times, the values of $M_{\rm dyn}/M_{\rm real}$ significantly differ from 1 and display a large scatter, indicating that TDGs are still far from dynamical equilibrium. After a few 100 Myr, however, the scatter substantially decrease and the values of $M_{\rm dyn}/M_{\rm real}$ gets closer and closer to 1. This suggests that the simulated TDGs may have reached dynamical equilibrium in less than an orbital time.

For collisionless stellar systems, it is known that the process of relaxation takes a few orbital times, since it occurs through the energy exchange between individual particles and the total gravitational potential evolving over orbital timescales \citep{LyndenBell67}. For collisional gas-dominated systems (such as TDGs), the energy can be exchanged and radiated away on a local crossing time of the order of $\sim$10 Myr \citep[][]{MacLow99}. Thus, one may speculate that in collisional gas-rich systems the process of relaxation takes less than one orbital time and is significantly faster than in collisionless ones. Further theoretical studies are needed to understand these processes. It is clear, however, that TDGs provide a unique opportunity to study galaxy relaxation and represent a local benchmark for understanding disc formation in gas-rich galaxies at high redshifts.

\subsection{Implications for the DM distribution}

Assuming that the TDGs in our sample are in dynamical equilibrium, we deduce that they are devoid of DM as expected from numerical simulations where tidal forces segregate disc material from halo material \citep{Barnes92, Bournaud06, Wetzstein07}. This provides an indirect but empirical confirmation that most DM must be distributed in pressure-supported halos. 

Using cosmological simulations, \citet{Read08, Read09} and \citet{Pillepich14} argued that spiral galaxies like the Milky Way should possess a rotating disc of non-baryonic DM. This forms because merging satellites are preferentially dragged towards the disc plane by dynamical friction and then disrupted by tides, depositing both stars and DM into the disc. \citet{Ruchti14, Ruchti15} used a chemo-dynamical template to search for accreted stars in the Milky Way and found \textit{no} evidence for an accreted stellar disc, implying that near-plane mergers have been very rare in our Galaxy during the past $\sim$10~Gyr. Nonetheless, a ``dark disk'' may help to explain the vertical kinematics of stars in the solar neighborhood \citep{Bienayme14}. Moreover, \citet{Fan13} and \citet{McCullough13} speculated that exotic forms of self-interacting DM can form rotating dark discs as massive as the baryonic ones. The values of $M_{\rm dyn}/M_{\rm bar}$ in TDGs shed new light on these putative dark discs, given that any kind of rotation-supported material in the progenitor galaxies may form tidal tails and be accreted in TDGs. Our results imply that either these putative dark discs are significantly less massive than the observed baryonic discs or they are too dynamically-hot to be accreted by TDGs.

\subsection{Implications for dark gas}

Several observations suggest that a fraction of gas in galaxy discs is not accounted for by \hi or CO observations. In the Milky Way, this ``dark gas'' has been inferred from (i) $\gamma$-ray emission due to the collision of cosmic rays with interstellar nucleons \citep{Grenier05, Abdo10}, and (ii) a measured excess of thermal dust emission with respect to a canonical dust-to-gas ratio, as indicated by FIR/submm maps from IRAS \citep{Reach94}, COBE \citep{Reach98}, and Planck \citep{Planck11a}. Both techniques indicate that, in the Milky Way, the dark gas mass is comparable to the CO-traced molecular mass. The relative mass of dark gas may be even larger in the LMC \citep{Galliano11}. This dark gas may consist of optically thick \hi and/or H$_2$ in the outer regions of molecular clouds, where CO is photodissociated \citep[][]{Wolfire10, Glover10}. TDGs offer a unique way to probe dark gas in galaxies outside the Local Group, given that this undetected baryonic component would give values of $M_{\rm dyn}/M_{\rm bar}>1$. Unfortunately, our current estimates of $M_{\rm dyn}/M_{\rm bar}$ are not precise enough to identify any potential contribution from dark gas at the expected level.

\citet{Pfenniger94a} and \citet{Pfenniger94b} proposed very cold molecular gas as an alternative to non-baryonic DM in galaxies. This putative dark gas would consist of small ``clumpuscules'' that do not emit any detectable line and are not associated with star formation. Intriguingly, the observed rotation curves of disc galaxies can be well reproduced simply by scaling up the observed contributions of stars and gas \citep{Hoekstra01, Swaters12}. This very cold molecular gas has also been supported by other dynamical arguments such as the existence of spiral patterns and long-lived warps in the outer regions of \hi discs \citep{Bureau99, Masset03, Revaz04, Revaz09}. These H$_2$ clumpuscules should end up in TDGs and be detected as a significant excess of dynamical mass over the baryonic one. In contrast with B07, our current results do not support the existence of large numbers of cold H$_2$ clumpuscules in excess of the more modest expectation for ``standard'' dark gas.

\begin{figure}
\centering
\includegraphics[width=0.45\textwidth]{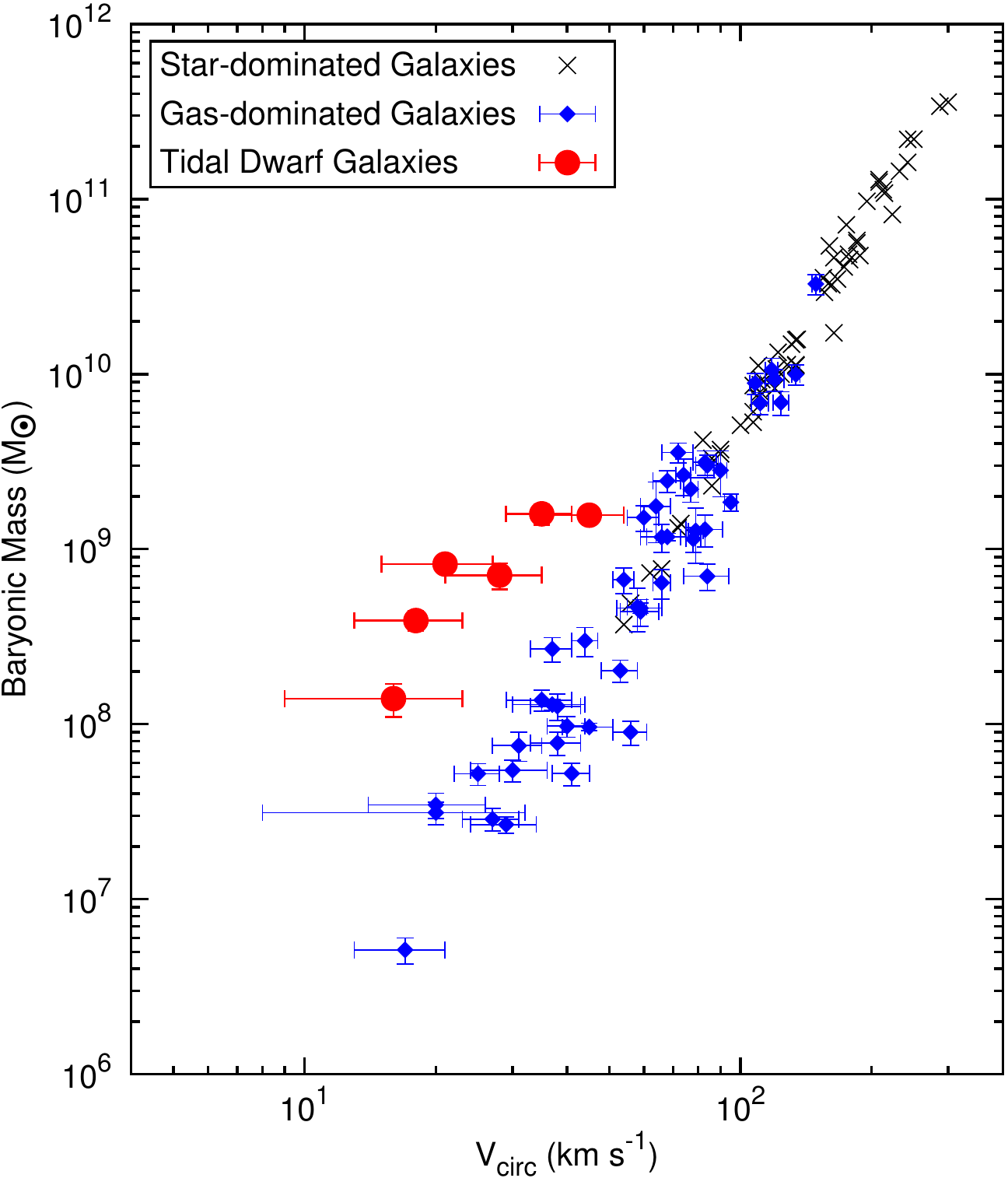}
\caption{The location of TDGs on the BTFR. $V_{\rm circ}$ is the circular velocity after correcting for pressure support (Eq.~\ref{eq:drift}). For star-dominated galaxies \citep[][]{McGaugh05} and gas-dominated ones \citep[][]{McGaugh12}, $V_{\rm circ}$ is measured along the flat part of the rotation curve. For TDGs, the shape of the rotation curve is uncertain, thus $V_{\rm circ}$ may not correspond to the velocity along its flat part.}
\label{fig:BTFR}
\end{figure}

\begin{table*}[t!]
\centering
\caption{MOND analysis assuming dynamical equilibrium.}
\begin{tabular}{lcccccccc}
\hline
Object     & $D_{\rm p}$  & $g_{\rm Ne}/a_{0}$ & $g_{\rm Ni}/a_{0}$ & $V_{\rm ISO(1)}$ & $V_{\rm ISO(2)}$ & $V_{\rm EFE(1)}$ & $V_{\rm EFE(2)}$ & $V_{\rm circ}$\\
           & ($\kpc$)     &                    &                    & ($\kms$)          & ($\kms$)          & ($\kms$)          & ($\kms$)& ($\kms$)\\
\hline
NGC~5291N  & 65           & 0.021 & 0.073 & 77$\pm$2 & 73$\pm$2 & 62$\pm$7 & 57$\pm$7 &45$\pm$9\\
NGC~5291S  & 60           & 0.025 & 0.033 & 76$\pm$3 & 73$\pm$2 & 54$\pm$8 & 49$\pm$8 &35$\pm$6\\
NGC~5291SW & 75           & 0.016 & 0.033 & 62$\pm$3 & 59$\pm$3 & 46$\pm$7 & 43$\pm$7 &28$\pm$7\\
NGC~7252E  & 50           & 0.044 & 0.019 & 52$\pm$2 & 50$\pm$2 & 30$\pm$5 & 28$\pm$5 &18$\pm$5\\
NGC~7252NW & 80           & 0.017 & 0.015 & 63$\pm$2 & 61$\pm$2 & 41$\pm$6 & 39$\pm$6 &21$\pm$6\\
VCC~2062   & 20           & 0.024 & 0.022 & 41$\pm$2 & 39$\pm$2 & 27$\pm$5 & 25$\pm$5 &16$\pm$7\\
\hline
\end{tabular}
\tablefoot{$D_{\rm p}$ is the projected distance of the TDG to the central galaxy remnant, which is used to estimate $g_{\rm Ne}$.}
\label{tab:MOND}
\end{table*}
\subsection{The baryonic Tully-Fisher relation and MOND}
\label{sec:MOND}
Fig.~\ref{fig:BTFR} shows the location of TDGs on the BTFR. We use data from \citet{McGaugh05} for star-dominated galaxies and \citet{McGaugh12} for gas-dominated ones. TDGs are systematically shifted to the left side of the BTFR. This is not fully surprising since these TDGs seem to have smaller DM content than typical dwarf galaxies (Fig.~\ref{fig:mass}), hence they have lower rotation velocities for a given $M_{\rm bar}$. Note that \citet{McGaugh05, McGaugh12} used the asymptotic velocity along the flat part of the rotation curve ($V_{\rm flat}$), which is known to minimize the scatter on the BTFR \citep[][]{Verheijen01, Noordermeer07}. For the TDGs in our sample, the shape of the rotation curve is uncertain, thus it remains unclear whether we are really tracing $V_{\rm flat}$. The \hi data, however, are consistent with a flat rotation curve (Sect.~\ref{sec:3Dmodels}). 

The BTFR is one of the major successful predictions of MOND \citep[e.g.][]{McGaugh05, McGaugh12}. According to MOND, disc galaxies can deviate from the BTFR in only two circumstances: (i) if they are out of dynamical equilibrium \citep[e.g.][]{McGaugh10} and/or (ii) if they are affected by the external-field effect \citep[EFE,][]{Bekenstein84}. The numerical simulations discussed in Sect.~\ref{sec:simu} suggest that TDGs may reach dynamical equilibrium after only a few 100~Myr (see Fig.~\ref{fig:simu}), but it remains unclear whether this would still hold in MOND. Hydrodynamical simulations in MOND are needed to clarify this issue. 

If MOND is interpreted as modified gravity, the strong equivalence principle is violated and the internal dynamics of a system can be affected by an external gravitational field \citep{Bekenstein84}. We define $g_{\rm Ni} = G M_{\rm bar}/R_{\rm out}^2$ as the internal Newtonian acceleration of the TDG. We also define $g_{\rm Ne} = G M_{\rm host}/D_{\rm p}^2$ as the external Newtonian acceleration, where $M_{\rm host}$ is the baryonic mass of the central galaxy remnant and $D_{\rm p}$ is the projected distance to the TDGs. $M_{\rm host}$ is estimated as $L_{\rm K} \times \Upsilon^{\rm K}_{*}$, where $L_{\rm K}$ is given in Table~\ref{tab:sample} and $\Upsilon^{\rm K}_{*}=0.6$ $M_{\odot}/L_{\odot}$ \citep{McGaugh15}. The values of $g_{\rm Ni}$ and $g_{\rm Ne}$ are given in Table~\ref{tab:MOND}. They are roughly comparable, hence the EFE may be important.

To a first approximation, the EFE can be described by Eq.~60 of \citet{Famaey12}\footnote{This equation derives from the quasi-linear formulation of MOND \citep[][]{Milgrom10}. We have also considered the non-linear formulation of \citet{Bekenstein84} using Eq.~59 of \citet{Famaey12} and adopting the simple interpolation function. We find consistent results within $\sim$1 $\kms$.}:
\begin{equation}\label{eq:MOND}
a_{i} = g_{\rm Ni} \nu\bigg(\dfrac{g_{\rm Ni}+g_{\rm Ne}}{a_0}\bigg) + g_{\rm Ne}\bigg[ \nu\bigg(\dfrac{g_{\rm Ni}+g_{\rm Ne}}{a_0}\bigg) -\nu\bigg(\dfrac{g_{\rm Ne}}{a_0}\bigg) \bigg],
\end{equation}
where $a_{\rm i}=V^2/R$ is the internal acceleration of the TDG, $a_{0}\simeq1.3\times10^{-8}$ cm~s$^{-2}$ is the MOND acceleration scale, and $\nu(y)$ is the interpolation function. Following \citet{McGaugh08}, we assume $\nu_{n}(y) = 1/[1 - \exp(-y^{n/2})]^{1/n}$, where the cases $n=1$ and $n=2$ nearly correspond to the so-called ``simple'' and ``standard'' functions, respectively. In Table~\ref{tab:MOND} we provide the velocities expected in MOND for $n=1$ and $n=2$, considering both the isolated case ($V_{\rm ISO}$ with $g_{\rm Ne} = 0$) and the EFE ($V_{\rm EFE}$). Formal errors are estimated considering the uncertainties on $M_{\rm bar}$ and allowing $M_{\rm host}$ to vary by a factor of 2. The MOND velocities are decreased by the EFE, but they are still \textit{systematically} higher than the observed ones. For individual TDGs, the tension between MOND and observations is between $\sim$1 and $\sim$2$\sigma$.

We stress that Eq.~\ref{eq:MOND} grossly simplifies the EFE because it formally applies to the one-dimensional case and neglects the direction of the external field. We are also neglecting the time evolution of the EFE, which may be relevant here because TDGs start to form closer to the host galaxy (where the EFE is stronger) and are subsequently ejected to larger distances at high speeds. Since TDGs have long orbital timescales, they may respond in a non-adiabatic fashion to the rapidly varying EFE, keeping memory of the initial, stronger EFE. It would be interesting to investigate whether a more rigorous treatment of the EFE (using numerical simulations) could further decrease the velocities expected in MOND.

Finally, we note that MOND can also be interpreted as modified inertia \citep{Milgrom94, Milgrom06}. In this case, the theory becomes strongly non-local and the force acting on a test particle depends on the full orbital history. A simple solution can be obtained in the case of stable, stationary, circular orbits \citep{Milgrom94}, but it is unclear what to expect for discs that have undergone less than one rotation. In this respect, TDGs greatly differ from typical disc galaxies and could, in principle, provide a test for such non-local theories.
 
\section{Conclusions}
\label{sec:conclusions}

We presented a systematic study of the gas dynamics in a sample of six TDGs around three interacting/merging systems: NGC~4694, NGC~5291, and NGC~7252 (``Atoms for Peace''). For NGC~4694 and NGC~5291, we re-analysed existing \hi datacubes from D07 and B07, respectively. For NGC~7252, we presented new \hi observations from the Jansky VLA, IFU observations from VLT-GIRAFFE, and long-slit spectroscopy from VLT-FORS1. Our results can be summarized as follows:
\begin{enumerate}
\item The putative TDGs near NGC~7252 have near solar metallicities (12+log(O/H)$\simeq$8.6), indicating that they are currently forming out of pre-enriched tidal material. Moreover, they are associated with steep \hi velocity gradients that represent kinematic discontinuities along the tails, pointing to local potential wells. Similar results have been found for three TDG candidates near NGC~5291 (B07) and one near NGC~4694 (D07). This confirms that the six galaxies in our sample are bona-fide TDGs.
\item The H$\alpha$ kinematics of NGC~7252NW is complex and characterized by very broad line profiles. Most likely, we are observing \HII regions dominated by turbulence and/or non-circular motions due to stellar feedback.
\item The \hi emission associated with TDGs can be described by 3D disc models. These discs, however, have orbital times ranging from $\sim$1 to $\sim$3 Gyr, which are significantly longer than the TDG formation timescales ($\lesssim$1~Gyr). This raises the question as to whether TDGs have had enough time to reach dynamical equilibrium. TDGs appear to be rotation-supported galaxies that are forming at $z\simeq0$ and, thus, are key objects to study the processes of disc building and relaxation.
\item If one assumes that TDGs are in dynamical equilibrium (as suggested by numerical simulations), the inferred dynamical masses are fully consistent with the observed baryonic masses. This implies that TDGs are devoid of DM, as predicted by theoretical models where tidal forces segregate baryons in the disc from DM in the halo. This also puts constraints on putative dark discs (either baryonic or non-baryonic) in the progenitor galaxies. Moreover, TDGs seem to systematically deviate from the baryonic Tully-Fisher relation. These results provide a challenging test for theories like MOND.
\end{enumerate}

\begin{acknowledgements}
We thank Benoit Famaey, Ralf Kotulla, Pavel Kroupa, Chris Mihos, Moti Milgrom, Marcel Pawlowski, and Florent Renaud for stimulating discussions and suggestions. The work of FL and SSM is made possible through the support of a grant from the John Templeton Foundation. FB acknowledges support from the EC through grant ERC-StG-257720. UL has been supported by research projects AYA2007-67625-C02-02 and AYA2011-24728 from the Spanish Ministerio de Ciencia y Educaci\'on and the Junta de Andaluc\'ia (Spain) grants FQM108.
\end{acknowledgements}

\bibliographystyle{aa}
\bibliography{bibliography}

\begin{thebibliography}{112}
\expandafter\ifx\csname natexlab\endcsname\relax\def\natexlab#1{#1}\fi

\bibitem[{{Abdo} {et~al.}(2010){Abdo}, {Ackermann}, {Ajello}, {Baldini},
  {Ballet}, {Barbiellini}, {Bastieri}, {Baughman}, {Bechtol}, {Bellazzini},
  {Berenji}, {Bloom}, {Bonamente}, {Borgland}, {Bregeon}, {Brez}, {Brigida},
  {Bruel}, {Burnett}, {Buson}, {Caliandro}, {Cameron}, {Caraveo}, {Casandjian},
  {Cecchi}, {{\c C}elik}, {Chekhtman}, {Cheung}, {Chiang}, {Ciprini}, {Claus},
  {Cohen-Tanugi}, {Cominsky}, {Conrad}, {Dermer}, {de Palma}, {Digel}, {Silva},
  {Drell}, {Dubois}, {Dumora}, {Farnier}, {Favuzzi}, {Fegan}, {Focke},
  {Fortin}, {Frailis}, {Fukazawa}, {Funk}, {Fusco}, {Gargano}, {Gehrels},
  {Germani}, {Giavitto}, {Giebels}, {Giglietto}, {Giordano}, {Glanzman},
  {Godfrey}, {Grenier}, {Grondin}, {Grove}, {Guillemot}, {Guiriec}, {Harding},
  {Hayashida}, {Horan}, {Hughes}, {Jackson}, {J{\'o}hannesson}, {Johnson},
  {Johnson}, {Kamae}, {Katagiri}, {Kataoka}, {Kawai}, {Kerr}, {Kn{\"o}dlseder},
  {Kuss}, {Lande}, {Latronico}, {Lemoine-Goumard}, {Longo}, {Loparco}, {Lott},
  {Lovellette}, {Lubrano}, {Makeev}, {Mazziotta}, {McEnery}, {Meurer},
  {Michelson}, {Mitthumsiri}, {Mizuno}, {Monte}, {Monzani}, {Morselli},
  {Moskalenko}, {Murgia}, {Nolan}, {Norris}, {Nuss}, {Ohsugi}, {Okumura},
  {Omodei}, {Orlando}, {Ormes}, {Paneque}, {Pelassa}, {Pepe}, {Pesce-Rollins},
  {Piron}, {Porter}, {Rain{\`o}}, {Rando}, {Razzano}, {Reimer}, {Reimer},
  {Reposeur}, {Rodriguez}, {Ryde}, {Sadrozinski}, {Sanchez}, {Sander}, {Saz
  Parkinson}, {Sgr{\`o}}, {Siskind}, {Smith}, {Spandre}, {Spinelli}, {Starck},
  {Strickman}, {Strong}, {Suson}, {Takahashi}, {Tanaka}, {Thayer}, {Thayer},
  {Thompson}, {Tibaldo}, {Torres}, {Tosti}, {Tramacere}, {Uchiyama}, {Usher},
  {Vasileiou}, {Vilchez}, {Vitale}, {Waite}, {Wang}, {Winer}, {Wood}, {Ylinen},
  {Ziegler}, \& {Fermi/LAT Collaboration}}]{Abdo10}
{Abdo}, A.~A., {Ackermann}, M., {Ajello}, M., {et~al.} 2010, \apj, 710, 133

\bibitem[{{Baade} {et~al.}(1999){Baade}, {Meisenheimer}, {Iwert}, {Alonso},
  {Augusteijn}, {Beletic}, {Bellemann}, {Benesch}, {B{\"o}hm}, {B{\"o}hnhardt},
  {Brewer}, {Deiries}, {Delabre}, {Donaldson}, {Dupuy}, {Franke}, {Gerdes},
  {Gilliotte}, {Grimm}, {Haddad}, {Hess}, {Ihle}, {Klein}, {Lenzen}, {Lizon},
  {Mancini}, {M{\"u}nch}, {Pizarro}, {Prado}, {Rahmer}, {Reyes}, {Richardson},
  {Robledo}, {Sanchez}, {Silber}, {Sinclaire}, {Wackermann}, \&
  {Zaggia}}]{Baade99}
{Baade}, D., {Meisenheimer}, K., {Iwert}, O., {et~al.} 1999, The Messenger, 95,
  15

\bibitem[{{Barnes} \& {Hernquist}(1992)}]{Barnes92}
{Barnes}, J.~E. \& {Hernquist}, L. 1992, \nat, 360, 715

\bibitem[{{Bekenstein} \& {Milgrom}(1984)}]{Bekenstein84}
{Bekenstein}, J. \& {Milgrom}, M. 1984, \apj, 286, 7

\bibitem[{{Bienaym{\'e}} {et~al.}(2014){Bienaym{\'e}}, {Famaey}, {Siebert},
  {Freeman}, {Gibson}, {Gilmore}, {Grebel}, {Bland-Hawthorn}, {Kordopatis},
  {Munari}, {Navarro}, {Parker}, {Reid}, {Seabroke}, {Siviero}, {Steinmetz},
  {Watson}, {Wyse}, \& {Zwitter}}]{Bienayme14}
{Bienaym{\'e}}, O., {Famaey}, B., {Siebert}, A., {et~al.} 2014, \aap, 571, A92

\bibitem[{{Boquien} {et~al.}(2010){Boquien}, {Duc}, {Galliano}, {Braine},
  {Lisenfeld}, {Charmandaris}, \& {Appleton}}]{Boquien10}
{Boquien}, M., {Duc}, P.-A., {Galliano}, F., {et~al.} 2010, \aj, 140, 2124

\bibitem[{{Boquien} {et~al.}(2009){Boquien}, {Duc}, {Wu}, {Charmandaris},
  {Lisenfeld}, {Braine}, {Brinks}, {Iglesias-P{\'a}ramo}, \& {Xu}}]{Boquien09}
{Boquien}, M., {Duc}, P.-A., {Wu}, Y., {et~al.} 2009, \aj, 137, 4561

\bibitem[{{Boquien} {et~al.}(2011){Boquien}, {Lisenfeld}, {Duc}, {Braine},
  {Bournaud}, {Brinks}, \& {Charmandaris}}]{Boquien11}
{Boquien}, M., {Lisenfeld}, U., {Duc}, P.-A., {et~al.} 2011, \aap, 533, A19

\bibitem[{{Bournaud} \& {Duc}(2006)}]{Bournaud06}
{Bournaud}, F. \& {Duc}, P.-A. 2006, \aap, 456, 481

\bibitem[{{Bournaud} {et~al.}(2004){Bournaud}, {Duc}, {Amram}, {Combes}, \&
  {Gach}}]{Bournaud04}
{Bournaud}, F., {Duc}, P.-A., {Amram}, P., {Combes}, F., \& {Gach}, J.-L. 2004,
  \aap, 425, 813

\bibitem[{{Bournaud} {et~al.}(2007){Bournaud}, {Duc}, {Brinks}, {Boquiem},
  {Amram}, {Lisenfeld}, {Koribalski}, {Walter}, \& {Charmadaris}}]{Bournaud07}
{Bournaud}, F., {Duc}, P.-A., {Brinks}, E., {et~al.} 2007, Science, 316, 1186

\bibitem[{{Braine} {et~al.}(2001){Braine}, {Duc}, {Lisenfeld}, {Charmandaris},
  {Vallejo}, {Leon}, \& {Brinks}}]{Braine01}
{Braine}, J., {Duc}, P.-A., {Lisenfeld}, U., {et~al.} 2001, \aap, 378, 51

\bibitem[{{Briggs}(1995)}]{Briggs95}
{Briggs}, D.~S. 1995, in Bulletin of the American Astronomical Society,
  Vol.~27, American Astronomical Society Meeting Abstracts, 1444

\bibitem[{{Bureau} {et~al.}(1999){Bureau}, {Freeman}, {Pfitzner}, \&
  {Meurer}}]{Bureau99}
{Bureau}, M., {Freeman}, K.~C., {Pfitzner}, D.~W., \& {Meurer}, G.~R. 1999,
  \aj, 118, 2158

\bibitem[{{Chien} \& {Barnes}(2010)}]{Chien10}
{Chien}, L.-H. \& {Barnes}, J.~E. 2010, \mnras, 407, 43

\bibitem[{{Cornwell}(2008)}]{Cornwell08}
{Cornwell}, T.~J. 2008, IEEE Journal of Selected Topics in Signal Processing,
  2, 793

\bibitem[{{Dabringhausen} \& {Kroupa}(2013)}]{Dabringhausen13}
{Dabringhausen}, J. \& {Kroupa}, P. 2013, \mnras, 429, 1858

\bibitem[{{Denicol{\'o}} {et~al.}(2002){Denicol{\'o}}, {Terlevich}, \&
  {Terlevich}}]{Denicolo02}
{Denicol{\'o}}, G., {Terlevich}, R., \& {Terlevich}, E. 2002, \mnras, 330, 69

\bibitem[{{Duc} {et~al.}(2000){Duc}, {Brinks}, {Springel}, {Pichardo},
  {Weilbacher}, \& {Mirabel}}]{Duc00}
{Duc}, P., {Brinks}, E., {Springel}, V., {et~al.} 2000, \aj, 120, 1238

\bibitem[{{Duc} {et~al.}(2004){Duc}, {Bournaud}, \& {Masset}}]{Duc04}
{Duc}, P.-A., {Bournaud}, F., \& {Masset}, F. 2004, \aap, 427, 803

\bibitem[{{Duc} {et~al.}(2007){Duc}, {Braine}, {Lisenfeld}, {Brinks}, \&
  {Boquien}}]{Duc07}
{Duc}, P.-A., {Braine}, J., {Lisenfeld}, U., {Brinks}, E., \& {Boquien}, M.
  2007, \aap, 475, 187

\bibitem[{{Duc} \& {Mirabel}(1998)}]{Duc98b}
{Duc}, P.-A. \& {Mirabel}, I.~F. 1998, \aap, 333, 813

\bibitem[{{Duc} {et~al.}(2014){Duc}, {Paudel}, {McDermid}, {Cuillandre},
  {Serra}, {Bournaud}, {Cappellari}, \& {Emsellem}}]{Duc14}
{Duc}, P.-A., {Paudel}, S., {McDermid}, R.~M., {et~al.} 2014, \mnras, 440, 1458

\bibitem[{{Dupraz} {et~al.}(1990){Dupraz}, {Casoli}, {Combes}, \&
  {Kazes}}]{Dupraz90}
{Dupraz}, C., {Casoli}, F., {Combes}, F., \& {Kazes}, I. 1990, \aap, 228, L5

\bibitem[{{Elmegreen} {et~al.}(1993){Elmegreen}, {Kaufman}, \&
  {Thomasson}}]{Elmegreen93}
{Elmegreen}, B.~G., {Kaufman}, M., \& {Thomasson}, M. 1993, \apj, 412, 90

\bibitem[{{Famaey} \& {McGaugh}(2012)}]{Famaey12}
{Famaey}, B. \& {McGaugh}, S.~S. 2012, Living Reviews in Relativity, 15, 10

\bibitem[{{Fan} {et~al.}(2013){Fan}, {Katz}, {Randall}, \& {Reece}}]{Fan13}
{Fan}, J., {Katz}, A., {Randall}, L., \& {Reece}, M. 2013, Physical Review
  Letters, 110, 211302

\bibitem[{{Ferguson} {et~al.}(1998){Ferguson}, {Gallagher}, \&
  {Wyse}}]{Ferguson98}
{Ferguson}, A.~M.~N., {Gallagher}, J.~S., \& {Wyse}, R.~F.~G. 1998, \aj, 116,
  673

\bibitem[{{Ferrarese} {et~al.}(2012){Ferrarese}, {C{\^o}t{\'e}}, {Cuillandre},
  {Gwyn}, {Peng}, {MacArthur}, {Duc}, {Boselli}, {Mei}, {Erben}, {McConnachie},
  {Durrell}, {Mihos}, {Jord{\'a}n}, {Lan{\c c}on}, {Puzia}, {Emsellem},
  {Balogh}, {Blakeslee}, {van Waerbeke}, {Gavazzi}, {Vollmer}, {Kavelaars},
  {Woods}, {Ball}, {Boissier}, {Courteau}, {Ferriere}, {Gavazzi},
  {Hildebrandt}, {Hudelot}, {Huertas-Company}, {Liu}, {McLaughlin}, {Mellier},
  {Milkeraitis}, {Schade}, {Balkowski}, {Bournaud}, {Carlberg}, {Chapman},
  {Hoekstra}, {Peng}, {Sawicki}, {Simard}, {Taylor}, {Tully}, {van Driel},
  {Wilson}, {Burdullis}, {Mahoney}, \& {Manset}}]{Ferrarese12}
{Ferrarese}, L., {C{\^o}t{\'e}}, P., {Cuillandre}, J.-C., {et~al.} 2012, \apjs,
  200, 4

\bibitem[{{Freeman}(1970)}]{Freeman70}
{Freeman}, K.~C. 1970, \apj, 160, 811

\bibitem[{{Galliano} {et~al.}(2011){Galliano}, {Hony}, {Bernard}, {Bot},
  {Madden}, {Roman-Duval}, {Galametz}, {Li}, {Meixner}, {Engelbracht},
  {Lebouteiller}, {Misselt}, {Montiel}, {Panuzzo}, {Reach}, \&
  {Skibba}}]{Galliano11}
{Galliano}, F., {Hony}, S., {Bernard}, J.-P., {et~al.} 2011, \aap, 536, A88

\bibitem[{{Gentile} {et~al.}(2010){Gentile}, {Baes}, {Famaey}, \& {van
  Acoleyen}}]{Gentile10}
{Gentile}, G., {Baes}, M., {Famaey}, B., \& {van Acoleyen}, K. 2010, \mnras,
  406, 2493

\bibitem[{{Gentile} {et~al.}(2007{\natexlab{a}}){Gentile}, {Famaey}, {Combes},
  {Kroupa}, {Zhao}, \& {Tiret}}]{Gentile07b}
{Gentile}, G., {Famaey}, B., {Combes}, F., {et~al.} 2007{\natexlab{a}}, \aap,
  472, L25

\bibitem[{{Gentile} {et~al.}(2007{\natexlab{b}}){Gentile}, {Salucci}, {Klein},
  \& {Granato}}]{Gentile07a}
{Gentile}, G., {Salucci}, P., {Klein}, U., \& {Granato}, G.~L.
  2007{\natexlab{b}}, \mnras, 375, 199

\bibitem[{{Glover} {et~al.}(2010){Glover}, {Federrath}, {Mac Low}, \&
  {Klessen}}]{Glover10}
{Glover}, S.~C.~O., {Federrath}, C., {Mac Low}, M.-M., \& {Klessen}, R.~S.
  2010, \mnras, 404, 2

\bibitem[{{Greisen}(2003)}]{Greisen03}
{Greisen}, E.~W. 2003, Information Handling in Astronomy - Historical Vistas,
  285, 109

\bibitem[{{Greisen} {et~al.}(2009){Greisen}, {Spekkens}, \& {van
  Moorsel}}]{Greisen09}
{Greisen}, E.~W., {Spekkens}, K., \& {van Moorsel}, G.~A. 2009, \aj, 137, 4718

\bibitem[{{Grenier} {et~al.}(2005){Grenier}, {Casandjian}, \&
  {Terrier}}]{Grenier05}
{Grenier}, I.~A., {Casandjian}, J.-M., \& {Terrier}, R. 2005, Science, 307,
  1292

\bibitem[{{Hibbard} {et~al.}(1994){Hibbard}, {Guhathakurta}, {van Gorkom}, \&
  {Schweizer}}]{Hibbard94}
{Hibbard}, J.~E., {Guhathakurta}, P., {van Gorkom}, J.~H., \& {Schweizer}, F.
  1994, \aj, 107, 67

\bibitem[{{Hibbard} \& {Mihos}(1995)}]{Hibbard95}
{Hibbard}, J.~E. \& {Mihos}, J.~C. 1995, \aj, 110, 140

\bibitem[{{Hoekstra} {et~al.}(2001){Hoekstra}, {van Albada}, \&
  {Sancisi}}]{Hoekstra01}
{Hoekstra}, H., {van Albada}, T.~S., \& {Sancisi}, R. 2001, \mnras, 323, 453

\bibitem[{{Huchtmeier}(1997)}]{Huchtmeier97}
{Huchtmeier}, W.~K. 1997, \aap, 319, 401

\bibitem[{{Hunter} {et~al.}(2000){Hunter}, {Hunsberger}, \& {Roye}}]{Hunter00}
{Hunter}, D.~A., {Hunsberger}, S.~D., \& {Roye}, E.~W. 2000, \apj, 542, 137

\bibitem[{{Kennicutt} \& {Evans}(2012)}]{Kennicutt12}
{Kennicutt}, R.~C. \& {Evans}, N.~J. 2012, \araa, 50, 531

\bibitem[{{Kewley} {et~al.}(2010){Kewley}, {Rupke}, {Zahid}, {Geller}, \&
  {Barton}}]{Kewley10}
{Kewley}, L.~J., {Rupke}, D., {Zahid}, H.~J., {Geller}, M.~J., \& {Barton},
  E.~J. 2010, \apjl, 721, L48

\bibitem[{{Kroupa}(2012)}]{Kroupa12}
{Kroupa}, P. 2012, \pasa, 29, 395

\bibitem[{{Kroupa} \& {Weidner}(2003)}]{Kroupa03}
{Kroupa}, P. \& {Weidner}, C. 2003, \apj, 598, 1076

\bibitem[{{Kuzio de Naray} {et~al.}(2004){Kuzio de Naray}, {McGaugh}, \& {de
  Blok}}]{KdN04}
{Kuzio de Naray}, R., {McGaugh}, S.~S., \& {de Blok}, W.~J.~G. 2004, \mnras,
  355, 887

\bibitem[{{Leitherer} {et~al.}(1999){Leitherer}, {Schaerer}, {Goldader},
  {Delgado}, {Robert}, {Kune}, {de Mello}, {Devost}, \&
  {Heckman}}]{Leitherer99}
{Leitherer}, C., {Schaerer}, D., {Goldader}, J.~D., {et~al.} 1999, \apjs, 123,
  3

\bibitem[{{Lelli} {et~al.}(2010){Lelli}, {Fraternali}, \& {Sancisi}}]{Lelli10}
{Lelli}, F., {Fraternali}, F., \& {Sancisi}, R. 2010, \aap, 516, A11

\bibitem[{{Lelli} {et~al.}(2014{\natexlab{a}}){Lelli}, {Fraternali}, \&
  {Verheijen}}]{Lelli14a}
{Lelli}, F., {Fraternali}, F., \& {Verheijen}, M. 2014{\natexlab{a}}, \aap,
  563, A27

\bibitem[{{Lelli} {et~al.}(2014{\natexlab{b}}){Lelli}, {Verheijen}, \&
  {Fraternali}}]{Lelli14}
{Lelli}, F., {Verheijen}, M., \& {Fraternali}, F. 2014{\natexlab{b}}, \aap,
  566, A71

\bibitem[{{Lelli} {et~al.}(2012{\natexlab{a}}){Lelli}, {Verheijen},
  {Fraternali}, \& {Sancisi}}]{Lelli12a}
{Lelli}, F., {Verheijen}, M., {Fraternali}, F., \& {Sancisi}, R.
  2012{\natexlab{a}}, \aap, 537, A72

\bibitem[{{Lelli} {et~al.}(2012{\natexlab{b}}){Lelli}, {Verheijen},
  {Fraternali}, \& {Sancisi}}]{Lelli12b}
{Lelli}, F., {Verheijen}, M., {Fraternali}, F., \& {Sancisi}, R.
  2012{\natexlab{b}}, \aap, 544, A145

\bibitem[{{Leroy} {et~al.}(2008){Leroy}, {Walter}, {Brinks}, {Bigiel}, {de
  Blok}, {Madore}, \& {Thornley}}]{Leroy08}
{Leroy}, A.~K., {Walter}, F., {Brinks}, E., {et~al.} 2008, \aj, 136, 2782

\bibitem[{{Lynden-Bell}(1967)}]{LyndenBell67}
{Lynden-Bell}, D. 1967, \mnras, 136, 101

\bibitem[{{Mac Low}(1999)}]{MacLow99}
{Mac Low}, M.-M. 1999, \apj, 524, 169

\bibitem[{{Malphrus} {et~al.}(1997){Malphrus}, {Simpson}, {Gottesman}, \&
  {Hawarden}}]{Malphrus97}
{Malphrus}, B.~K., {Simpson}, C.~E., {Gottesman}, S.~T., \& {Hawarden}, T.~G.
  1997, \aj, 114, 1427

\bibitem[{{Masset} \& {Bureau}(2003)}]{Masset03}
{Masset}, F.~S. \& {Bureau}, M. 2003, \apj, 586, 152

\bibitem[{{McCullough} \& {Randall}(2013)}]{McCullough13}
{McCullough}, M. \& {Randall}, L. 2013, \jcap, 10, 58

\bibitem[{{McGaugh}(1991)}]{McGaugh91}
{McGaugh}, S.~S. 1991, \apj, 380, 140

\bibitem[{{McGaugh}(1994)}]{McGaugh94}
{McGaugh}, S.~S. 1994, \apj, 426, 135

\bibitem[{{McGaugh}(2005)}]{McGaugh05}
{McGaugh}, S.~S. 2005, \apj, 632, 859

\bibitem[{{McGaugh}(2008)}]{McGaugh08}
{McGaugh}, S.~S. 2008, \apj, 683, 137

\bibitem[{{McGaugh}(2012)}]{McGaugh12}
{McGaugh}, S.~S. 2012, \aj, 143, 40

\bibitem[{{McGaugh} \& {Schombert}(2015)}]{McGaugh15}
{McGaugh}, S.~S. \& {Schombert}, J.~M. 2015, \apj, 802, 18

\bibitem[{{McGaugh} {et~al.}(2000){McGaugh}, {Schombert}, {Bothun}, \& {de
  Blok}}]{McGaugh00}
{McGaugh}, S.~S., {Schombert}, J.~M., {Bothun}, G.~D., \& {de Blok}, W.~J.~G.
  2000, \apjl, 533, L99

\bibitem[{{McGaugh} \& {Wolf}(2010)}]{McGaugh10}
{McGaugh}, S.~S. \& {Wolf}, J. 2010, \apj, 722, 248

\bibitem[{{Mei} {et~al.}(2007){Mei}, {Blakeslee}, {C{\^o}t{\'e}}, {Tonry},
  {West}, {Ferrarese}, {Jord{\'a}n}, {Peng}, {Anthony}, \& {Merritt}}]{Mei07}
{Mei}, S., {Blakeslee}, J.~P., {C{\^o}t{\'e}}, P., {et~al.} 2007, \apj, 655,
  144

\bibitem[{{Milgrom}(1994)}]{Milgrom94}
{Milgrom}, M. 1994, Annals of Physics, 229, 384

\bibitem[{{Milgrom}(2006)}]{Milgrom06}
{Milgrom}, M. 2006, in EAS Publications Series, Vol.~20, EAS Publications
  Series, ed. G.~A. {Mamon}, F.~{Combes}, C.~{Deffayet}, \& B.~{Fort}, 217--224

\bibitem[{{Milgrom}(2007)}]{Milgrom07}
{Milgrom}, M. 2007, \apjl, 667, L45

\bibitem[{{Milgrom}(2010)}]{Milgrom10}
{Milgrom}, M. 2010, \mnras, 403, 886

\bibitem[{{Montuori} {et~al.}(2010){Montuori}, {Di Matteo}, {Lehnert},
  {Combes}, \& {Semelin}}]{Montuori10}
{Montuori}, M., {Di Matteo}, P., {Lehnert}, M.~D., {Combes}, F., \& {Semelin},
  B. 2010, \aap, 518, A56

\bibitem[{{Noordermeer} \& {Verheijen}(2007)}]{Noordermeer07}
{Noordermeer}, E. \& {Verheijen}, M.~A.~W. 2007, \mnras, 381, 1463

\bibitem[{{Oh} {et~al.}(2015){Oh}, {Hunter}, {Brinks}, {Elmegreen}, {Schruba},
  {Walter}, {Rupen}, {Young}, {Simpson}, {Johnson}, {Herrmann}, {Ficut-Vicas},
  {Cigan}, {Heesen}, {Ashley}, \& {Zhang}}]{Oh15}
{Oh}, S.-H., {Hunter}, D.~A., {Brinks}, E., {et~al.} 2015, \aj, 149, 180

\bibitem[{{Pasquini} {et~al.}(2002){Pasquini}, {Avila}, {Blecha}, {Cacciari},
  {Cayatte}, {Colless}, {Damiani}, {de Propris}, {Dekker}, {di Marcantonio},
  {Farrell}, {Gillingham}, {Guinouard}, {Hammer}, {Kaufer}, {Hill}, {Marteaud},
  {Modigliani}, {Mulas}, {North}, {Popovic}, {Rossetti}, {Royer}, {Santin},
  {Schmutzer}, {Simond}, {Vola}, {Waller}, \& {Zoccali}}]{Pasquini02}
{Pasquini}, L., {Avila}, G., {Blecha}, A., {et~al.} 2002, The Messenger, 110, 1

\bibitem[{{Pfenniger} \& {Combes}(1994)}]{Pfenniger94b}
{Pfenniger}, D. \& {Combes}, F. 1994, \aap, 285, 94

\bibitem[{{Pfenniger} {et~al.}(1994){Pfenniger}, {Combes}, \&
  {Martinet}}]{Pfenniger94a}
{Pfenniger}, D., {Combes}, F., \& {Martinet}, L. 1994, \aap, 285, 79

\bibitem[{{Pillepich} {et~al.}(2014){Pillepich}, {Kuhlen}, {Guedes}, \&
  {Madau}}]{Pillepich14}
{Pillepich}, A., {Kuhlen}, M., {Guedes}, J., \& {Madau}, P. 2014, \apj, 784,
  161

\bibitem[{{Pilyugin}(2000)}]{Pilyugin00}
{Pilyugin}, L.~S. 2000, \aap, 362, 325

\bibitem[{{Planck Collaboration} {et~al.}(2011){Planck Collaboration}, {Ade},
  {Aghanim}, {Arnaud}, {Ashdown}, {Aumont}, {Baccigalupi}, {Balbi}, {Banday},
  {Barreiro}, \& et~al.}]{Planck11a}
{Planck Collaboration}, {Ade}, P.~A.~R., {Aghanim}, N., {et~al.} 2011, \aap,
  536, A19

\bibitem[{{Ploeckinger} {et~al.}(2014){Ploeckinger}, {Hensler}, {Recchi},
  {Mitchell}, \& {Kroupa}}]{Ploeckinger14}
{Ploeckinger}, S., {Hensler}, G., {Recchi}, S., {Mitchell}, N., \& {Kroupa}, P.
  2014, \mnras, 437, 3980

\bibitem[{{Ploeckinger} {et~al.}(2015){Ploeckinger}, {Recchi}, {Hensler}, \&
  {Kroupa}}]{Ploeckinger15}
{Ploeckinger}, S., {Recchi}, S., {Hensler}, G., \& {Kroupa}, P. 2015, \mnras,
  447, 2512

\bibitem[{{Reach} {et~al.}(1994){Reach}, {Koo}, \& {Heiles}}]{Reach94}
{Reach}, W.~T., {Koo}, B.-C., \& {Heiles}, C. 1994, \apj, 429, 672

\bibitem[{{Reach} {et~al.}(1998){Reach}, {Wall}, \& {Odegard}}]{Reach98}
{Reach}, W.~T., {Wall}, W.~F., \& {Odegard}, N. 1998, \apj, 507, 507

\bibitem[{{Read} {et~al.}(2008){Read}, {Lake}, {Agertz}, \&
  {Debattista}}]{Read08}
{Read}, J.~I., {Lake}, G., {Agertz}, O., \& {Debattista}, V.~P. 2008, \mnras,
  389, 1041

\bibitem[{{Read} {et~al.}(2009){Read}, {Mayer}, {Brooks}, {Governato}, \&
  {Lake}}]{Read09}
{Read}, J.~I., {Mayer}, L., {Brooks}, A.~M., {Governato}, F., \& {Lake}, G.
  2009, \mnras, 397, 44

\bibitem[{{Recchi} {et~al.}(2007){Recchi}, {Theis}, {Kroupa}, \&
  {Hensler}}]{Recchi07}
{Recchi}, S., {Theis}, C., {Kroupa}, P., \& {Hensler}, G. 2007, \aap, 470, L5

\bibitem[{{Revaz} \& {Pfenniger}(2004)}]{Revaz04}
{Revaz}, Y. \& {Pfenniger}, D. 2004, \aap, 425, 67

\bibitem[{{Revaz} {et~al.}(2009){Revaz}, {Pfenniger}, {Combes}, \&
  {Bournaud}}]{Revaz09}
{Revaz}, Y., {Pfenniger}, D., {Combes}, F., \& {Bournaud}, F. 2009, \aap, 501,
  171

\bibitem[{{Richter} {et~al.}(1994){Richter}, {Sackett}, \&
  {Sparke}}]{Richter94}
{Richter}, O.-G., {Sackett}, P.~D., \& {Sparke}, L.~S. 1994, \aj, 107, 99

\bibitem[{{Ruchti} {et~al.}(2014){Ruchti}, {Read}, {Feltzing}, {Pipino}, \&
  {Bensby}}]{Ruchti14}
{Ruchti}, G.~R., {Read}, J.~I., {Feltzing}, S., {Pipino}, A., \& {Bensby}, T.
  2014, \mnras, 444, 515

\bibitem[{{Ruchti} {et~al.}(2015){Ruchti}, {Read}, {Feltzing}, {Serenelli},
  {McMillan}, {Lind}, {Bensby}, {Bergemann}, {Asplund}, {Vallenari},
  {Flaccomio}, {Pancino}, {Korn}, {Recio-Blanco}, {Bayo}, {Carraro}, {Costado},
  {Damiani}, {Heiter}, {Hourihane}, {Jofre}, {Kordopatis}, {Lardo}, {de
  Laverny}, {Monaco}, {Morbidelli}, {Sbordone}, {Worley}, \&
  {Zaggia}}]{Ruchti15}
{Ruchti}, G.~R., {Read}, J.~I., {Feltzing}, S., {et~al.} 2015, ArXiv e-prints

\bibitem[{{Rupke} {et~al.}(2010{\natexlab{a}}){Rupke}, {Kewley}, \&
  {Barnes}}]{Rupke10b}
{Rupke}, D.~S.~N., {Kewley}, L.~J., \& {Barnes}, J.~E. 2010{\natexlab{a}},
  \apjl, 710, L156

\bibitem[{{Rupke} {et~al.}(2010{\natexlab{b}}){Rupke}, {Kewley}, \&
  {Chien}}]{Rupke10a}
{Rupke}, D.~S.~N., {Kewley}, L.~J., \& {Chien}, L.-H. 2010{\natexlab{b}}, \apj,
  723, 1255

\bibitem[{{Schweizer}(1982)}]{Schweizer82}
{Schweizer}, F. 1982, \aj, 252, 455

\bibitem[{{Swaters}(1999)}]{Swaters99}
{Swaters}, R.~A. 1999, PhD thesis, , Rijksuniversiteit Groningen, (1999)

\bibitem[{{Swaters} {et~al.}(2009){Swaters}, {Sancisi}, {van Albada}, \& {van
  der Hulst}}]{Swaters09}
{Swaters}, R.~A., {Sancisi}, R., {van Albada}, T.~S., \& {van der Hulst}, J.~M.
  2009, \aap, 493, 871

\bibitem[{{Swaters} {et~al.}(2012){Swaters}, {Sancisi}, {van der Hulst}, \&
  {van Albada}}]{Swaters12}
{Swaters}, R.~A., {Sancisi}, R., {van der Hulst}, J.~M., \& {van Albada}, T.~S.
  2012, \mnras, 425, 2299

\bibitem[{{Swaters} {et~al.}(2002){Swaters}, {van Albada}, {van der Hulst}, \&
  {Sancisi}}]{Swaters02}
{Swaters}, R.~A., {van Albada}, T.~S., {van der Hulst}, J.~M., \& {Sancisi}, R.
  2002, \aap, 390, 829

\bibitem[{{Sweet} {et~al.}(2014){Sweet}, {Drinkwater}, {Meurer}, {Bekki},
  {Dopita}, {Kilborn}, \& {Nicholls}}]{Sweet14}
{Sweet}, S.~M., {Drinkwater}, M.~J., {Meurer}, G., {et~al.} 2014, \apj, 782, 35

\bibitem[{{Toomre}(1977)}]{Toomre77}
{Toomre}, A. 1977, in Evolution of Galaxies and Stellar Populations, ed. B.~M.
  {Tinsley} \& R.~B.~G. {Larson}, D.~Campbell, 401

\bibitem[{{Toomre} \& {Toomre}(1972)}]{Toomre72}
{Toomre}, A. \& {Toomre}, J. 1972, \apj, 178, 623

\bibitem[{{Verheijen}(2001)}]{Verheijen01}
{Verheijen}, M.~A.~W. 2001, \apj, 563, 694

\bibitem[{{Vogelaar} \& {Terlouw}(2001)}]{Vogelaar01}
{Vogelaar}, M.~G.~R. \& {Terlouw}, J.~P. 2001, in Astronomical Society of the
  Pacific Conference Series, Vol. 238, Astronomical Data Analysis Software and
  Systems X, ed. F.~R. {Harnden}, Jr., F.~A. {Primini}, \& H.~E. {Payne}, 358

\bibitem[{{Weilbacher} {et~al.}(2003){Weilbacher}, {Duc}, \&
  {Fritze-v.~Alvensleben}}]{Weilbacher03}
{Weilbacher}, P.~M., {Duc}, P.-A., \& {Fritze-v.~Alvensleben}, U. 2003, \aap,
  397, 545

\bibitem[{{Weilbacher} {et~al.}(2000){Weilbacher}, {Duc}, {Fritze
  v.~Alvensleben}, {Martin}, \& {Fricke}}]{Weilbacher00}
{Weilbacher}, P.~M., {Duc}, P.-A., {Fritze v.~Alvensleben}, U., {Martin}, P.,
  \& {Fricke}, K.~J. 2000, \aap, 358, 819

\bibitem[{{Werk} {et~al.}(2011){Werk}, {Putman}, {Meurer}, \&
  {Santiago-Figueroa}}]{Werk11}
{Werk}, J.~K., {Putman}, M.~E., {Meurer}, G.~R., \& {Santiago-Figueroa}, N.
  2011, \apj, 735, 71

\bibitem[{{Wetzstein} {et~al.}(2007){Wetzstein}, {Naab}, \&
  {Burkert}}]{Wetzstein07}
{Wetzstein}, M., {Naab}, T., \& {Burkert}, A. 2007, \mnras, 375, 805

\bibitem[{{Wolfire} {et~al.}(2010){Wolfire}, {Hollenbach}, \&
  {McKee}}]{Wolfire10}
{Wolfire}, M.~G., {Hollenbach}, D., \& {McKee}, C.~F. 2010, \apj, 716, 1191

\bibitem[{{Zwicky}(1956)}]{Zwicky56}
{Zwicky}, F. 1956, Ergebnisse der exakten Naturwissenschaften, 29, 344

\end{thebibliography}

\newpage
\appendix

\section{Comparison with archival VLA data and single-dish observations for NGC~7252}
\label{app:ArchObs}
For a sanity check, we retrieved from the VLA archive the \hi observations of NGC~7252 from \citet{Hibbard94} and compare them with the new ones. We reduced the archival observations following standard procedures in AIPS and created a datacube with spatial resolution of $25''$ (robust weights) and velocity resolution of $10.6\,\kms$. The D-array data were interpolated to match the spectral resolution of the C-array data.

\begin{figure}[thb]
\centering
\includegraphics[width=0.475\textwidth]{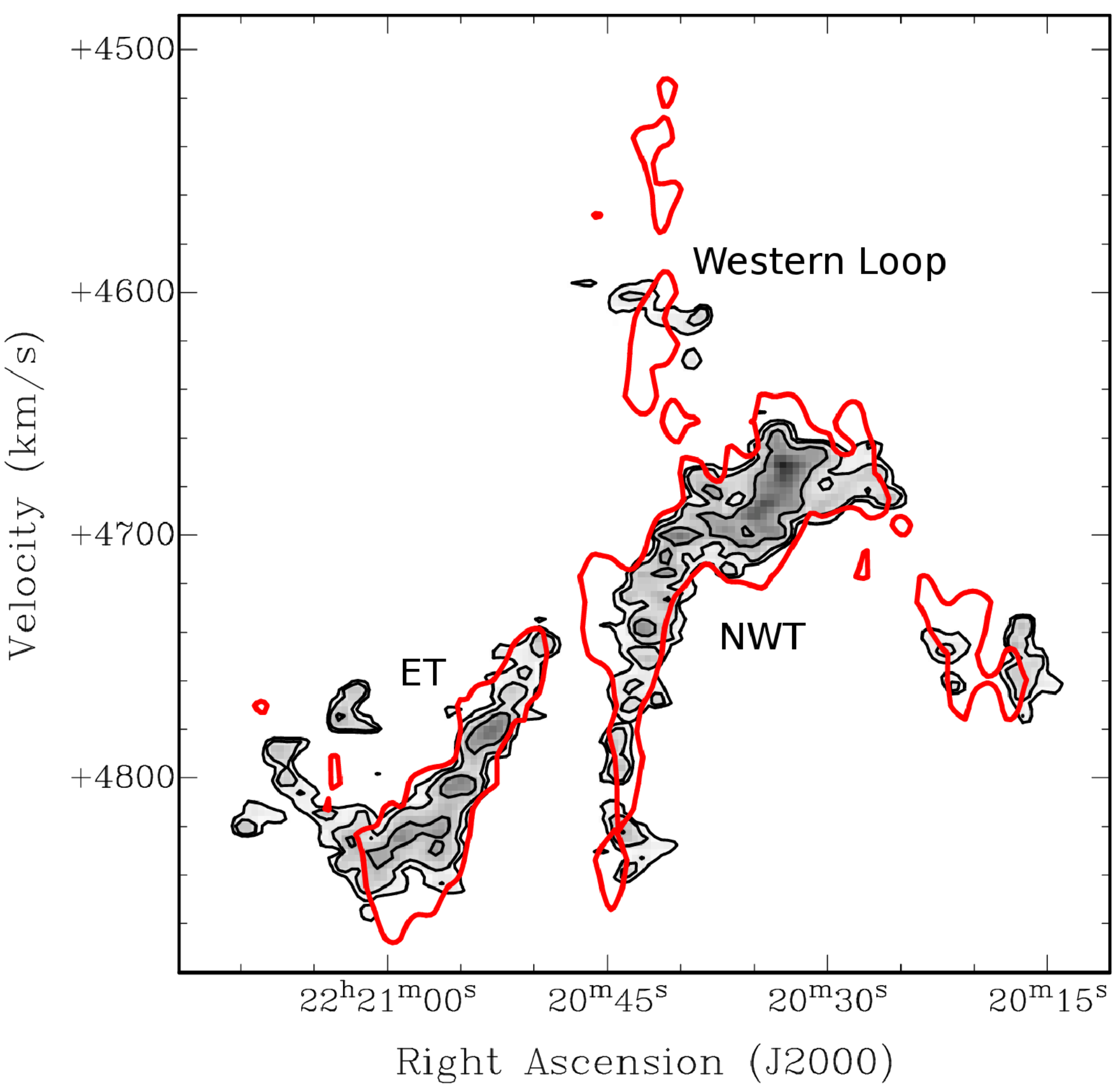}
\caption{PV-diagram for NGC~7252 obtained by integrating the \hi emission along the declination axis. The new data are shown in grayscale. Contours are at 18, 36, and 72 mJy/beam. The archival data \citep[published by][]{Hibbard94} are represented with a red contour at 18 mJy/beam. Both datasets were spatially smoothed to 25$''$ resolution.}
\label{completeness}
\end{figure}
Fig.~\ref{completeness} shows a PV-diagram obtained by collapsing the final datacube along the declination axis, giving a complete overview of the kinematic structure of the system. The comparison between the new (grey scale) and archival (red contour) data shows some differences on small spatial scales due to different noise levels in the two data sets.
Overall the new VLA data are consistent with the archival ones.

From the new VLA data we derive a total \hi flux of 5.0 Jy~$\kms$. This is in good agreement with the single-dish measurement by \citet{Richter94}, who find a total \hi flux of 4.6 Jy~$\kms$ using the Green Bank Telescope ($\rm{FWHM}=21'\times21'$). Other single-dish measurements (who rely on single pointings) reported lower fluxes: \citet{Huchtmeier97} find 3.8 Jy~$\kms$ using the Effelsberg telescope ($\rm{FWHM}=9'\times9'$), while \citet{Dupraz90} find 3.6 Jy~$\kms$ using the Nan\c cay telescope ($\rm{FWHM}=4'\rm{\,[E-W]}\times 22' \rm{\,[N-S]}$). Most likely, the latter single-dish observations miss part of the \hi emission in the tidal tails, extending for $\sim$12$'$ in the E-W direction. We conclude that no significant emission is missing from the new VLA data.

\section{Comparison with B07 for NGC~5291}
\label{app:compaB07}
\begin{figure*}[thb]
\centering
\includegraphics[width=\textwidth]{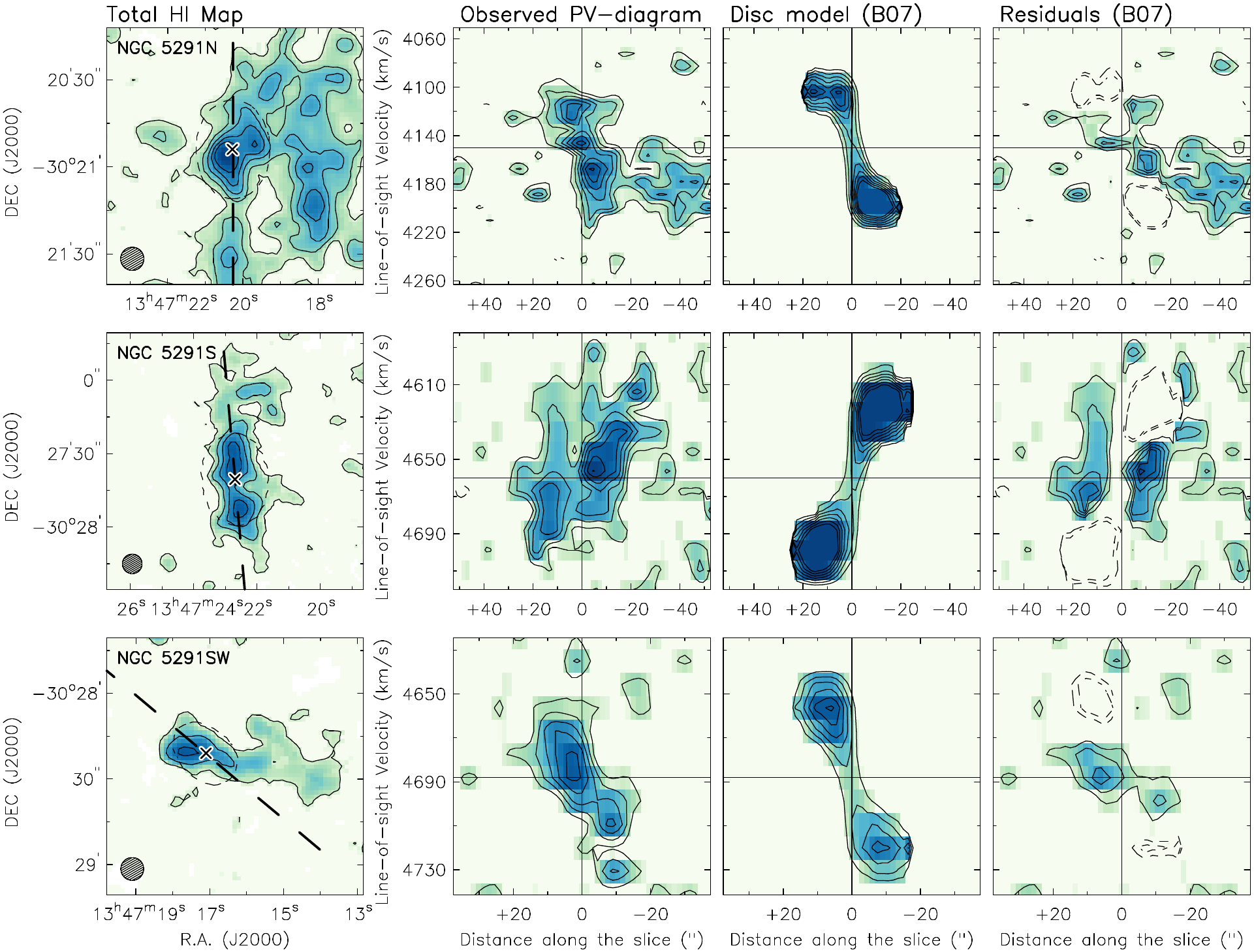}
\caption{Disc models for NGC~5291N (\textit{top}), NGC~5291S (\textit{middle}), and NGC~5291SW (\textit{bottom}) adopting the parameters from B07. \textit{Left panels}: total \hi maps from the observed cube (solid contours) and the model-cube (dashed ellipse). The cross and dashed line illustrate the kinematical centre and major axis, respectively. The circle to the bottom-left corner shows the \hi beam. \textit{Right panels}: PV diagrams along the major axis obtained from the observed cube, model-cube, and residual-cube. Solid contours range from 2$\sigma$ to 8$\sigma$ in steps of $1\sigma$. Dashed contours range from $-$2$\sigma$ to $-$4$\sigma$ in steps of $-$1$\sigma$. The horizontal and vertical lines correspond to the systemic velocity and dynamical centre, respectively.}
\label{fig:compaB07}
\end{figure*}
For the TDGs around NGC~5291, we find significantly lower values of $M_{\rm dyn}/M_{\rm bar}$ than B07 because we estimate both higher values of $M_{\rm gas}$ (by factors of $\sim$1.5 to $\sim$2) and lower values of $V_{\rm rot}$ (by a factor of $\sim$2). There are two reasons for the different estimates of $M_{\rm gas}$: (i) we integrate the \hi flux out to larger radii than B07, according to the results of our 3D disc models; this is the dominant effect; and (ii) we consider all the \hi emission that is consistent with a rotating disc, whereas B07 assumed that some \hi gas at the high/low velocity end was in the background/foreground (instead of being associated with the TDG potential well). The different estimates of $V_{\rm rot}$ are also driven by two separate reasons: (i) we find that line-of-sight velocities at the edges of the PV-diagrams (employed by B07) tend to over-estimate the ``true'' rotation velocity because they do not consider the line-broadening due to turbulent motions; and (ii) we find higher disc inclinations than assumed by B07. Hereafter, we discuss these two effects in details.

B07 derived rotation curves using the envelope-tracing method, which considers line-of-sight velocities near the edges of PV-diagrams. Specifically, for a given radius along the major axis, they took the velocity towards the far edge of the PV-diagram at 50$\%$ of the \hi peak level. For poorly-resolved rotating discs, this is a rough way to consider beam-smearing effects, which are well-known to artificially broaden the emission-line profiles and systematically skew them towards the systemic velocity. The details of beam-smearing effects, however, depend on the intrinsic gas distribution, rotation curve shape, gas velocity dispersion, and disc inclination. Thus, full 3D models are required to take them into account \citep{Swaters09, Lelli10}. In particular, the velocity at 50$\%$ of the \hi peak may systematically over-estimate the rotation velocity in discs with low values of $V_{\rm rot}/\sigma_{\hi}$ (as dwarf galaxies), given that the broadening of the line profile is driven by both $\sigma_{\hi}$ and resolution effects. Our 3D disc models take all these observational effects into account and suggest that the rotation velocities of these TDGs are closer to the \hi peak than adopted by B07. This is demonstrated in Fig.~\ref{fig:compaB07}.

For the three TDGs near NGC~5291, we built additional 3D models following the same procedures of Sect.~\ref{sec:3Dmodels} and adopting the kinematical parameters of B07 ($i=45^{\circ}$, $\sigma_{\hi}=10$ km~s$^{-1}$, and $V_{\rm rot}$ from 50 to 70 km~s$^{-1}$ depending on the TDG). The right panels of Fig.~~\ref{fig:compaB07} compare PV-diagrams along the major axis obtained from the observed cubes and these model-cubes. The kinematic parameters from B07 produce PV-diagrams that are too extended in the velocity direction. This leads to large negative residuals at high/low line-of-sight velocities and large positive residuals near the systemic velocity, indicating that $V_{\rm rot} \sin(i)$ is overestimated and $\sigma_{\hi}$ is underestimated. It is clear that our new kinematic parameters provide a much better description of the data (cf. with Figs.~\ref{fig:N5291N}, \ref{fig:N5291S}, and \ref{fig:N5291SW}).

Regardless of the method employed to estimate the rotation velocities, these projected quantities must be corrected for inclination by multiplying by $1/\sin(i)$. B07 assumed that the TDGs around NGC~5291 have the same inclination angle as the collisional \hi ring ($i = 45^{\circ}$), which was estimated using numerical simulations. B07 pointed out that this value of $i$ is uncertain because (i) the modelled ring is not exactly circular and its morphology depends on the details of the collision, and (ii) some simulated TDGs show a discrepancy between their rotation axis and the ring axis (up to 18$^{\circ}$ in a few cases). Using 3D disc models, we directly estimate $i$ by comparing observed \hi maps with model \hi maps (before pixel-to-pixel renormalization). In left panels of Fig.~\ref{fig:compaB07}, the dashed ellipses correspond to disc models with $i=45^{\circ}$. This value of $i$ is roughly acceptable for NGC~5291N and NGC~5291SW, whereas it is clearly underestimated for NGC~5291S. The dashed ellipses in the top-middle panels of Figs.~\ref{fig:N5291N}, \ref{fig:N5291S}, and \ref{fig:N5291SW} illustrate our adopted inclination, which are slightly larger than $45^{\circ}$ (from 55$^{\circ}$ to 65$^{\circ}$, as given in Table~\ref{tab:TDGkin}). Thus our estimates of $V_{\rm rot}$ are further decreased with respect to those of B07 by a factor $\sin(i=45^{\circ})/\sin(i_{\rm new})$, i.e. by $\sim$15$\%$ to $\sim$30$\%$. Clearly, inclination plays a minor role here.

\section{Channel Maps of individual TDGs}\label{sec:ChanMaps}

\begin{figure*}[thb]
\centering
\includegraphics[width=\textwidth]{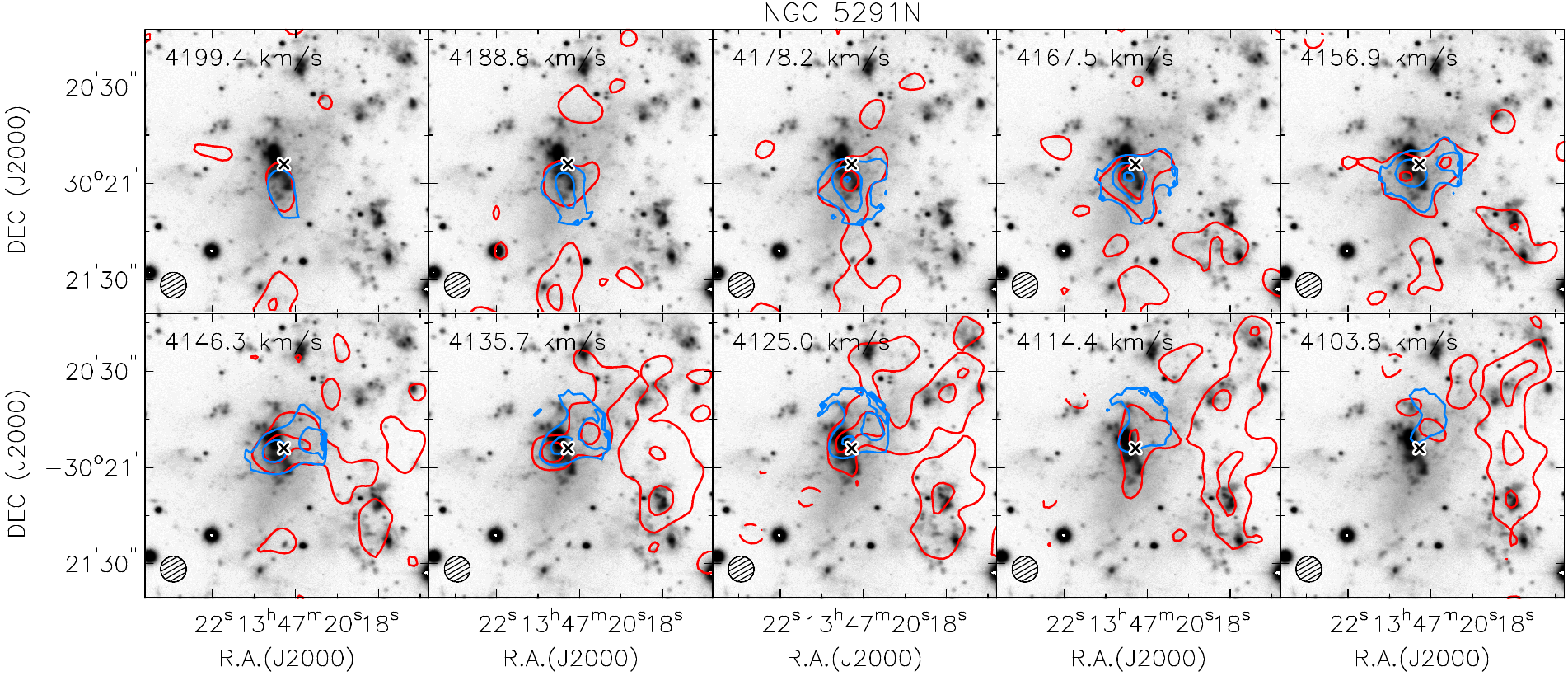}\vspace{0.1cm}
\includegraphics[width=\textwidth]{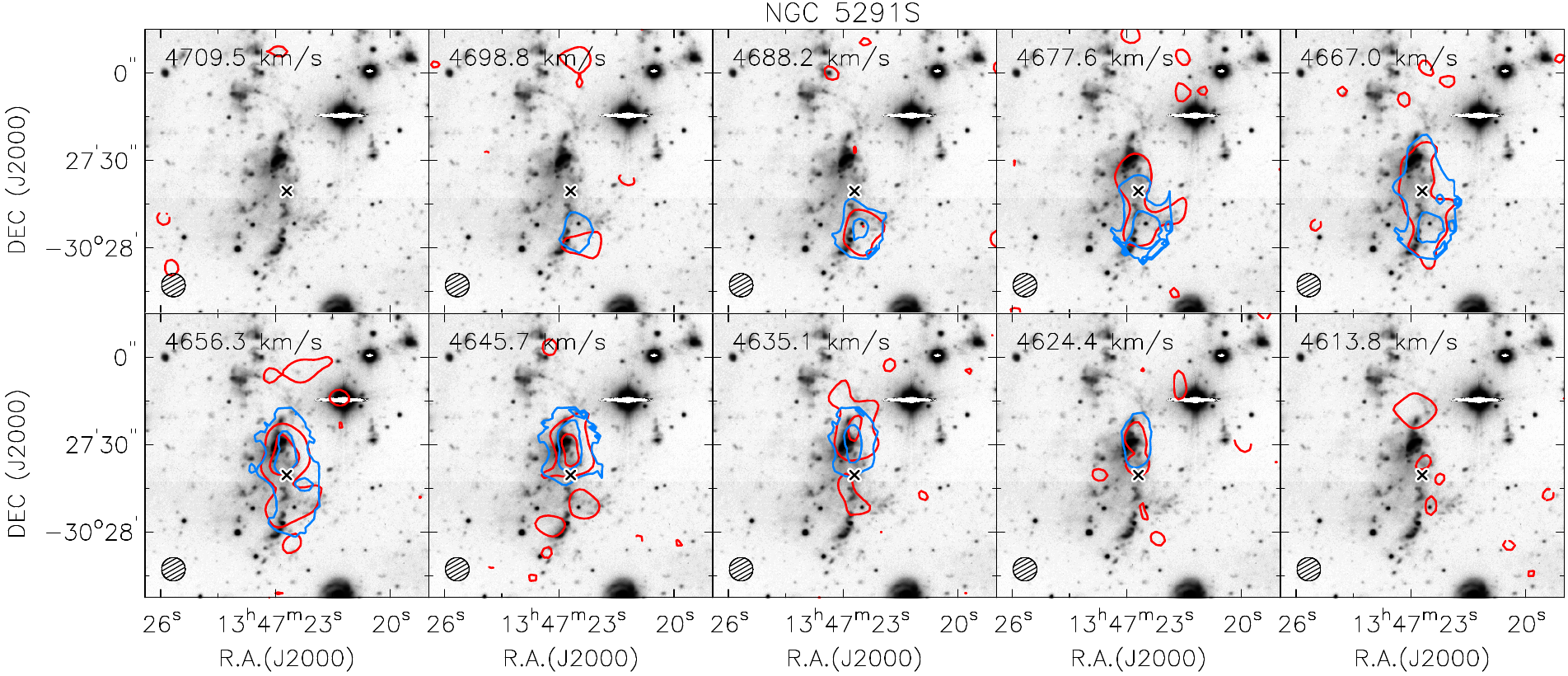}\vspace{0.1cm}
\includegraphics[width=\textwidth]{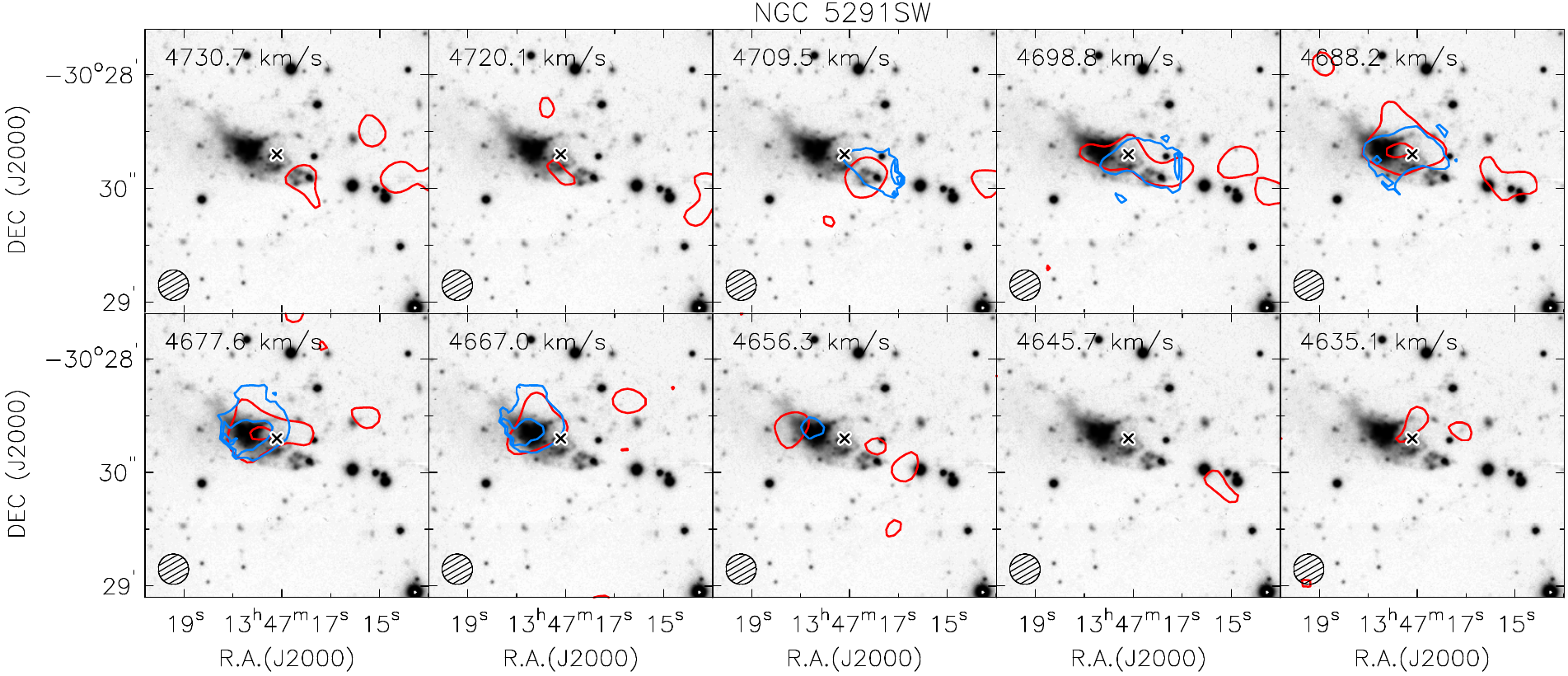}
\caption{\hi channel maps for NGC~5291N (top), NGC~5291S (middle), and NGC~5291SW (bottom). See Appendix \ref{sec:ChanMaps} for a detailed description of this image.}
\label{fig:CM1}
\end{figure*}

\begin{figure*}[thb]
\centering
\includegraphics[width=\textwidth]{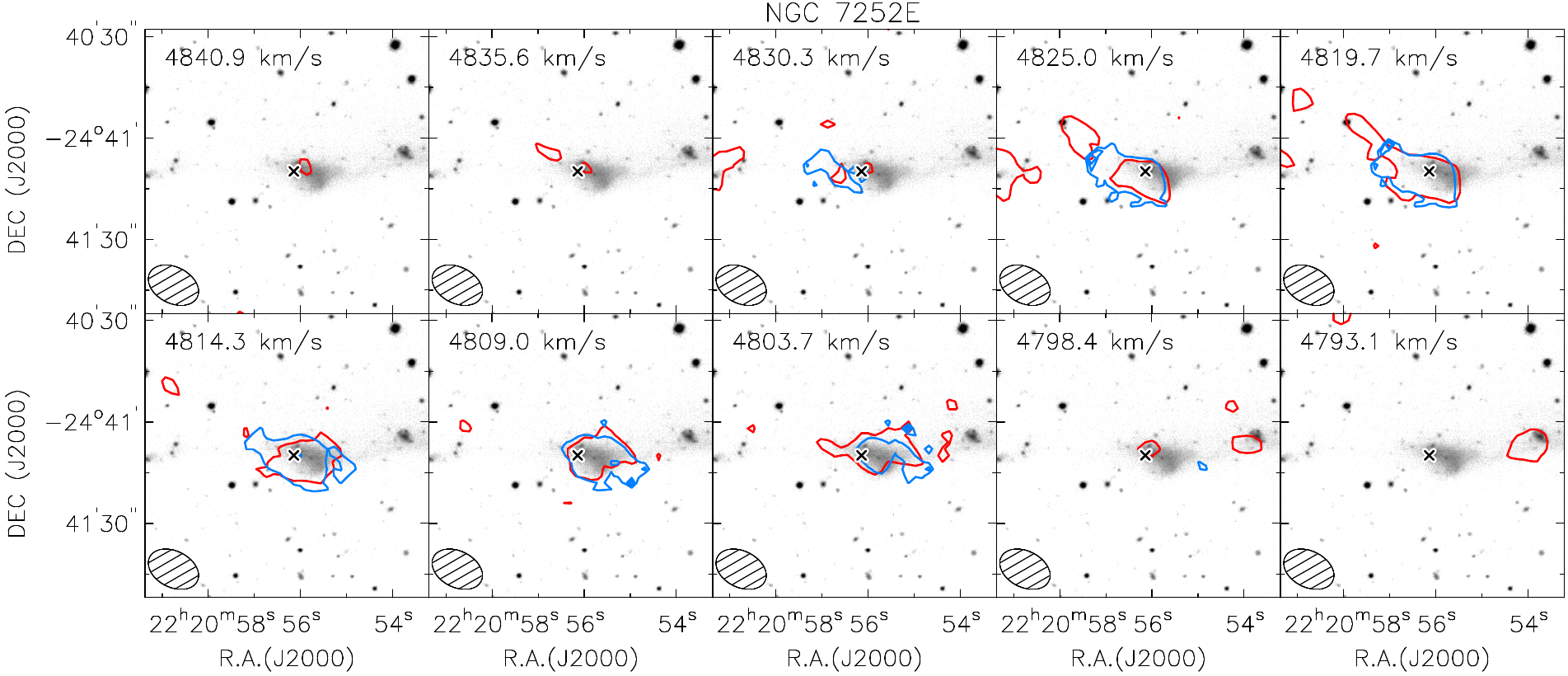}\vspace{0.1cm}
\includegraphics[width=\textwidth]{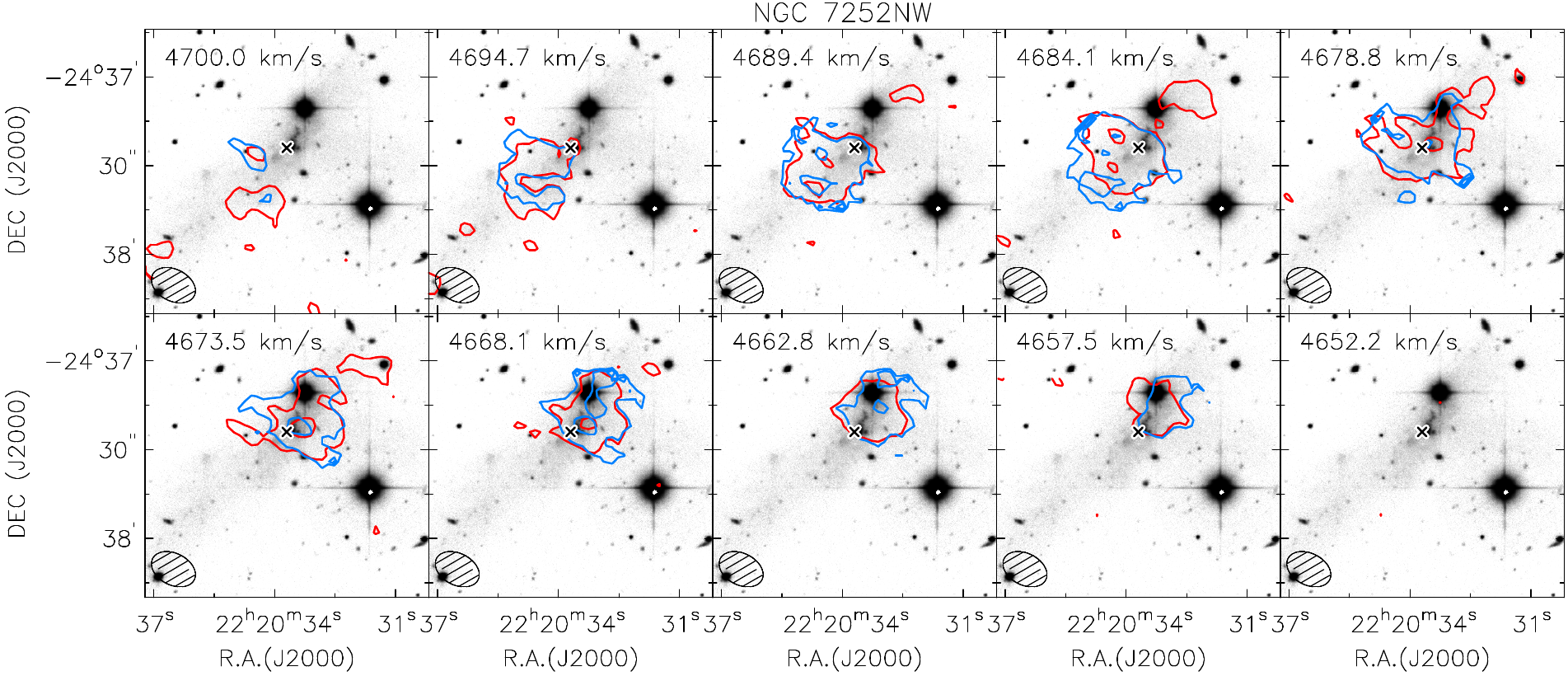}\vspace{0.1cm}
\includegraphics[width=\textwidth]{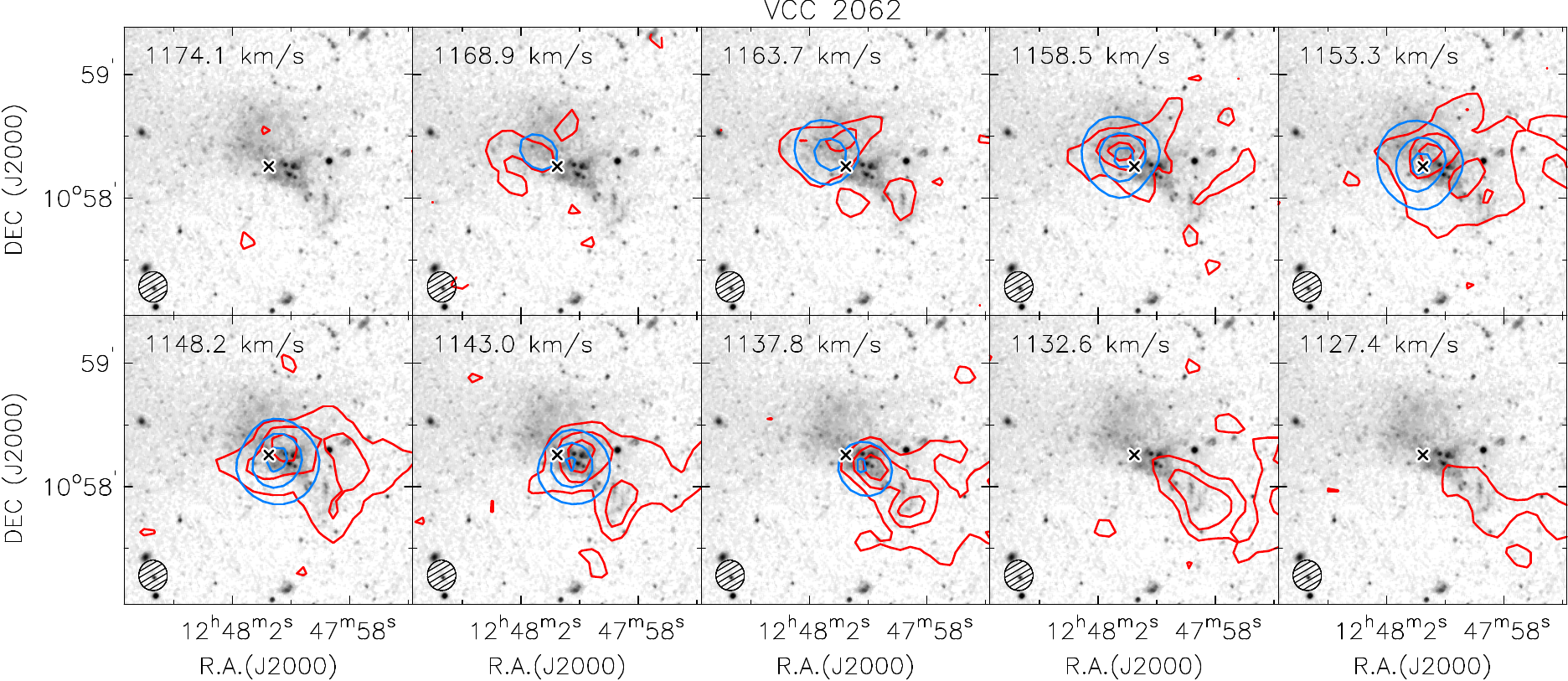}
\caption{\hi channel maps for NGC~7252E (middle), and NGC~7252NW (bottom), and VCC~2062 (top). See Appendix \ref{sec:ChanMaps} for a detailed description of this image.}
\label{fig:CM2}
\end{figure*}

Figures \ref{fig:CM1} and \ref{fig:CM2} show \hi channel maps from the observed cubes (red contours) and model-cubes (blue contours). Contours are at -3$\sigma$ (dashed), 3$\sigma$, 6$\sigma$, and 9$\sigma$. The channel maps are superimposed on an optical image of the TDG. The cross marks the kinematical center. Line-of-sight velocities are indicated to the top-left corner. The \hi beam is shown to the bottom-left corner. The model-cubes are described in detail in Sect.~\ref{sec:3Dmodels}. Here we provide a concise description for each TDG.\\
NGC~5291N is surrounded by \hi emission belonging to the underlying tidal debris. At $V_{\rm l.o.s.}\simeq4145$ to 4100~$\kms$, there is extended \hi emission to the West, which may correspond to another, more uncertain TDG candidate.\\
NGC~5291S is nicely reproduced by our disc model. At approaching velocities ($V_{\rm l.o.s.}\simeq4645-4635$~$\kms$), there is some anomalous gas on the receding side of the disc.\\
NGC~5291SW is a complex case due to the low signal-to-noise ratio. The \hi emission at $V_{\rm l.o.s.}\simeq4730-4720$~$\kms$ does not show a continuous, coherent kinematical structure, hence it has not been considered in the disc model.\\
NGC~7252E is well reproduced by our disc model despite it is a poorly resolved case with low signal-to-noise ratio.\\
NGC~7252NW is closely reproduced by our disc model.\\
VCC~2062 is characterized by two distinct kinematic components, which overlap in both space and velocity (see PV-diagrams in Fig.~\ref{fig:VCC2062}). The second, irregular component extends to the South-West at $V_{\rm l.o.s.}\simeq1155-1125$~$\kms$ and does not show a coherent velocity structure, hence it is not considered in our disc model.

\section{Additional 3D disc models}\label{sec:solidbody}

\begin{figure*}[thb]
\centering
\includegraphics[width=\textwidth]{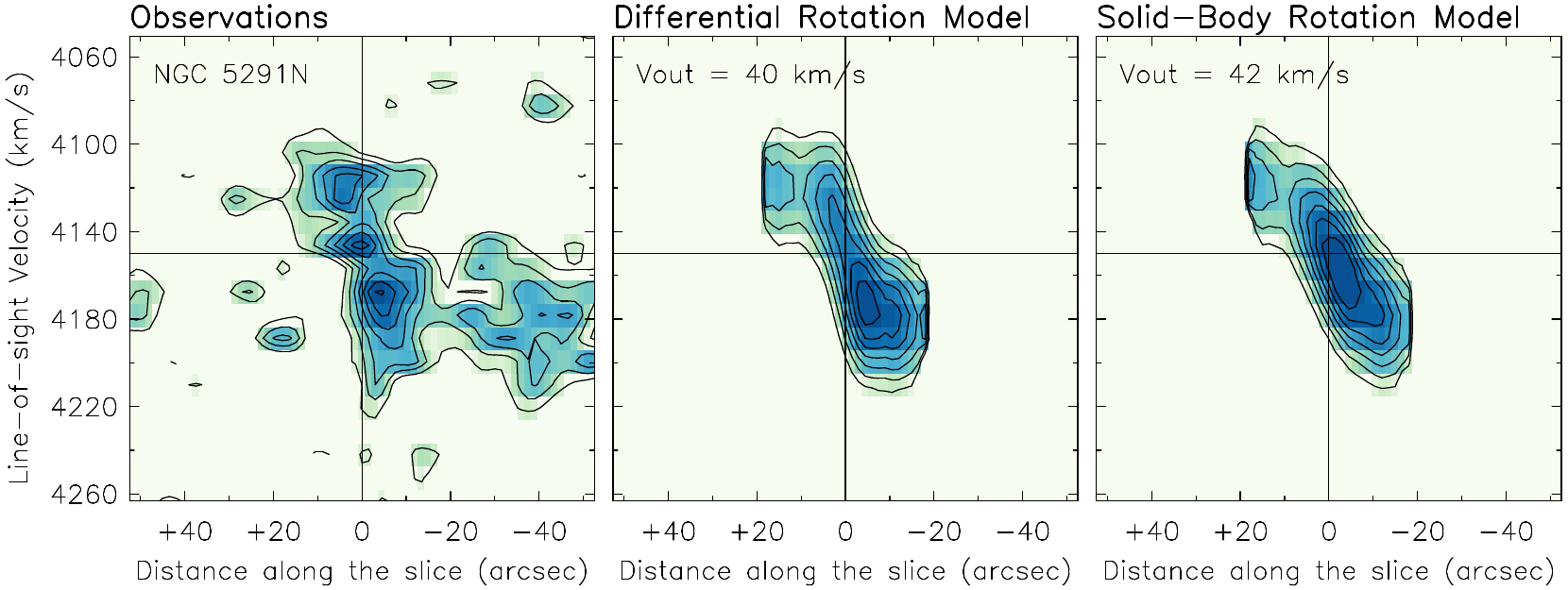}\vspace{0.2cm}
\includegraphics[width=\textwidth]{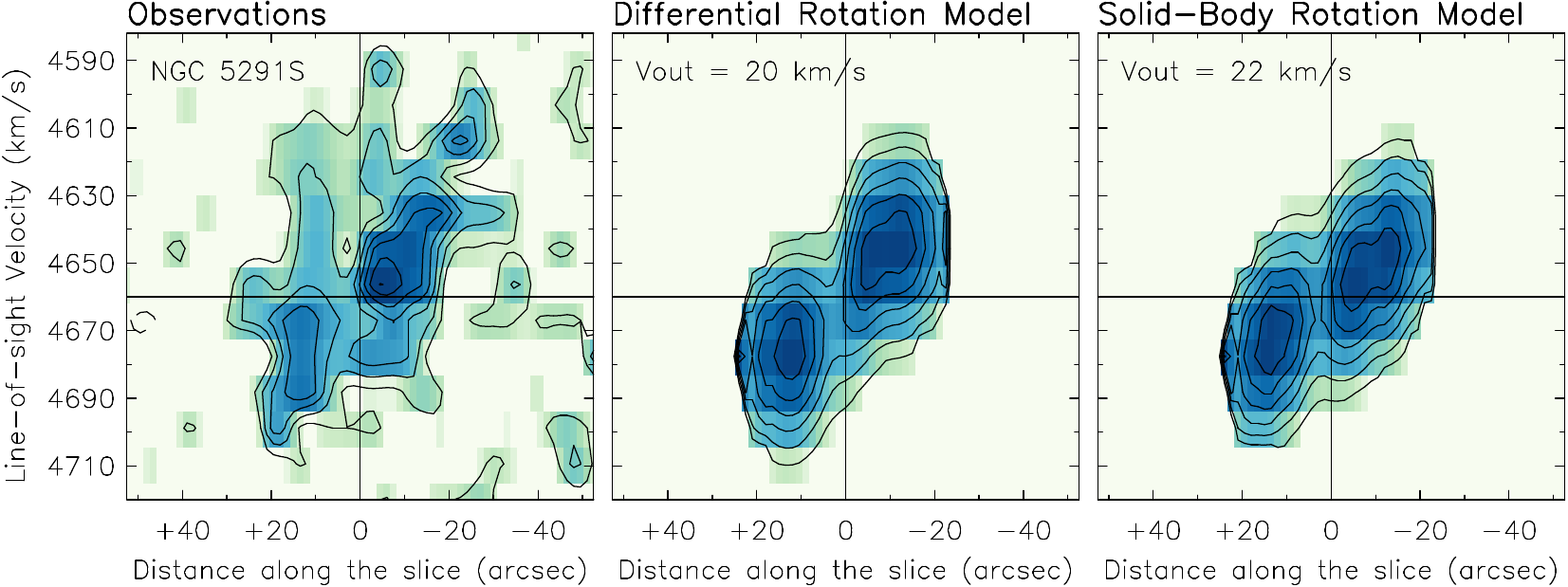}\vspace{0.2cm}
\includegraphics[width=\textwidth]{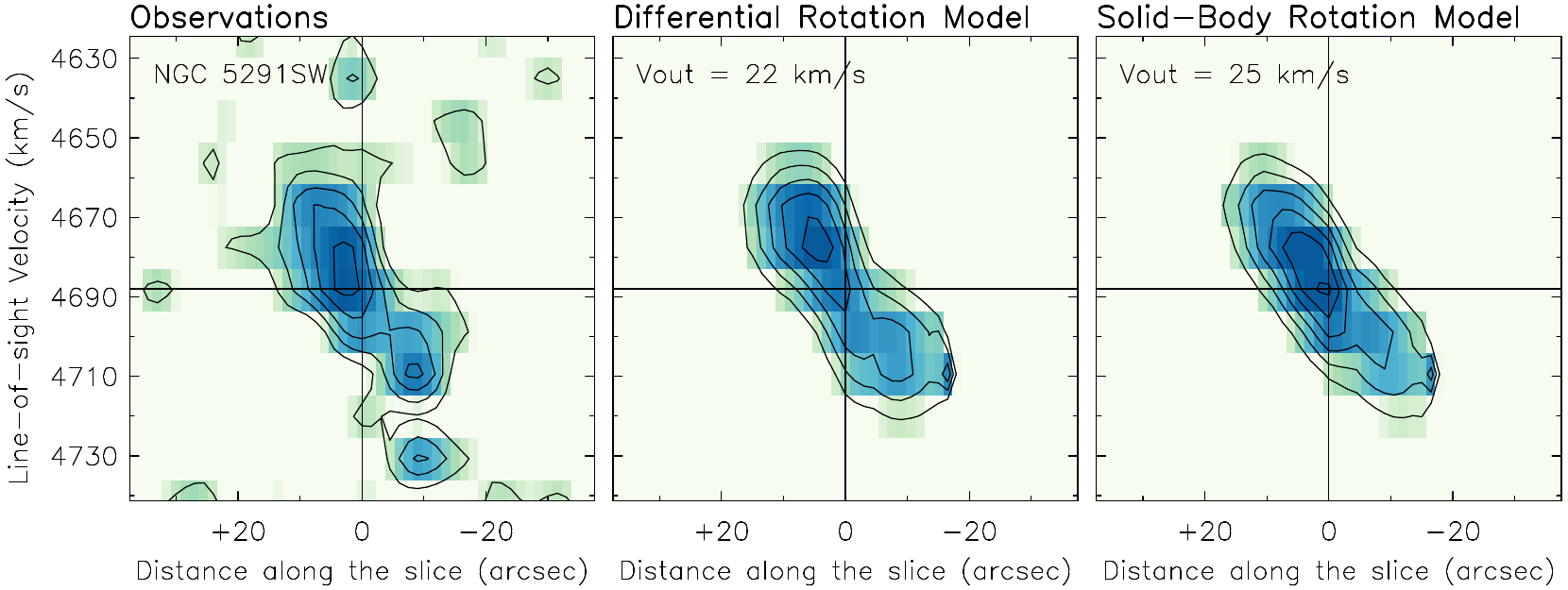}
\caption{Major axis PV-diagrams for NGC~5291N (top), NGC~5291S (middle), and NGC~5291SW (bottom). See Appendix \ref{sec:solidbody} for a detailed description of this image.}
\label{fig:SB1}
\end{figure*}

\begin{figure*}[thb]
\centering
\includegraphics[width=\textwidth]{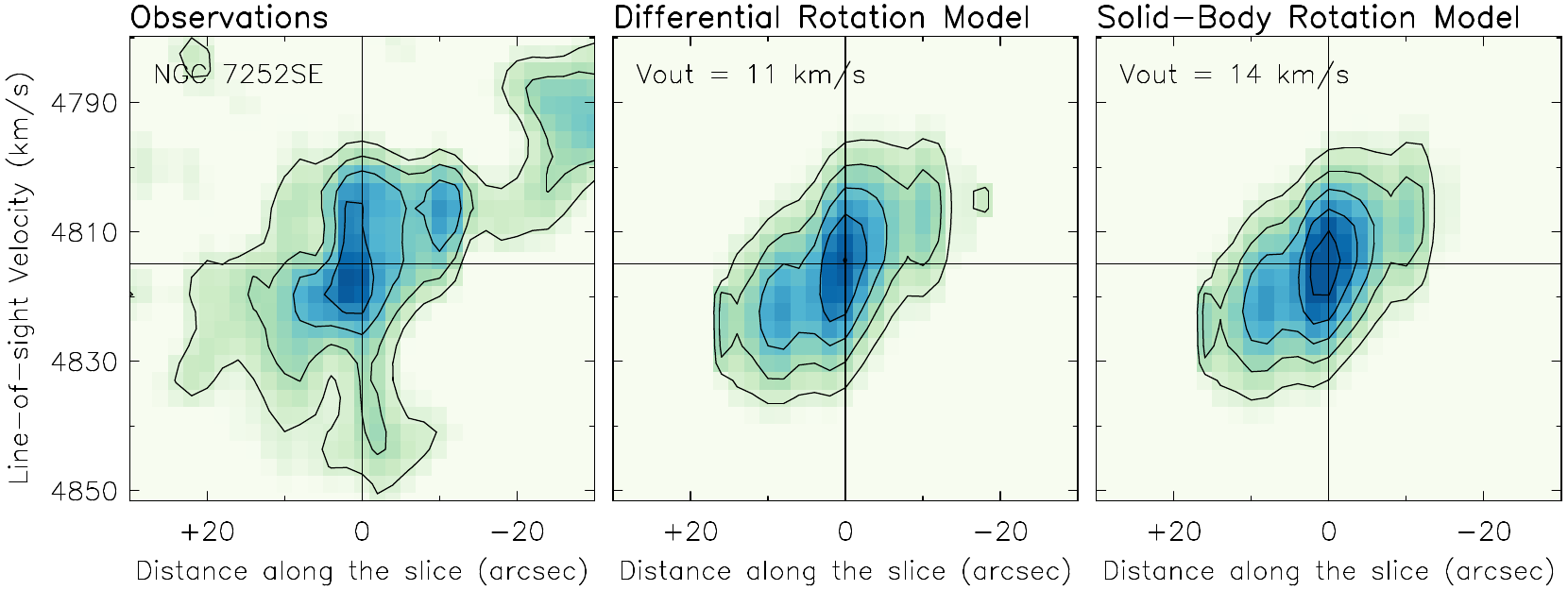}\vspace{0.2cm}
\includegraphics[width=\textwidth]{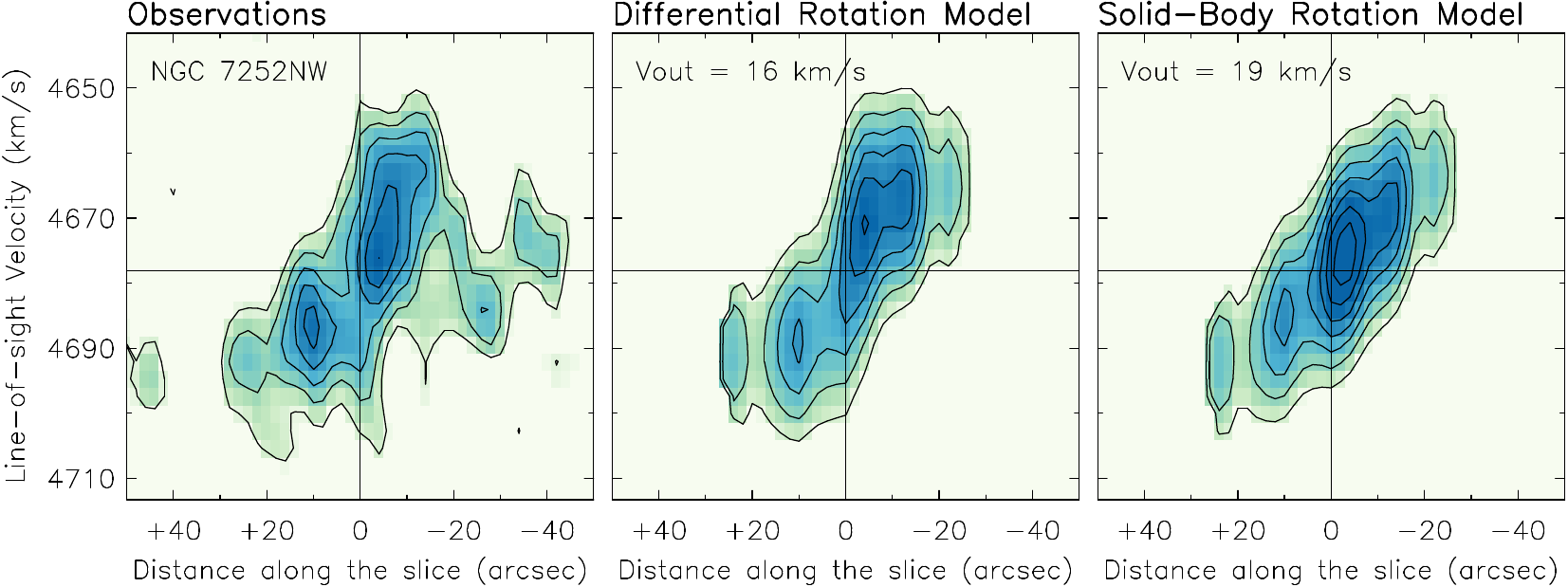}\vspace{0.2cm}
\includegraphics[width=\textwidth]{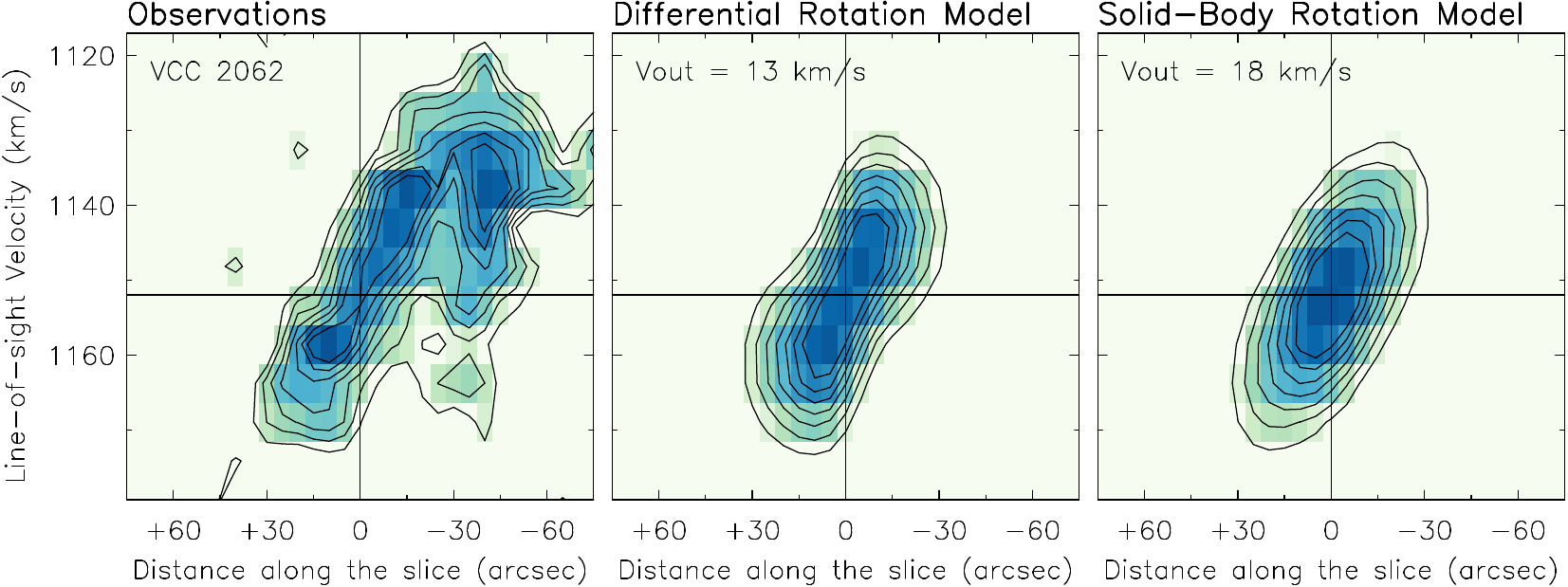}
\caption{Major axis PV-diagrams for NGC~7252E (top), NGC~7252NW (middle), and VCC~2062 (bottom). See Appendix \ref{sec:solidbody} for a detailed description of this image.}
\label{fig:SB2}
\end{figure*}

Figures \ref{fig:SB1} and \ref{fig:SB2} compare 3D disc models based on a flat rotation curve (middle panel) and a solid-body rotation curve (right panel). The models are described in detail in Sect.~\ref{sec:3Dmodels}. The PV-diagrams are obtained along the disc major axis. Solid contours range from 2$\sigma$ to 8$\sigma$ in steps of $1\sigma$. Dashed contours range from $-$2$\sigma$ to $-$4$\sigma$ in steps of $-$1$\sigma$. The horizontal and vertical lines correspond to the systemic velocity and dynamical centre, respectively.

\end{document}